\documentclass[11pt,english]{article}
\usepackage{graphicx}
\usepackage{xcolor}
\usepackage{amstext}
\usepackage{caption}
\usepackage{etoolbox}
\usepackage{makeidx}
\usepackage{amsfonts}
\usepackage{authblk} 
\usepackage{sectsty}
\usepackage{amsmath,amssymb,epsfig}
\usepackage{amscd}
\usepackage{amsthm}
\usepackage{mathrsfs}
\usepackage{dsfont}
\usepackage[applemac]{inputenc}
\usepackage[english]{babel}
\usepackage{enumitem} 
\usepackage[]{latexsym}
\usepackage[labelfont=bf]{caption}
\usepackage{subfigure}
\usepackage{hyperref}
\usepackage{blindtext}
\usepackage{verbatim}
\usepackage{cancel}
\usepackage{cases}
\usepackage{cite}

\hypersetup{
	pdftitle={},%
	pdfauthor={},%
	pdfsubject={},%
	pdfkeywords={},%
	colorlinks=true,%
	linkcolor=blue,%
	citecolor=crimson,%
	linktocpage=true,%
	hyperfootnotes=true,%
	pageanchor=true%
}

\usepackage{scalerel,stackengine}
\stackMath
\newcommand\reallywidehat[1]{%
\savestack{\tmpbox}{\stretchto{%
  \scaleto{%
    \scalerel*[\widthof{\ensuremath{#1}}]{\kern-.6pt\bigwedge\kern-.6pt}%
    {\rule[-\textheight/2]{1ex}{\textheight}}
  }{\textheight}%
}{0.5ex}}%
\stackon[1pt]{#1}{\tmpbox}%
}

\definecolor{crimson}{rgb}{0.7, 0.08, 0.24}

\newcommand{\partdif}[2]{\frac{\partial #1}{\partial #2}}

\newcommand*{\affaddr}[1]{#1}
\newcommand*{\affmark}[1][*]{\textsuperscript{#1}}

\newtheorem*{proof*}{Proof}

\newcommand{\be}{\begin{equation}}

\newcommand{\ee}{\end{equation}}
\def\beqa{\begin{eqnarray}}
\def\eeqa{\end{eqnarray}}
\def\bean{\begin{eqnarray*}}
\def\eean{\end{eqnarray*}}

\newcommand{\N}{\mathbb{N}}

\newcommand{\dd}{\mathrm{d}}

\renewenvironment{thebibliography}[1]
         {\section*{References}\frenchspacing\small
          \begin{list}{[\arabic{enumi}]}
         {\usecounter{enumi}\parsep=2pt\topsep 0pt
         \settowidth{\labelwidth}{[#1]}
         \leftmargin=\labelwidth\advance\leftmargin\labelsep
         \rightmargin=0pt\itemsep=1pt\sloppy}}{\end{list}}

 \numberwithin{equation}{section}

\usepackage[normalem]{ulem}

\newcommand{\Poisson}[2]{\left\{#1 \,,\, #2\right\}}

\newcommand{\fidmetric}[2]{\mathring{q}_{#1#2}}

\newcommand{\ket}[1]{\left| #1 \right\rangle}

\newcommand{\braket}[2]{\left\langle \vphantom {#1 #2} #1 \hphantom{|} \right| \left. \vphantom {#1 #2} #2 \right\rangle}

\newcommand{\braopket}[3]{\left\langle \vphantom {#1 #2 #3} #1 \hphantom{|} \right| #2 \left| \hphantom{|} \vphantom {#1 #2 #3} #3 \right\rangle}

\usepackage{titlesec}
\titleformat{\section}
  {\normalfont\sffamily\Large\bfseries}
  {\thesection}{1em}{}
  \titleformat{\subsection}
  {\normalfont\sffamily\large\bfseries}
  {\thesubsection}{1em}{}
  \titleformat{\subsubsection}
  {\normalfont\sffamily\large\bfseries}
  {\thesubsubsection}{1em}{}

\usepackage{fancyhdr}

\fancypagestyle{plain}{%
\fancyhf{} 
\fancyfoot[C]{\sffamily\fontsize{11pt}{11pt}\selectfont\thepage} 

}
\title{\textbf{\textsf{On the Role of Fiducial Structures in Minisuperspace Reduction and Quantum Fluctuations in LQC}}\vspace{0.35cm}}

\author{
\textsf{Fabio M. Mele\affmark[1,2]\footnote{\texttt{fmele@uwo.ca}} $\,$ and$\,$ Johannes M\"unch\affmark[3]\footnote{\texttt{johannes.muench@cpt.univ-mrs.fr}}}\\
\affaddr{\affmark[1]\textsf{Okinawa Institute of Science and Technology,}}\\
\affaddr{\textsf{1919-1 Tancha, Onna-son, Okinawa 904-0495, Japan}\vspace{0.35cm}}\\
\affaddr{\affmark[2]\textsf{Department of Physics \& Astronomy, Western University,}}\\
\affaddr{\textsf{N6A3K7, London ON, Canada}
\vspace{0.35cm}}\\
\affaddr{\affmark[2]\textsf{Aix-Marseille Universit\'e, Universit\'e de Toulon, CNRS, CPT,}}\\
\affaddr{\textsf{13288 Marseille, France}}
}

\begin{document}

\begin{sffamily}
\maketitle

\vspace{-0.75cm}
\begin{abstract}
\noindent
In spatially non-compact homogeneous minisuperpace models, spatial integrals in the Hamiltonian and symplectic form must be regularised by confining them to a finite volume $V_o$, known as the \emph{fiducial cell}.~As this restriction is unnecessary in the complete field theory before homogeneous reduction, the physical significance of the fiducial cell has been largely debated, especially in the context of (loop) quantum cosmology.~Understanding the role of $V_o$ is in turn essential for assessing the minisuperspace description's validity and its connection to the full theory.~In this work we present a systematic procedure for the field theory reduction to spatially homogeneous and isotropic minisuperspaces within the canonical framework and apply it to both a massive scalar field theory and gravity.~Our strategy consists in implementing spatial homogeneity via second-class constraints for the discrete field modes over a partitioning of the spatial slice into countably many disjoint cells.~The reduced theory's canonical structure is then given by the corresponding Dirac bracket.~Importantly, the latter can only be defined on a finite number of cells homogeneously patched together.~This identifies a finite region, the fiducial cell, whose physical size acquires then a precise meaning already at the classical level as the scale over which homogenenity is imposed.~Additionally, the procedure allows us to track the information lost during homogeneous reduction and how the error depends on $V_o$.~We then move to the quantisation of the classically reduced theories, focusing in particular on the relation between the theories for different $V_o$, and study the implications for statistical moments, quantum fluctuations, and semiclassical states.~In the case of a quantum scalar field, a subsector of the full quantum field theory where the results from the ``first reduced, then quantised" approach can be reproduced is identified and the conditions for this to be a good approximation are also determined.
\end{abstract}
\end{sffamily}

\newpage
\begin{sffamily}
\vspace{-5mm}\centering\rule{\linewidth}{0.3mm} 
\tableofcontents
\vspace{3mm}\centering\rule{\linewidth}{0.3mm}
\end{sffamily}
\newpage

\section{Introduction}

Symmetry-reduced gravitational models provide us with ideal testbeds for (quantum) gravitational theories and their possible predictions for our Universe.~The high energy densities necessary to induce strong quantum gravity effects are reached e.g.~near cosmological or black hole singularities so that understanding symmetry reduced sectors of quantum gravity theories is often enough for highlighting departures from classical general relativity.~In the context of loop quantum gravity (LQG) \cite{ThiemannBook,RovelliQuantumGravity,RovelliBook2}, the application of LQG quantisation techniques to the cosmological context has generated a vast field of research commonly referred to as loop quantum cosmology (LQC), see e.g.~\cite{BojowaldAbsenceOfSingularity,AshtekarMathematicalStructureOf,AshtekarQuantumNatureOf} for seminal papers and~\cite{BojowaldLoopQuantumCosmology,AshtekarLoopquantumcosmology:astatusreport,Agullo:2016tjh,BodendorferAnElementaryIntroduction} for reviews.~Remarkably, the field has matured enough that several points of contact with cosmological observations have been proposed (see e.g.~\cite{Ashtekar:2021kfp} for a recent review addressed to non-experts).~These are however predictions of models within LQC and not direct consequences of full LQG.~Despite the progress made in the quantisation of symmetry-reduced spacetimes, filling the gap between the resulting quantum mini- and midi-superspace models and the full theory remains in fact a thorny open issue (see e.g.~\cite{BojowaldSymmetryReductionFor,KoslowskiReductionofa,EngleQFTandits,EngleRelatingLoopQuantum,EngleEmbeddingLoopQuantum,AlesciQuantum-ReducedLoop,AlesciLoopQuantumCosmology,BodendorferQuantumReductionTo,BodendorferAnEmbeddingOf,BodendorferStateRefinementsAnd,BeetleDiffeomorphismInvariantCosmological,DaporCosmologicaleffectiveHamiltonian,HanEffectivedynamicsfromcoherent,BeetleQuantumisotropyand,Bruno:2024vxx} for a sample of ongoing efforts in connecting LQC and full LQG).~Its resolution is thus of crucial importance for bridging between Planck scale physics and observations, and potentially test the underlying fundamental quantum theory of gravity.~This is a renowned challenging problem where multiple aspects come together and several questions need to be addressed.~Some of these are for example:~What kind of approximations are involved in the symmetry-reduction? In which regimes can these be trusted? Can we identify a proper quantum sector of the full theory and understand the symmetry reduction from the full theory point of view? Would some coarse-graining and renormalisation procedure be required in going from the small scales where QG effects dominate to the the large scales of today's universe?

Given this situation, to keep track of the ingredients and approximations involved in the symmetry-reduction procedure is of key importance, both at the classical and quantum level.~A closely related point in this respect concerns the introduction in symmetry-reduced models with non-compact spatial topology of a \textit{fiducial cell} regularising the otherwise divergent spatial integrals resulting from some mini- or midi-superspace ansatz for the field content of the theory under consideration.~For example, considering a $(D+1)$-dimensional spacetime manifold $\mathcal M$ which, for the purposes of the canonical analysis, is often assumed to be globally hyperbolic $\mathcal M\simeq\Sigma\times\mathbb R$, and denoting the canonical fields of the (classical) field theory under consideration and their conjugate momenta collectively by $\Phi(t,x)$ and $\Pi(t,x)$, the spatially homogeneous theory is typically described by fewer degrees of freedom, say $Q(t)$ and $P(t)$, which only depend on the time coordinate $t$ and whose dynamics is governed by a minisuperspace action of the form of a particle-mechanical system.~Schematically,
\be\label{eq:minisymmred}
\resizebox{.9 \textwidth}{!}
     {%
        $S=\int_{\mathbb R}\dd t\int_{\Sigma}\dd^Dx\bigl[\Pi(t,x)\dot\Phi(t,x)-\mathcal H(\Phi,\Pi)\bigr]\quad\xrightarrow[\text{reduction}]{\text{symmetry}}\quad S=V_o\int_{\mathbb R}\dd t\bigl[P(t)\dot Q(t)-\mathcal H(Q,P)\bigr]$%
     }
\ee

\noindent
where $V_o$ denotes the coordinate volume of a finite region over which the otherwise divergent integral over the non-compact spatial slice $\Sigma$ is restricted.~A similar regularisation will then also occur in the (pre-)symplectic potential and Poisson brackets.~A prototype situation in which this volume regularisation is introduced is homogeneous and isotropic cosmology where the use of a FLRW metric ansatz reduces the Einstein-Hilbert action for general relativity to a ``mechanical system'' whose degrees of freedom, the scale factor and its conjugate momentum, only depend on the time coordinate and all volume integrals like those occurring in the expressions of the action, the Hamiltonian, and the symplectic structure are restricted to a so-called fiducial cell, which due to the symmetries of the model is assumed to be cubical \cite{AshtekarLoopquantumcosmology:astatusreport,Ashtekar:BianchiCosmologies}.

As the restriction to a finite region seems to be only needed to make sense of the symmetry-reduced theory without being a priori required at the full theory level, it is natural to ask whether such an additional structure is just an auxiliary regulator devoid of any physical meaning or not.~This originated two opposite point of views in the LQC community and the interpretation of the fiducial cell has been long debated.~One point of view \cite{AshtekarQuantumNatureOf,AshtekarRobustnessOfKey,Agullo:2016tjh,AshtekarLoopquantumcosmology:astatusreport} is to consider the fiducial cell as an infrared regulator of purely auxiliary nature so that physical results should not depend on its choice.~This might seem reasonable at the classical level, after all the physical results of a \emph{local} classical theory such as the dynamics of observables do not depend on the region $V_o\subset\Sigma$ over which the fields are assumed to be spatially constant\footnote{Here and in the rest of the manuscript we shall refer to both the integration region and its coordinate volume as $V_o$ unless otherwise specified as required from the context to avoid confusion between the two.} which can be then removed by taking the limit $V_o\to\infty$ once the Poisson brackets are evaluated.~More subtle is the situation at the quantum level where \emph{non-local} quantum features such as flutuations and correlations might be affected by the choice of $V_o$.~Expanding on this, studies based on effective quantum field theory techniques \cite{BojowaldCanonicalderivationof,BojowaldMinisuperspacemodelsas,BojowaldTheBKLscenario,BojowaldEffectiveEquationsof,BojowaldEffectiveFieldTheory,BojowaldCriticalEvaluationof,Bojowald:2006ww} have suggested an interpretation for the size of the region $V_o$ not as mere regulator but rather as the infrared scale of perturbative inhomogeneities.~This scale should thus evolve along with the evolution of the universe and is expected to be subject to an infrared renormalisation.~In particular, close to the classical singularity, this scale is expected to become microscopic \cite{BojowaldTheBKLscenario,BojowaldCriticalEvaluationof}.~From this point of view, the limit $V_o \rightarrow \infty$, if possible, should be related to a renormalisation group flow and the size of the fiducial cell acquires physical relevance as a renormalisation scale.~The interpretation of this fiducial structure has in turn profound consequences on the validity of the \textit{effective equations} often exploited in LQC which, by capturing the qualitative behavior of quantum states sharply peaked onto classical trajectories at late times/large $V_o$, aim to incorporate the relevant quantum corrections into a smooth spacetime obeying effective modified dynamics.~The common belief in the literature \cite{TaverasCorrectionstothe,RovelliWhyAreThe,Corichi:2011sd,CorichiCoherentSemiclassicalStates} is that the effective dynamics is applicable and quantum fluctuations are negligible all the way from late to early times, close to the resolved classical singularity, which is at odds with the picture resulting from the second viewpoint mentioned above.

Focusing on the relation between a field theory and its spatially homogeneous minisuperspace reduction, the aim of the present paper is to contribute to the above discussion by developing a systematic construction of the classical symmetry-reduction procedure in the canonical Hamiltonian formalism and thoroughly studying its consequences for the quantisation of the resulting homogeneous theories.~This allows us to explicitly keep track of how the fiducial cell enters the various steps of the symmetry-reduction and of the approximations and assumptions involved, a necessary first step to inquire for the validity of the minisuperspace description.~Our symmetry-reduction procedure can be summarised into the following steps:
\begin{enumerate}
    \item first, we identify the relevant field theory smeared functionals which, for the homogeneous setting we are interested in, are obtained by averaging the canonical fields over a spatial region;
    \item spatial homogeneity is then implemented by introducing second-class constraints, which roughly speaking demand the spatial derivatives of the fields to locally vanish (local Killing equations for the metric in the case of gravity), and can be equivalently translated into sets of mutually commuting constraints for the field modes;
    \item the Hamiltonian and canonical structure of the homogeneous theory can be obtained by constructing the associated Dirac bracket and imposing the constraints strongly;
    \item finally, we compare the physical output of the resulting mechanical mini-superspace model and the full theory by e.g.~looking (if possible) at the dynamics of the non-symmetry-reduced smeared observables and the dynamics generated by the mini-superspace Hamiltonian.
    \end{enumerate}

We first apply this procedure to the simple case of a massive scalar field theory in Minkowski spacetime, and then to general relativity coupled with a real, massless scalar field used as a clock, highlighting the key differences that arise from the presence or absence of a background metric.~In both cases, the construction of the Dirac bracket cannot be achieved via a continuous Fourier decomposition.~We therefore first decompose the spatial hypersurface into the disjoint union of countably many cells and then decompose the fields into discrete Fourier modes in each cell.~In the case of gravity, unlike the scalar field case on Minkowski spacetime where the presence of a background metric allows us to canonically associate dual field modes via $L^2$-pairing, the absence of a fixed background geometry requires us to introduce a further fiducial structure, namely a \emph{fiducial metric} $\fidmetric{a}{b}$ associated with the local coordinate axes along the edges of the cells.~The homogeneity constraints can be thus translated into two sets of mutually commuting second-class constraints for the field modes.~The first kind of constraints demands all modes with non-zero wave-number to vanish, while the second kind sets the zero-modes to be equal across the different boxes.~Importantly, the construction of the Dirac bracket to implement these constraints strongly is possible only for a finite number $d<\infty$ of cells.~As a result, spatial homogeneity can \emph{only} be imposed on the finite region resulting from grouping together $d$ adjacent cells and constraining the field modes therein.~This is the fiducial cell $V_o$ appearing in \eqref{eq:minisymmred} which thus acquires physical meaning already at the classical level as setting the scale over which homogeneity is imposed.~The symmetry-reduction so constructed allows then to pinpoint the approximations and truncations leading to the homogeneuous model.~The latter is in fact the result of a twofold procedure:~first, modes with wavelength smaller than the fiducial cell size $L$ are neglected and the dynamical fields are approximated with their zero modes within the region $V_o$; second, the remaining modes with wavelength larger than the cell size are truncated.~This also reflects into the homogeneous Hamiltonian where the imposition of the homogeneity constraints give rise to boundary terms encoding the interactions between neighbouring cells and which are also neglected by specifying for instance certain boundary conditions.~We further show that the error made in such a truncation is of order $\mathcal O(1/kL)$ and is thus more or less drastic depending on the size of $V_o$.~For gravity, this is measured w.r.t.~the fiducial metric $\fidmetric{a}{b}$ but can be translated into a physical length scale, at least in the homogeneous setting. 

The Dirac bracket and Hamiltonian of the resulting homogeneous classical minisuperspaces depend on the region $V_o$ and have well-defined scaling behaviours under an active rescaling of it.~Specifically, the Dirac bracket of the homogeneous zero modes scales with the inverse of the volume of $V_o$ and agrees with the full theory Poisson brackets of the volume averaged fields.~The Hamiltonian scales instead linearly with $V_o$.~This ensures the equations of motion and dynamics of classical observables to be independent of $V_o$ as expected for a local classical theory.~Nevertheless, the set of full theory quantities captured by the homogeneous description depends on the smearing region.~To judge how accurate such a description is then requires us to compare the resulting dynamics with the full field theory.~This can be done explicitly for the case of a scalar field of mass $m$ and it turns out that the homogeneous dynamics computed via the Dirac bracket differs from the full theory only by a surface term due to the contribution of inhomogeneities and can be neglected for $L\gg1/m$.~In the gravitational case, the non-linear coupled nature of the dynamical equations prevents us from a straightforward comparison with the full theory dynamics.~Nevertheless, our framework allows us to account for the inclusion of the first small momentum/large wavelength inhomogeneous modes providing some hint to them contributing to dynamics. 

The $V_o$-dependence of the Dirac bracket has immediate consequences also for the quantisation of the homogeneous theory.~In fact, an active transformation of the fiducial cell is not a canonical transformation and the homogeneous minisuperspace model consists then of an entire family of canonically inequivalent theories, each identified by the different region over which homogeneity has been imposed.~Correspondingly, the quantum representation of the elementary operators and the canonical commutation relations come to be $V_o$-dependent.~This ensures the correct scaling behaviour of the quantum commutators and is consistent with the fact that, even reabsorbing the $V_o$ factors via a redefinition of the canonical variables to make the classical bracket insensitive to the choice of the fiducial metric and passive coordinate rescaling of the cell, the canonical structure of the classically reduced theory \emph{does} scale under an active change of its physical volume \cite{AshtekarLoopquantumcosmology:astatusreport}.~Different $V_o$ correspond then to different, yet isomorphic Hilbert spaces carrying the representation of the quantum operators which eventually exhibit different scaling properties compared to their classical counterparts.~As outlined for LQC in our previous work \cite{MeleThePhysicalRelavance}, an explicit isomorphism between the different Hilbert spaces can be constructed as a mapping between states with equivalent dynamics and provides us with a quantum implementation of an active cell rescaling.~In the present work, we detail its construction both for the case of a real, massive (free and interacting) scalar field minisuperspace theory in the standard Schr\"odinger quantisation, and for the polymer quantisation of homogeneous and isotropic cosmology.~Interestingly, in the interacting theory, the volume of the fiducial cell combines with the coupling of the interactions.~Moreover, the realisation of the above dynamics-preserving isomorphism requires the mass and the coupling constant to scale with $V_o$ thus suggesting a possible role for it in renormalisation and resonating with the above mentioned literature based on effective field theory techniques.

We then analyse the consequences of the proposed isomorphism between the quantum theories corresponding to different $V_o$ for expectation values, statistical moments, uncertainty relations, and coherent states.~In particular, states saturating the uncertainties in the theory for a given $V_o$ are mapped into states saturating them also in the theory with a different $V_o$ provided that the classical value on which they are peaked and their width scale accordingly.~Moreover, for generic states, we observe that as a result of the non-local nature of the quantum theory, the quantum fluctuations of the smeared operators over a region $V\subset V_o$ depend on the ratio $\frac{V}{V_o}$, that is on the (inverse) number of subcells $V$ homogeneously patched together into $V_o$.~As homogeneity can only be imposed over finite regions $V_o$, this indicates that for sufficiently small regions quantum fluctuations are not negligible.~This nicely aligns with recent work on dynamical symmetries in gravitational minisuperspace models where the fiducial volume $V_o$, or more precisely its ratio with the Planck volume, turns out to be related to the central charge of the Schr\"odinger symmetry algebra of such systems \cite{BenAchourSchroedingerSymmetry}.~In analogy with the role played by the average number of microscopic constituents in the hydrodynamic description of quantum many-body systems sharing the same symmetry, it was then suggested that $V_o$ should in fact play a role in setting the scale for how classical or quantum is the system \cite{BenAchourSchroedingerSymmetry}.

The rest of the paper is organised as follows.~Sec.~\ref{Sec:Warmupscalarfield} and \ref{Sec:homogscalarfieldquantisation} are respectively devoted to the classical symmetry-reduction of a scalar field theory and the canonical quantisation of the resulting (family of) homogeneous theories.~In Sec.~\ref{sec:symredscalarQFT}, we discuss a quantum reduction procedure for the massive scalar quantum field theory and identify a subsector of the full QFT where the results and scaling behaviours determined in the ``first reduced, then quantised'' $V_o$-labeled quantum minisuperspace theories are reproduced.~The conditions for this to hold true with good approximation are also spelled out and agree with the classical results obtained by comparing the classically symmetry-reduced and the full theory dynamics.~In Sec.~\ref{Sec:Comsology}, we move then to present the classical set-up used for the gravitational case with a minimally coupled real massless scalar field and the implementation of the symmetry-reduction constraints.~The quantisation of the resulting homogeneous and isotropic cosmological theories is discussed in Sec.~\ref{Sec:QuantumCosmology} where also the consequences for the uncertainty relations and quantum fluctuations are analysed.~The inclusion of the first inhomogeneous modes is reported in Sec.~\ref{outlook:beyondhom}.~We close then in Sec.~\ref{sec:conclusion} with a summary and discussion of the results, and an outlook on future directions.~Two appendices complement the main body of the paper and respectively contain the details of the mode decomposition (App.~\ref{app:modedecomposition}) and the computation of the scaling properties of semiclassical states in LQC (App.~\ref{app:CS}).

\section{Warm-up: Symmetry Reduction for Scalar Field Theory}\label{Sec:Warmupscalarfield}

To get a better intuition and develop a systematic strategy to symmetry reduce and pose questions about the fiducial cell, in this section we start by analysing a real scalar field theory as a toy model example.~The reader interested in the discussion of cosmology can in principle skip this and the next two sections to move directly to Sec.~\ref{Sec:Comsology}.~However, for pedagogical reasons and to illustrate in the simplest way possible the main steps for implementing spatial homogeneity within the canonical framework of the field theory under consideration and its consequences for the resulting field mode decomposition without diving yet into the subtleties posed by the absence of a background geometry in the gravitational setting, we prefer to first discuss in details the symmetry-reduction procedure for the simpler case of a free massive real scalar field theory in a given background spacetime.~For concreteness and simplicity, the latter is chosen to be a 4-dimensional Minkowski spacetime.~Such a simple example will also allow us to discuss, on the one hand, possible extensions of the strategy developed in this paper to the interacting case (Sec.~\ref{Sec:homogscalarfieldquantisation}) and, on the other hand, the implementation of the symmetry-reduction directly at the quantum field theory level rather than first symmetry-reducing the classical theory and then quantising the resulting spatially homogeneous theory, thus allowing us to perform a comparison between the two strategies (Sec.~\ref{sec:symredscalarQFT}).  

\subsection{Full Theory and Observables}\label{sec:scalarfieldsetup}

Let us consider a real scalar field theory on a fixed Minkowski background spacetime, whose Hamiltonian is given by
\begin{equation}\label{eq:scalarfieldHamiltonian}
	H = \int_{\Sigma_t} \dd^3 x \frac{1}{2}\left(\pi(x)^2 + \partial_a \phi(x) \partial^a \phi(x) + m^2 \phi(x)^2\right) \;,
\end{equation}

\noindent
where $\Sigma_t$ is a flat Cauchy surface, $\partial_a = \partdif{}{x^a}$, $a = 1,2,3$ the spatial derivatives, and $\pi= \dot{\phi}$ the canonical momentum conjugate to the field $\phi$, with dots denoting time derivatives.
The Poisson structure is defined by the only non-trivial equal time brackets
\begin{equation}
	\Poisson{\phi(x)}{\pi(y)} = \delta(x-y)\;,
\end{equation}

\noindent
where $\delta(x-y)$ is the Dirac-$\delta$-distribution.
Consequently, the equations of motion are given by
\begin{subequations}\label{eq:FTEoM}
	\begin{align}
		\dot{\phi}(x) =& \Poisson{\phi(x)}{H} = \pi(x)\;,
		\\
		\dot{\pi}(x) =& \Poisson{\pi(x)}{H} = \partial_a \partial^a \phi(x) -m^2 \phi(x) \;,
	\end{align}
\end{subequations}

\noindent
which can be combined to give the well-known Klein-Gordon equation.~Let us further note that $\phi$ is a spacetime scalar (desnity weight 0) and thus $\pi$ is a scalar density of weight 1.~This is not important for this simple system on a flat background, but it is worth keeping track of the density weight properties already at this stage for the later purposes of generalisation to gravity.

Before symmetry reducing to a fully homogeneous system, let us first discuss the observables of the theory.~As this is a field theory, observables are only regular if they are smeared against certain test functions.\footnote{Tipically, for a scalar field in Minkowski spacetime, these are Schwartz functions.~These functions prevent divergencies when integrating on non-compact spatial slices and allow to identify the appropriate quantum configuration space for Schr\"odinger functional quantization of the Klein-Gordon field as the space of tempered distributions on $\Sigma_t$ \cite{GlimmJaffebook,ReedBook1}.}
Denoting such a test function by $f$, we thus define the smeared fields 
\begin{equation}
	\phi[f;t] = \int_{\Sigma_t} \dd^3x f(x) \phi(t,x) \qquad , \qquad \pi[f;t] = \int_{\Sigma_t} \dd^3x f(x) \pi(t,x)\;,
\end{equation}

\noindent
where we note that, given the different density weight of $\phi$ and $\pi$, $f$ is a scalar density of weight 1 for (the smearing of) $\phi$, while it is a scalar 0-density for $\pi$.~Consequently, we have the regular equal time Poisson brackets
\begin{equation}
	\Poisson{\phi[f;t]}{\pi[g;t]} = \int_{\Sigma_t}\dd^3 x f(x) g(x) =: \left< f, g \right>\;.
\end{equation}

\noindent
If we were interested only in the homogeneous degrees of freedom, we could probe them by using as test function $f$ the characteristic function of a certain region $V \subset \Sigma_t$, defined by
\begin{equation}
	\chi_V(x) = \begin{cases}
		1 ,& x \in V\\
		0 ,& \text{else}
	\end{cases} \;.
\end{equation}

\noindent
The averaged field observables then read as
\begin{equation}\label{eq:smearedphipi}
	\phi(V) = \frac{1}{\text{vol}(V)} \int_V \dd^3x\, \phi(x) \qquad , \qquad \pi(V) = \int_V\dd^3x\, \pi(x) \qquad \text{with} \qquad \text{vol}(V) = \int_V\dd^3x \,.
\end{equation}

\noindent
Note that here the different density weight properties of $\phi$ and $\pi$ as respectively being a density 0 and 1 object becomes evident.
Indeed, proper spatial averaging of the local scalar quantity $\phi(x)$ requires a division by the volume $\text{vol}(V)$ of the region $V$ over which it is averaged.
The smeared observable $\phi(V)$ is thus intensive.
In contrast, $\pi(x)$ is a density 1 object and can naturally be integrated over a volume without averaging.\footnote{This becomes evident from the discussion of the mode decomposition in Appendix~\ref{app:modedecomposition}, which will be important in the next section. Choosing as modes simply $f_V = \chi_V$, we can naturally define $\pi(V) := \pi\left[f_V\right]$, which is well-defined as $\pi(x)$ is a density and thus can be integrated over a volume. The dual-functional is then defined as $F_V[g]:= N\int_{\Sigma_t} \dd^3x \,\sqrt{q}\, \chi_V(x) g(x)$, which requires knowledge of the background metric in order to be explicitly written. The normalisation constant $N$ can be chosen to be $1/\text{vol}(V)$ such that $F_V\left[f_V\right] = 1$. Then we have the natural definition $\phi(V) := F_V\left[\phi\right]$. However, we could e.g.~define the functions $f_V$ with an additional factor $1/\text{vol}(V)$ and choose $N=1$, which is mathematically equivalent, but changes the extensive/intensive properties.}
The density property is then reflected by the fact that $\pi(V)$ is extensive, i.e. $\pi(V_1) + \pi(V_2) = \pi\left(V_1 \cup V_2\right)$ for any two disjoint regions $V_1, V_2\subset\Sigma_t$. 
As it will be discussed in details in the following sections, after imposing spatial homogeneity over a certain region $V_o$, the above smeared fields reduce to $\phi(V_o)=\phi(x)|_{x\in V_o}$ and $\pi(V_o)=\text{vol}(V_o)\pi(x)|_{x\in V_o}$ (see Eqs. \eqref{reducedandfullphi}, \eqref{reducedandfullpi}, and \eqref{eq:FTobssymred} below).
Therefore, the above-mentioned intensive and extensive nature of the (smeared) configuration field and its conjugate momentum can be thought of as the full theory counterpart of the scaling behaviours at the symmetry reduced level according to which, under a rescaling $\text{vol}(V_o)\mapsto\beta\text{vol}(V_o)$ with $\beta\in\mathbb R$, we have $\phi\mapsto\phi$ and $\pi\mapsto\beta\pi$.\footnote{In the cosmology case, where the configuration field is given by the spatial volume $v$ itself and its conjugate momentum $b$ is (related to) the Hubble rate, the situation would be exactly the opposite in terms of extensive and intensive scaling properties. We will come back on this point later in Sec.~\ref{Sec:gravityDBtruncatedtheory} where such considerations will lead to the correct scaling behaviours as expected from LQC literature \cite{BojowaldLoopQuantumCosmology,AshtekarLoopquantumcosmology:astatusreport,BodendorferAnElementaryIntroduction}.}

\noindent
The Poisson bracket of the smeared fields \eqref{eq:smearedphipi} reads then as
\begin{equation}\label{eq:smearedPB}
	\Poisson{\phi(V)}{\pi(V')} = \frac{\text{vol}\left(V \cap V'\right)}{\text{vol}\left(V\right)} \;.
\end{equation}

\noindent
We will recover these relations later on at the symmetry reduced level.
Though physically obvious, it is important to note that Eq. \eqref{eq:smearedPB} only yields a non-trivial Poisson bracket when the two volumes actually intersect.
In particular, the following limiting cases are of interest.
First of all, for any given finite region $V'$, if $V$ is assumed to become larger and larger, it will completely enclose $V'$ at some point so that
\begin{equation}\label{eq:smearedPBinf}
	\Poisson{\phi(V)}{\pi(V')} \stackrel{V' \subset V}{=} \frac{\text{vol}\left(V'\right)}{\text{vol}\left(V\right)} \xrightarrow{\text{vol}(V) \rightarrow \infty} 0\;.
\end{equation}

\noindent
For large volumes $V$ correlations with $\pi(V')$ become negligible.
This is physically plausible as $\phi$ and $\pi$ have only non-trivial Poisson brackets on a finite volume $V'$, which is negligible when it is averaged over the whole spatial slice.
On the other hand, letting $V'$ grow, the Poisson bracket \eqref{eq:smearedPB} becomes simply $1$ as soon as $V \subset V'$ independently of how large $V'$ is chosen.
Finally, for $V = V'$ this bracket is simply $1$ and $\phi(V)$ and $\pi(V)$ are canonically conjugate.

As a last point, we can even track the dynamics of these averaged observables.
Assuming the volume we want to track remains fixed over time, we find the equation
\begin{equation}\label{eq:smeareddynamics}
	\frac{\dd^2}{\dd t^2}\phi(V) = -m^2 \phi(V) + \frac{1}{\text{vol}(V)} \int_{\partial V} \dd S \,n_a \partial^a\phi(x) \,,
\end{equation} 

\noindent
where $n_a$ is the unit normal on the boundary $\partial V$ and $\dd S$ the induced surface element.~All spatial dependencies are smeared out and interactions are introduced via boundary terms.
Note that this is only possible as there are no interaction terms, e.g. $\propto \phi^4$ involved, in which case Eq.~\eqref{eq:smeareddynamics} would still be dependent on local degrees of freedom.
Further, the boundary terms and in fact them being neglected play an important role for the validity of the symmetry-reduced theory.
In the following, we focus on the symmetry reduction by implementing spatial homogeneity constraints and we will come back on boundary terms in the following sections while reserving further comments on their possible role beyond the homogeneous setting to the outlook of the concluding section \ref{sec:conclusion} once gravity and cosmology are also discussed.

\subsection{Homogeneity: Constraints and Implementation}\label{sec:implementinghom}\label{sec:constraintimplementation}

In the following, we are interested in the symmetry-reduced scalar field theory containing only the spatially homogeneous fields.~As we will show in this section, such a symmetry-reduced theory can be obtained from the full theory by imposing spatial homogeneity through the implementation of suitable constraints following Dirac's constraint theory \cite{DiracGeneralizedHamiltonian,DiracLecturesonquantum}.~This is a brute force modification of the original theory to throw out all spatial dependencies.~It is in fact only an approximation and, as it was already pointed out in the late '80s and early '90s (see e.g. \cite{KucharRyanProceeding,KucharIsMinisuperspace,SinhaValidityofthe}), it violates the uncertainty principle as it will be clear later in modes representation where spatial homogeneity amounts to equally specifying all space-dependent non-zero modes to vanish for both configuration fields and their conjugate momenta.~Such an approximation is therefore valid as long as the backreaction of higher modes can be neglected \cite{KucharIsMinisuperspace,SinhaValidityofthe} (see also \cite{Hu:1993cg,Calzetta:1999zr} and more recently \cite{Brahma:2021mng} where the homogeneous modes on which the minisuperspace approximation is based are regarded as forming an open quantum system and the validity of the minisuperspace description can be investigated by studying the correlations and decoherence induced by the environmental inhomogeneous modes).

The first step to implement spatial homogeneity in our scalar field theory is then to introduce the following constraints on the configuration field
\begin{equation}\label{eq:phiconstraint}
	\Psi^\phi_a (x)= \partial_a \phi(x) \approx 0 \;,
\end{equation}

\noindent
with $a$ running over the spatial directions.
From the stability algorithm $\dot \Psi^\phi_a(x) = \{\Psi^\phi_a(x),H\} = \partial_a \pi(x) \approx 0$ follows the second set of homogeneity constraints for the conjugate momentum field
\begin{equation}\label{eq:piconstraint}
	\Psi^\pi_a (x)= \partial_a \pi(x) \approx 0 \;.
\end{equation}

\noindent
As it can be easily checked by direct computation, the two sets of constraints \eqref{eq:phiconstraint}, \eqref{eq:piconstraint} are second-class as they do not Poisson-commute among themselves
\begin{equation}\label{eq:secondclassconstr}
	\Poisson{\Psi^\phi_a(x)}{\Psi^\pi_b(y)} = \partdif{}{x^a} \partdif{}{y^b} \delta(x-y) \neq 0 \;.
\end{equation}

\noindent
No further constraints arise from stability of \eqref{eq:piconstraint} and the constraint algorithm terminates then with the total Hamiltonian $H_T = H$.

The strategy to obtain the homogeneous theory is to implement the above second-class constraints strongly by using the Dirac bracket.
To this aim, we decompose the fields in modes, with the zero mode being spatially constant.\footnote{This gives a systematic way of going beyond the homogeneous approximation by including space-dependent modes into the canonical analysis. As our main interest in this work is the application to cosmology, we will come back on the discussion of higher modes in Sec.~\ref{outlook:beyondhom} where a strategy to go beyond homogeneous cosmology is outlined.} This is necessary as for the construction of the Dirac bracket we have to invert the matrix $C_{(\phi,a),(\pi,b)} = \{\Psi^\phi_a(x),\Psi^\pi_b(y)\}$ generated by Eq.~\eqref{eq:secondclassconstr}, which is a differential operator. The details of the mode decomposition, especially its rephrasing in a language suitable for a background independent framework as needed for gravity, are worked out in Appendix~\ref{app:modedecomposition}.
For the scalar field case under consideration, most of such details would technically overload the discussion as the background metric is fixed and very simple.

Before defining the decomposition, let us emphasise a general point.~Imposing the constraints is completely independent of the specific mode decomposition employed, though it might be simpler in a certain set than another.~In Fourier modes, for instance, spatial homogeneity simply amounts to keep the $\vec k=\mathbf 0$ modes only, while using e.g.~Hermite polynomials will be way more complicated.~This is due to the fact that the constraint matrix $C_{(\phi,a),(\pi,b)}$ still has to be diagonalised in the given mode decomposition.~Of course, the full theory is also completely independent of the choice of modes as summing up all of them gives back the canonical fields without loss of information.~The only thing really dependent on the modes is a truncation of the mode expansion.~This is completely plausible as this is now a physical requirement.~The particular choice of modes should then be physically well motivated as keeping Fourier modes with $k < k_{max}$ is a physically different scenario than keeping Hermite polynomials with $n< n_{max}$.~Keeping only Fourier modes with $\vec k=\mathbf 0$ leads to a homogeneous setting, while keeping only the $n = 1$ Hermite polynomial is inhomogeneous.~Nevertheless, there might be physical scenarios where one would like to allow inhomogeneity, e.g. by keeping modes with $k < k_{max}$ (see Sec.~\ref{outlook:beyondhom}) or simply a very particular form of the scalar field in which the truncation of Hermite polynomials would be a well-suited approximation.~However, here we focus on spatial homogeneity and leave a more detailed discussion of inhomogeneities and different mode decompositions for future research.~With this being said, we can now move to the actual imposition of the constraints \eqref{eq:phiconstraint} and \eqref{eq:piconstraint}.

As anticipated above, a convenient choice of mode decomposition for the scalar field case would be Fourier modes.
However, although a Fourier decomposition would allow us to solve the Dirac bracket, a continuous Fourier transform of the fields turns out to give a purely trivial result.
Therefore, we will still use Fourier modes decompositions, but the discrete ones.
This requires some additional work to be done at first.
\begin{figure}[t!]
	\centering
	\includegraphics[scale=0.375]{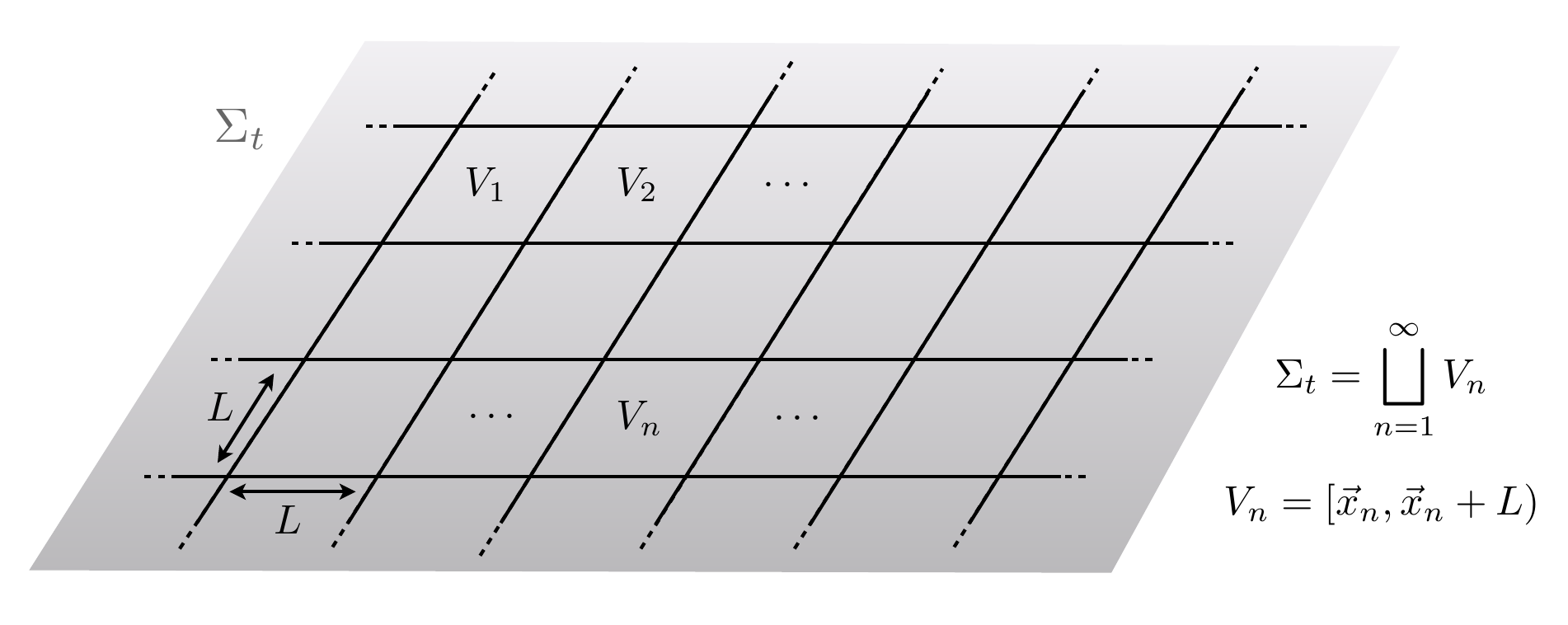}
	\caption{Partitioning the non-compact spatial slice $\Sigma_t$ into countably many disjoint identical cells $V_n = [\vec{x}_n, \vec{x}_n+L)$ of edge length $L$ with vector $\vec{x}_n$ pointing along the edges of the $n$-th cell. Such a partitioning allows us to construct the Dirac bracket for discrete Fourier modes of the dynamical fields and implement the second-class homogeneity constraints \eqref{eq:phiconstraint}, \eqref{eq:piconstraint} strongly.}
	\label{fig:celldecomp}
\end{figure}
As schematically depicted in Fig.~\ref{fig:celldecomp}, we partition the spatial slice $\Sigma_t$ into infinitely many identical boxes $V_n = [\vec{x}_n, \vec{x}_n+L)$ of edge length $L$\footnote{This is a purely topological construction at first. As the background metric is fixed, this coordinate length $L$ can be related to the physical edge length of the box measured w.r.t. the background metric, which is again $L$.}, and vector $\vec{x}_n \in \Sigma_t$ pointing to the edge of the box.
This decomposition has to be such that the boxes are disjoint, i.e. $V_n \cap V_m = \emptyset$ $\forall \,n,m$, and they cover the whole spatial slice, namely $\bigsqcup_n V_n = \Sigma_t$.
The total number of regions $V_n$ will be countably infinite as long as $\Sigma_t$ is assumed to be non-compact.
The fields (for simplicity and avoiding repeated equations, we restrict to $\phi$ here) can be decomposed as 
\begin{equation}\label{eq:phiboxes}
	\phi(x) = \sum_n \chi_{V_n}(x) \phi^n(x) \;.
\end{equation}

\noindent
Note that there is no information lost in the above decomposition.
On the contrary, there is redundant information in each $\phi^n(x)$ as only the part supported on $V_n$ contributes to $\phi$.
We can get rid of this redundancy by choosing $\phi^n(x \notin V_n) = 0$ and for instance demanding periodic boundary conditions or any other requirement.
In the following, we demand the $\phi^n$ to satisfy periodic boundary conditions.\footnote{\label{ftnte:aeequality}There is no problem with this even when $\phi$ does not satisfy these conditions. The function $\phi^n$ can be anything within $V_n$ and then can discontinuously jump to the correct boundary value. Therefore, even if we had chosen the boxes $V_n$ to be compact instead of half-opened, we have $\left.\phi\right|_{V_n} =_{a.e.} \phi^n$ (with the subscript a.e. standing for ``almost everywhere'').}
In this case, we can perform a discrete Fourier decomposition of each $\phi^n$, leading to
\be\label{eq:expansion}
		\phi^n(x) = \sum_{\vec k \in \mathbb{Z}^3} \tilde{\phi}^n_{\vec k} e^{+i\frac{2 \pi}{L} \vec k\cdot \vec x} \qquad,\qquad
		\tilde{\phi}_{\vec k}^n = \frac{1}{L^3} \int_{V_n} \dd^3 x\, \phi(x) e^{-i\frac{2 \pi}{L} \vec{k}\cdot \vec{x}}\,,
	\ee
	
\noindent
where $\vec k\cdot \vec x = k_1 x^1 + k_2 x^2 + k_3 x^3$.~In contrast to the continuous Fourier decomposition, this is now a countable number of modes.
For completeness, for the momentum $\pi$ conjugate to $\phi$, we can write
\begin{subequations}\label{eq:piexpansion}
	\begin{align}
		\pi(x) =&\; \sum_n \chi_{V_n}(x) \pi^n(x) \qquad , \qquad \pi^n(x) =\; \frac{1}{L^3}\sum_{\vec k \in \mathbb{Z}^3} \tilde{\pi}^n_{\vec k} e^{-i\frac{2 \pi}{L} \vec k\cdot \vec x} \,,\label{eq:pinexpansion}
		\\
		\tilde{\pi}_{\vec k}^n =&\;\int_{V_n} \dd^3 x\, \pi(x) e^{+i\frac{2 \pi}{L} \vec k\cdot \vec x}\,,
	\end{align}
\end{subequations}

\noindent
which -- due to the density property of $\pi$ -- is an expansion in terms of the dual-Fourier modes (see Appendix~\ref{app:modedecomposition} for details) as it becomes evident due to the additional $1/L^3$ factor in Eq.~\eqref{eq:pinexpansion} and the fact that the exponentials are complex conjugated.
The set of variables $\{\tilde{\phi}_{\vec{k}}^n\}, \{\tilde{\pi}_{\vec{k}}^n\}$ are related via a canonical transformation to the original fields and have non-vanishing Poisson bracket
\begin{equation}
	\Poisson{\tilde{\phi}_{\vec{k}}^n}{\tilde{\pi}_{\vec{p}}^m} = \delta_{nm} \,\delta_{\vec{k},\vec{p}} \;,
\end{equation}
\noindent
where $\delta_{nm}$ and $\delta_{\vec{k},\vec{p}}$ denote the Kronecker-$\delta$.

The homogeneity constraints \eqref{eq:phiconstraint}, \eqref{eq:piconstraint} on configuration fields and their conjugate momenta can be now rewritten in terms of their field modes as follows.
Combining the constraints \eqref{eq:phiconstraint} for the scalar field and the decomposition in boxes \eqref{eq:phiboxes}, we find
\begin{equation}\label{eq:constraintexpansion}
	\Psi_a^\phi = \partial_a \sum_n \chi_{V_n}(x) \phi^n(x) = \sum_n \chi_{V_n}(x) \partial_a \phi^n(x) + \sum_n \partial_a\left(\chi_{V_n}(x)\right) \phi^n(x) \approx 0 \;.
\end{equation}

\noindent
The two terms on the r.h.s. of Eq. \eqref{eq:constraintexpansion} can easily be interpreted.
The first one demands the fields $\phi^n$ to be homogeneous within each of the regions $V_n$, while the second term encodes the interaction between adjacent regions.
Indeed, $\partial_a\left(\chi_{V_n}(x)\right) \propto \delta_S(x-\partial V_n)$ is a boundary term, the latter taking care of the change of the fields $\phi^n$ across different boxes.
The three constraints per point $\Psi_a^\phi$ decompose therefore into two sets of constraints, one demanding the fields within each box to be homogeneous, the second type demanding the field to not change across one box and the other.
Additionally, making use of the Fourier-decomposition of the field, the first set of constraints becomes
$$
\partial_a \phi^n(x) = i \frac{2 \pi}{L}\sum_{\vec{k} \in \mathbb{Z}^3} k_a\, \tilde{\phi}^n_{\vec{k}}\,e^{+i\frac{2 \pi}{L} \vec{k}\cdot \vec{x}} \approx 0 \qquad \Leftrightarrow \qquad \tilde{\phi}^n_{\vec{k}} \approx 0 \qquad \forall\;\vec{k}\neq \mathbf 0\;.
$$
The same analysis applies to the momentum $\pi$, thus yielding the complete set of constraints
\begin{subequations}
	\begin{align}
		\xi_{n,\vec{k}}^\phi = \tilde{\phi}_{\vec{k}}^n \approx 0 \qquad &, \qquad \xi_{n,\vec{k}}^\pi = \tilde{\pi}_{\vec{k}}^n \approx 0 \qquad \forall\; \vec{k} \neq \mathbf 0\;,\label{eq:hominbox}
		\\
		\zeta_{n}^\phi = \tilde{\phi}_{\mathbf 0}^n - \tilde{\phi}_{\mathbf 0}^1 \qquad &, \qquad  \zeta_{n}^\pi = \tilde{\pi}_{\mathbf 0}^n - \tilde{\pi}_{\mathbf 0}^1\qquad \,\forall\; n \neq 1 \;.\label{eq:homacrossbox}
	\end{align}
\end{subequations}
As mentioned above, the constraints \eqref{eq:hominbox} demand that only the homogeneous zero-mode within a region $V_n$ is non-trivial.
All solutions satisfying these constraints are therefore homogeneous over $V_n$, but not yet necessarily outside of it.
The second kind of constraints Eq.~\eqref{eq:homacrossbox} demands that all zero modes are equal to a reference box, say $V_1$ as chosen here.
This enforces that the field cannot vary from box to box and thus the field is really homogeneous all over the space.
Here any other reference could be chosen.
In fact, zero-modes must have the same value only across neighbouring regions.
However, it is clear that if all neighbours have the same field value, then all are equivalent to one reference box.
We shall now proceed to solve these constraints strongly and construct the Dirac bracket as the new constrained Poisson structure for the resulting  symmetry-reduced theory. To this aim, we impose the two types of constraints \eqref{eq:hominbox}, \eqref{eq:homacrossbox} in separate steps.
This works as the above two different sets of constraints Poisson commute among each other. Let us start with the constraints \eqref{eq:hominbox} first.
The associated Dirac bracket is defined as
\begin{equation}\label{DBdef}
	\Poisson{f}{g}_{D,\xi} = \Poisson{f}{g} - \Poisson{f}{\Psi_\textsf{A}} C^{\textsf{AB}} \Poisson{\Psi_\textsf{B}}{g} \;.
\end{equation}
where $\Psi_\textsf{A}$ represents the set of all constraints (of the first type) and $\textsf{A} = (\phi,\pi, n, \vec k)$ is a multi-index. $C^{\textsf{AB}}$ is the inverse of the matrix $C_{\textsf{AB}} := \Poisson{\Psi_\textsf{A}}{\Psi_\textsf{B}}$.
The subscript $D,\xi$ indicates that we use the Dirac bracket for the first type of $\xi$-constraints only.
We find then
\be
	C_{\textsf{AB}} = \left(\begin{array}{c|c}
		0 & \{\xi_{n,\vec{k}}^\phi,\xi_{m,\vec p}^\pi\}\\
		\hline
		\{\xi_{m,\vec p}^\pi,\xi_{n,\vec k}^\phi\} & 0
	\end{array}
	\right)
	= \left(\begin{array}{c|c}
		0 & $\;\;$1 \\
		\hline
		-1 & $\;\;$0
	\end{array}\right)\delta_{nm}\,\delta_{\vec k, \vec p} \;,
\ee
and, consequently, its inverse reads as
\begin{equation}
	C^{\textsf{AB}} = \delta_{nm}\,\delta_{\vec k, \vec p} \left(\begin{array}{c|c}
		0 $\;\;$ & -1 \\
		\hline
		1$\;\;$ & 0
	\end{array}\right) \;.
\end{equation}
The Dirac brackets \eqref{DBdef} thus yields
\begin{equation}\label{eq:xiDB}
\begin{split}
	\Poisson{\tilde{\phi}_{\vec{k}}^n}{\tilde{\pi}_{\vec{p}}^m}_{D,\xi} =&\, \Poisson{\tilde{\phi}_{\vec{k}}^n}{\tilde{\pi}_{\vec{p}}^m} - \sum_{\vec{q},\vec{q'}\neq\mathbf 0, l,l'} \Poisson{\tilde{\phi}_{\vec{k}}^n}{\xi_{l,\vec{q}}^\pi} C^{(\pi,l,\vec{q})(\phi,l',\vec{q'})} \Poisson{\xi_{l',\vec{q'}}^\phi}{\tilde{\pi}_{\vec{p}}^m}
	\notag
	\\
	=&\, \delta_{nm}\,\delta_{\vec{k},\vec{p}} - \sum_{\vec{q},\vec{q'}\neq\mathbf 0, l,l'} \delta_{nl}\,\delta_{\vec{k},\vec{q}}\, \delta_{ll'}\, \delta_{\vec{q},\vec{q'}}\, \delta_{ml'}\,\delta_{\vec{q'},\vec{p}} 
	\notag
	\\
	=&\, \delta_{nm}\,\delta_{\vec{k},\vec{p}} - \sum_{\vec{q}\neq\mathbf 0} \delta_{nm}\,\delta_{\vec{k},\vec{q}}\,\delta_{\vec{q},\vec{p}} 
	\notag
	\\
	=&\; \delta_{nm}\, \delta_{\vec{k},\mathbf 0}\, \delta_{\vec{p},\mathbf 0} \;.
\end{split}
\end{equation}
As expected from the above discussion, only the $\vec k=\mathbf 0$ modes have non-trivial Dirac brackets.
Therefore, we can set all other modes strongly to zero, as long as we use this new bracket. Recalling the expressions \eqref{eq:expansion} and \eqref{eq:piexpansion}, it is straight forward to compute now the corresponding bracket in position space representation 
\begin{equation}\label{eq:positionrepxiDB}
	\Poisson{\phi^n(x)}{\pi^m(y)}_{D,\xi} = \frac{\delta_{nm}}{L^3} \;,
\end{equation}
which is now regular (unlike the $\delta(x-y)$-distributional bracket for non-smeared field densities we had before implementing the constraints). Note that the fields $\phi^n$ and their conjugate momenta $\pi^m$ influence each other only if they are supported in the same spatial box and, due to homogeneity, the exact point within the box does not matter.

\begin{figure}[t!]
	\centering
	\includegraphics[scale=0.375]{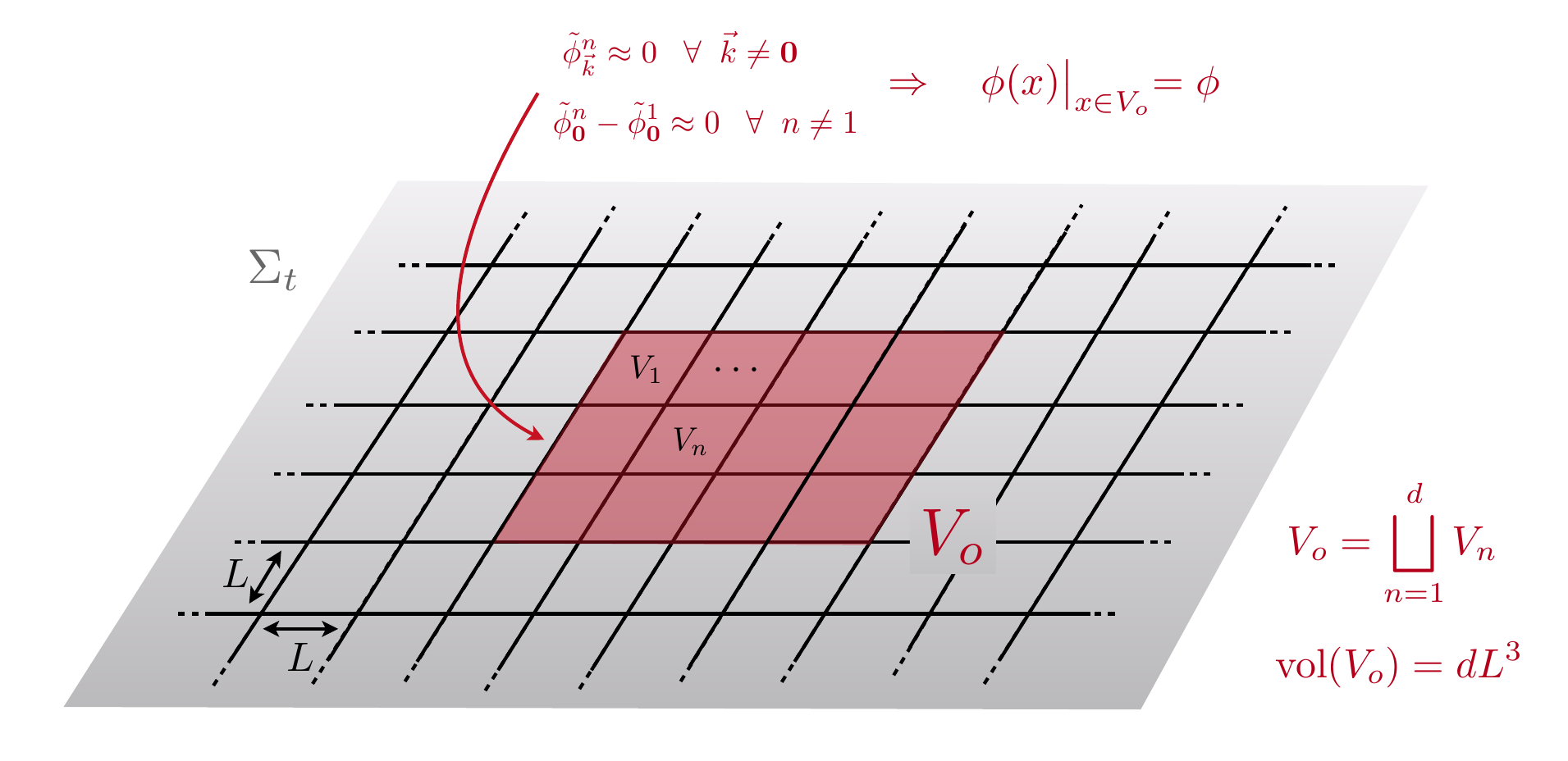}
	\caption{At the level of field modes, the homogeneity constraints \eqref{eq:phiconstraint}, \eqref{eq:piconstraint} translate into the two sets of constraints \eqref{eq:hominbox}, \eqref{eq:homacrossbox} which Poisson-commute among each other. The first set of constraints \eqref{eq:hominbox} sets all spatially inhomogeneous $\vec k\neq\mathbf 0$ modes within the cell $V_n$ to zero, thus keeping only the homogeneous zero modes within each cell. The second set of constraints \eqref{eq:homacrossbox} demands all zero modes to be equal across neighbouring boxes and ultimately to be all equal to that within a chosen reference box, say $V_1$. However, the implementation of the latter set of constraints via the Dirac bracket requires us to restrict ourselves only to a finite number $d<\infty$ of boxes so that spatial homogeneity cannot be imposed on the full non-compact spatial slice $\Sigma_t$ but only on a finite region $V_o=\bigsqcup_{n=1}^d V_n\subset\Sigma_t$ of volume $\text{vol}(V_o)=dL^3$ (red region).}
	\label{fig:implementinghomog}
\end{figure}

We can now implement the second kind of constraints Eq.~\eqref{eq:homacrossbox}.
We define the full Dirac bracket as
\begin{equation}\label{eq:fullDB}
	\Poisson{f}{g}_D = \Poisson{f}{g}_{D,\xi} - \Poisson{f}{\zeta_{\textsf{I}}}_{D,\xi} M^{\textsf{IJ}} \Poisson{\zeta_{\textsf{J}}}{g}_{D,\xi} \;,
\end{equation}
where again $\textsf I = (\phi,\pi,n)$ is a multi-index and $M^{\textsf{IJ}}$ is the inverse of the constraint matrix
\begin{equation}
	M_{\textsf{IJ}} := \left(\begin{array}{c|c}
		0 & \{\zeta_{n}^\phi,\zeta_{m}^\pi\}_{D,\xi} \\
		\hline
		\{\zeta_{n}^\pi,\zeta_{m}^\phi\}_{D,\xi} & 0
	\end{array}\right)
	=\left(\delta_{nm} + 1_{nm}\right) \left(\begin{array}{c|c}
		0 & $\;\;$1\\
		\hline
		-1 & $\;\;$0
	\end{array}\right)\;.
\end{equation}
Here $1_{nm}$ is the matrix which has a $1$ in every component.
For convenience, we define the matrix $Q_{nm} = \delta_{nm} + 1_{nm}$.
The inverse constraint matrix is thus given by
\begin{equation}\label{eq:Minverse}
	M^{\textsf{IJ}} = Q^{-1}_{nm} \left(\begin{array}{c|c}
		0$\;\;$ & -1 \\
		\hline
		1$\;\;$ & 0
	\end{array}\right) \;,
\end{equation}
and the remaining task is to invert $Q_{nm}$.
Assuming first $d$ volumes $V_n$, i.e. we have $d-1$ single constraints $\zeta_{n}^\phi$ $n=2,\dots,d$ as $n \neq 1$, and using Eqs. \eqref{eq:homacrossbox} and \eqref{eq:xiDB}, it is straight forward to prove that the inverse is given by
\begin{equation}\label{eq:Qinverse}
	Q^{-1}_{nm} = \delta_{nm} - \frac{1}{d}\,1_{nm} \quad,\quad n,m=2,\dots,d\;.
\end{equation}
Note that, if we want to impose homogeneity all over the non-compact spatial slice, we would need to take the $d \rightarrow \infty$ limit.
In such a limit, however, $Q^{-1}$ simply becomes the identity, which then obviously is not the inverse of $Q$ any more.~In other words, $Q$ cannot be invertible for $d \rightarrow \infty$.\footnote{This problem is avoided if we deal with compact spatial topology as e.g. in the case of spatial hyper-surfaces with 3-torus $\mathbb T^3$-topology. In this case, a finite number $d < \infty$ of elementary cells would be sufficient to impose full homogeneity all over the spatial manifolds.} Therefore, spatial homogeneity can only be imposed on a finite number $d$ of boxes (cfr. Fig. \ref{fig:implementinghomog}).~This is compatible with the need of introducing a fiducial cell in homogeneous and isotropic classical minisuperspace models to regularise the otherwise divergent integrals over the non-compact spatial slice such as the action and the symplectic structure.~As we shall see, the latter can be determined from the Dirac bracket for the $d$ boxes.

\noindent
Plugging Eqs. \eqref{eq:Minverse} and \eqref{eq:Qinverse} into the Dirac bracket \eqref{eq:fullDB} thus yields (for $n,m \in \left\{1,\dots,d\right\}$)
\begin{equation}\label{scalarfieldmodesfullDB}
\begin{split}
	\Poisson{\tilde{\phi}_{\vec{k}}^n}{\tilde{\pi}_{\vec{p}}^m}_D =&\; \Poisson{\tilde{\phi}_{\vec{k}}^n}{\tilde{\pi}_{\vec{p}}^m}_{D,\xi} - \sum_{l,l'=2}^{d} \Poisson{\tilde{\phi}_{\vec{k}}^n}{\zeta_l^\pi}_{D,\xi} Q^{-1}_{ll'} \Poisson{\zeta_{l'}^\phi}{\tilde{\pi}_{\vec{p}}^m}_{D,\xi} 
	\notag
	\\
	=&\; \delta_{{\vec{k}},\mathbf 0}\,\delta_{{\vec{p}},\mathbf 0}\left(\delta_{nm} - \sum_{l,l'=2}^{d}\left(\delta_{nl}-\delta_{n1}\right)Q^{-1}_{ll'}\left(\delta_{ml'} - \delta_{m1}\right)\right)
	\notag
	\\
	=&\; \delta_{{\vec{k}},\mathbf 0}\,\delta_{{\vec{p}},\mathbf 0} \begin{cases}
		1-\frac{d-1}{d}&,\;n=m=1 \\
		\delta_{nm} - Q^{-1}_{nm}&,\; n,m \neq 1\\
		\frac{1}{d}&,\; n=1,\, m \neq 1 \text{ or } m = 1, \, n \neq 1
	\end{cases}
	\notag
	\\
	=& \; \frac{\delta_{\vec{k},\mathbf 0}\,\delta_{{\vec{p}},\mathbf 0}}{d} \;, 
\end{split}
\end{equation}

\noindent
where we used the identities $\sum_{l=2}^d Q^{-1}_{ll'} = 1/d$ and $\sum_{l,l'=2}^d Q^{-1}_{ll'} = (d-1)/d$.
We see that the effect of the new constraints is that now each box has non-trivial Dirac bracket with each other one.
This is again expected as all boxes are equal, thus having the same information, and in fact have to know what is going on in neighbouring boxes to ensure homogeneity across the different boxes.
A similar computation yields the following Dirac bracket for position space representation (cfr. Eq. \eqref{eq:positionrepxiDB})
\begin{equation}\label{eq:Dbracketphixpiy}
	\Poisson{\phi(x)}{\pi(y)}_D = \frac{1}{dL^3}\;,
\end{equation}

\noindent
where we implicitly assume that $x,y \in \bigsqcup_{n=1}^d V_n$.
As we see from Eq. \eqref{eq:Dbracketphixpiy} above, the full bracket is divided by $dL^3$, which is the volume of the region over which homogeneity is imposed.
Again, for $dL^3 \rightarrow \infty$ this becomes zero and the Dirac bracket trivial.

We can interpret the above result by making contact with the full theory brackets given in Eq.~\eqref{eq:smearedPB}.
To this aim, let us define $V_o = \bigsqcup_{n=1}^d V_n$ which we call the \textit{fiducial cell} in analogy with the terminology often employed in the LQC literature that will be discussed later in Sec. \ref{Sec:Comsology}.
Using the expansion \eqref{eq:phiboxes} and \eqref{eq:expansion}, for fields restricted to the fiducial cell we find
\begin{align}\label{homogeneoussmearedphi}
	\left.\phi(x)\right|_{x \in V_o} &\,\,\,\;=\, \sum_{n=1}^d \chi_{V_n}(x) \sum_{\vec{k} \in \mathbb{Z}^3} \tilde{\phi}_{\vec{k}}^n e^{i\frac{2\pi}{L} \vec{k}\cdot \vec{x}}
	\notag
	\\
	&\stackrel{\xi_{n,\vec{k}}^\phi=0}{=}\, \sum_{n=1}^d \chi_{V_n}(x) \tilde{\phi}_\mathbf 0^n
	\notag
	\\
	&\,\stackrel{\zeta_{n}^\phi=0}{=}\, \tilde{\phi}_\mathbf 0^1 = \frac{1}{d} \sum_{n=1}^d \tilde{\phi}_\mathbf 0^n
	\notag
	\\
	&\;\,\,=\, \frac{1}{d L^3} \int_{V_o} \dd^3 x \phi(x)=\phi\left(V_o\right)\;.
\end{align}

\noindent
This tells us that the full theory field smeared over the fiducial cell $V_o$ coincides with the field $\phi(x)$ on the constraint surface $\xi_{n,\vec{k}}^\phi\approx0$ and $\zeta_{n}^\phi\approx0$. Similar conclusion holds for the momentum field $\pi$, namely
\begin{align}\label{reducedandfullphi}
	\pi[g] &\;\;\,\,= \sum_{n=1}^{d} \sum_{\vec{k} \in \mathbb{Z}^3} \tilde{\pi}^n_{\vec{k}} \frac{1}{L^3} \int_{V_n} \dd^3 x e^{-i\frac{2\pi}{L} \vec{k}\cdot \vec{x}} g(x)
	\notag
	\\
	&\,\stackrel{\xi_{n,\vec{k}}^\pi=0}{=}\, \sum_{n=1}^d \tilde{\pi}^n_\mathbf 0 \frac{1}{L^3} \int_{V_n} \dd^3 x g(x)
	\notag
	\\
	&\,\,\stackrel{\zeta_{n}^\pi=0}{=}\, \tilde{\pi}^1_\mathbf 0 \sum_{n=1}^d \frac{1}{L^3} \int_{V_n} \dd^3 x g(x)
	\notag
	\\
	&\;\;\,=\, d \cdot \tilde{\pi}^1_\mathbf 0 \frac{1}{d L^3} \int_{V_o} \dd^3 x g(x) = \bar{g} \sum_{n=1}^d \tilde{\pi}_\mathbf 0^n
	\notag
	\\
	&\;\;\,=\, \bar{g} \sum_{n=1}^d \int_{V_n} \dd^3 x \pi(x) = \bar{g} \pi\left(V_o\right)\;
\end{align}

\noindent
where $\bar{g} = \frac{1}{\text{vol}(V_o)} \int_{V_o} \dd^3\,x g(x)$ is the mean value of $g$ over the spatial region $V_o$.
We can reconstruct $\pi(x)$ by choosing $g(x) = \delta(x)$ and thus $\bar{g} = 1/dL^3$, from which it follows that
\be\label{reducedandfullpi}
\pi(x)|_{x\in V_o} = \pi[\delta(x)] = \frac{\pi\left(V_o\right)}{d\, L^3} \;. 
\ee

\noindent
Therefore, due to the constraints the local quantity $\pi(x)$ is the same as its average over the fiducial cell $\pi(V_o)/d\,L^3$.
Comparing with the full theory, we find (cfr. Eq.~\eqref{eq:smearedPB})
\begin{equation}
	\Poisson{\phi\left(V_o\right)}{\pi(x)} = \frac{1}{\text{vol}\left(V_o\right)} = \frac{1}{dL^3} = \Poisson{\phi(x)}{\pi(y)}_D \;.
\end{equation}

\noindent
Finally, we can also reconstruct the smeared observables \eqref{eq:smearedphipi}.
For $V \subset V_o$, we simply have
\begin{equation}\label{eq:FTobssymred}
	\phi(V) = \frac{1}{\text{vol}(V)} \int_V\dd^3 x \phi(x) \approx \phi(x) \quad , \quad \pi(V) = \int_V \dd^3 x \pi(x) \approx \text{vol}(V) \pi(x) \;.
\end{equation}

\noindent
Consistently, we find
\begin{equation}\label{eq:Diracsmeared}
	\Poisson{\phi(V)}{\pi(V')}_D = \frac{\text{vol}\left(V_o\cap V'\right)}{\text{vol}(V_o)} \stackrel{V'\subset V_o}{=} \frac{\text{vol}\left(V'\right)}{\text{vol}(V_o)}\;.
\end{equation}

\noindent
The above result does not depend on $V$ as the averaging of $\phi$ is the same for all volumes within the homogeneity region $V_o$.
Therefore, averaging over $V$ is the same as over $V_o$.
The result is then simply the same as the full theory result with the largest equivalent averaging volume, i.e. the volume of the whole region $V_o$ over which homogeneity is imposed.
On the other hand, the average of $\pi(x)$ over the volume $V'$ is independent of this volume, too.
Nevertheless, the smeared observable $\pi(V')\approx\text{vol}\left(V'\right) \pi(x)$ is extensive and as such it takes track of the total volume under consideration.
Note that we can formally send $V$ to infinity as in the full theory without troubles.
However, we are restricted to $V \subseteq V_o$ as else we have to consider new terms in the constraint analysis, i.e. larger $d$.

\subsection{Symmetry-Reduced and Truncated Theory}\label{sec:truncation}

Now that we have constructed the Dirac bracket and the homogeneity constraints can be imposed strongly, we can derive the reduced Hamiltonian for our constrained theory. However, as we do not impose the homogeneity across all the infinitely many boxes filling the spatial manifold but only across $d$ of them, there will be remaining boundaries.
As it was the case for Eq.~\eqref{eq:constraintexpansion}, after imposing homogeneity, the terms $\partial_a \phi$ will be proportional to $\partial_a \chi_{V_o}$.
The Hamiltonian \eqref{eq:scalarfieldHamiltonian} then reads as\footnote{The sum over ``all $V_o$'' is a sloppy notation for grouping together again $d$ neighbouring boxes to one larger region of the same shape as the individual boxes and call them $V_o$. This amounts to introducing again constraints similar to Eq.~\eqref{eq:homacrossbox} but using several different reference boxes, one for each $V_o$.}
\be\label{eq:scalarfieldHamiltonian2}
	H = \int_{\Sigma_t} \dd^3 x \frac{1}{2} \left(\pi(x)^2 + \partial_a \phi(x) \partial^a\phi(x) + m^2 \phi(x)^2\right)= \sum_{\text{all } V_o}\frac{d L^3}{2}\Bigl(\pi^2 + m^2 \phi^2 + \phi^2 \partial_a \left(\chi_{V_o}\right) \partial^a \left(\chi_{V_o}\right)\Bigr) \;. 
\ee

\noindent
Here $\phi = \phi(x)$ and $\pi = \pi(x)$ for any $x \in V_o$.
To make sense out of this expression one would have to regularise the terms $\partial_a \chi_{V_o}$.
This is equivalent to define proper boundary terms at $\partial V_o$ and defining how $\phi$ is changing across the fiducial cells.

There is an interesting relation to coarse-graining at this point.
Imposing the second kind of constraints $\zeta_{n}^\phi$ and $\zeta_{n}^\pi$ did not achieve fully homogeneous fields all over the space as we can only impose homogeneity for $d$ elementary boxes.
In fact, as discussed in the previous subsection, we collected the information of $d$ homogeneous boxes together and additionally enforced every inhomogeneity across the basic boxes to be zero.~As a result the description of a single box of volume $L^3$ is replicated over the $d$ boxes forming the homogeneity region $V_o$ of volume $d\,L^3$.~Fields are now decomposed into infinitely many boxes of size $d\,L^3$, while before the boxes were of size $L^3$.
All the changes are encoded in the interactions between the boxes and the boundary terms coming from $\partial_a \chi_{V_o}$.
As we will briefly discuss in the outlook of Sec.~\ref{sec:furtherfuturedirections}, taking care of all these boundary terms could lead to a proper renormalisation flow, which would be interesting to be studied in future work.
Here we leave it with this comment and refer the reader to similar work in the cosmological context \cite{BodendorferCoarseGrainingAs,BodendorferRenormalisationwithsu11,BodendorferPathIntegralRenormalization}.

Additionally, we can study how well the imposition of homogeneity worked and which modes remain.
To see this, let us perform a continuous Fourier transformation of the symmetry reduced field $\phi(x)$.
To fix some terminology, we denote the remaining cells, where $d$ of the initial cells are grouped together via the constraints $\zeta_{n}^\phi$ and $\zeta_{n}^\pi$, as fiducial cells $V_o^{(\vec{n})}$ with volume $V_o = d\cdot L^3$.
Note that $\vec{n}$ is a three dimensional vector in $\mathbb{Z}^3$.
Each of these fiducial volumes can be classified by a vector $\vec{x}_{\vec{n}}=V_o^{1/3}\vec{n}=V_o^{1/3}\cdot\left(n_x,n_y,n_z\right)$ pointing to one of the corners of the cells.
The field value in each cell can then be written as $\left.\phi(x)\right|_{V_o^{(\vec{n})}} = \tilde{\phi}^{\vec{n}}_\mathbf 0 = \tilde{\phi}_\mathbf 0^{(n_x,n_y,n_z)}$ and the Fourier transform then yields
\allowdisplaybreaks
\begin{align}
	\tilde{\phi}(k) &:= \frac{1}{\sqrt{2\pi}^3} \int \dd^3x \,\phi(x) e^{-i \vec{k}\cdot \vec{x}} = \frac{1}{\sqrt{2\pi}^3} \sum_{\vec{n}\in \mathbb{Z}^3} \tilde{\phi}^{\vec{n}}_\mathbf 0 \int \dd^3x \,\chi_{V_o^{(\vec{n})}}(x) e^{-i \vec{k}\cdot \vec{x}}
	\notag
	\\
	&= \sum_{\vec{n}\in \mathbb{Z}^3} \tilde{\phi}^{\vec{n}}_\mathbf 0 \prod_{\xi = {x,y,z}} \int_{V_o^{1/3}n_{\xi}}^{V_o^{1/3}\left(n_{\xi}+1\right)} \frac{\dd \xi}{\sqrt{2\pi}} e^{-i k_\xi \xi}
	\notag
	\\
	&= \sum_{\vec{n}\in \mathbb{Z}^3} \tilde{\phi}^{\vec{n}}_\mathbf 0 \prod_{\xi = {x,y,z}} \frac{2}{\sqrt{2\pi}} e^{-\frac{ik_\xi V_o^{1/3}}{2}} e^{-ik_\xi V_o^{1/3} n_\xi} \frac{\sin\left(\frac{k_\xi V_o^{(1/3)}}{2}\right)}{k_\xi}
	\notag
	\\
	&= \frac{8}{\sqrt{2\pi}^3} e^{-i \frac{V_o^{1/3} \left(k_x+k_y+k_z\right)}{2}} \frac{\sin\left(\frac{k_x V_o^{1/3}}{2}\right)}{k_x} \frac{\sin\left(\frac{k_y V_o^{1/3}}{2}\right)}{k_y} \frac{\sin\left(\frac{k_z V_o^{1/3}}{2}\right)}{k_z}\underbrace{\sum_{n\in \mathbb{Z}^3} \tilde{\phi}^{\vec{n}}_\mathbf 0 e^{-iV_o^{1/3} \vec{k} \cdot \vec{n}}}_{=:f(k_x,k_y,k_z)} \;.\label{scalrfieldmomprofile}
\end{align}

\noindent
Here, the function $f(k_x,k_y,k_z)$ is the result of a discrete Fourier series and therefore a function of period $2\pi/V_o^{1/3}$ in each direction.
This is true as long as the sum $\sum_{n\in \mathbb{Z}^3} \left|\tilde{\phi}^{(n_x,n_y,n_z)}_\mathbf 0\right|^2 < \infty$ converges\footnote{This again excludes the fully homogeneous situation where all of these contributions are equal.}.
Due to the the periodicity of this function, we can estimate
\begin{equation}
	|f|_{\infty}^2 \le \sum_{n\in \mathbb{Z}^3} \left|\tilde{\phi}^{(n_x,n_y,n_z)}_\mathbf 0\right|^2 =: C \;.
\end{equation}

\noindent
Finally, we find
\begin{align}\label{eq:remainingmodes}
	\left|\tilde{\phi}(k)\right|^2 = V_o^2 \frac{|f(k_x,k_y,k_z)|^2}{\left(2\pi\right)^3} \prod_{\xi = {x,y,z}} \frac{4}{V_o^\frac{2}{3}k^2_\xi} \sin\left(\frac{k_\xi V_o^{\frac{1}{3}}}{2}\right)^2  < \frac{V_o^2C}{\left(2\pi\right)^3} \prod_{\xi = {x,y,z}} \frac{4}{V_o^\frac{2}{3}k^2_\xi} \sin\left(\frac{k_\xi V_o^{\frac{1}{3}}}{2}\right)^2 \;.
\end{align}

\begin{figure}[t!]
	\centering
	\includegraphics[height=6.5cm]{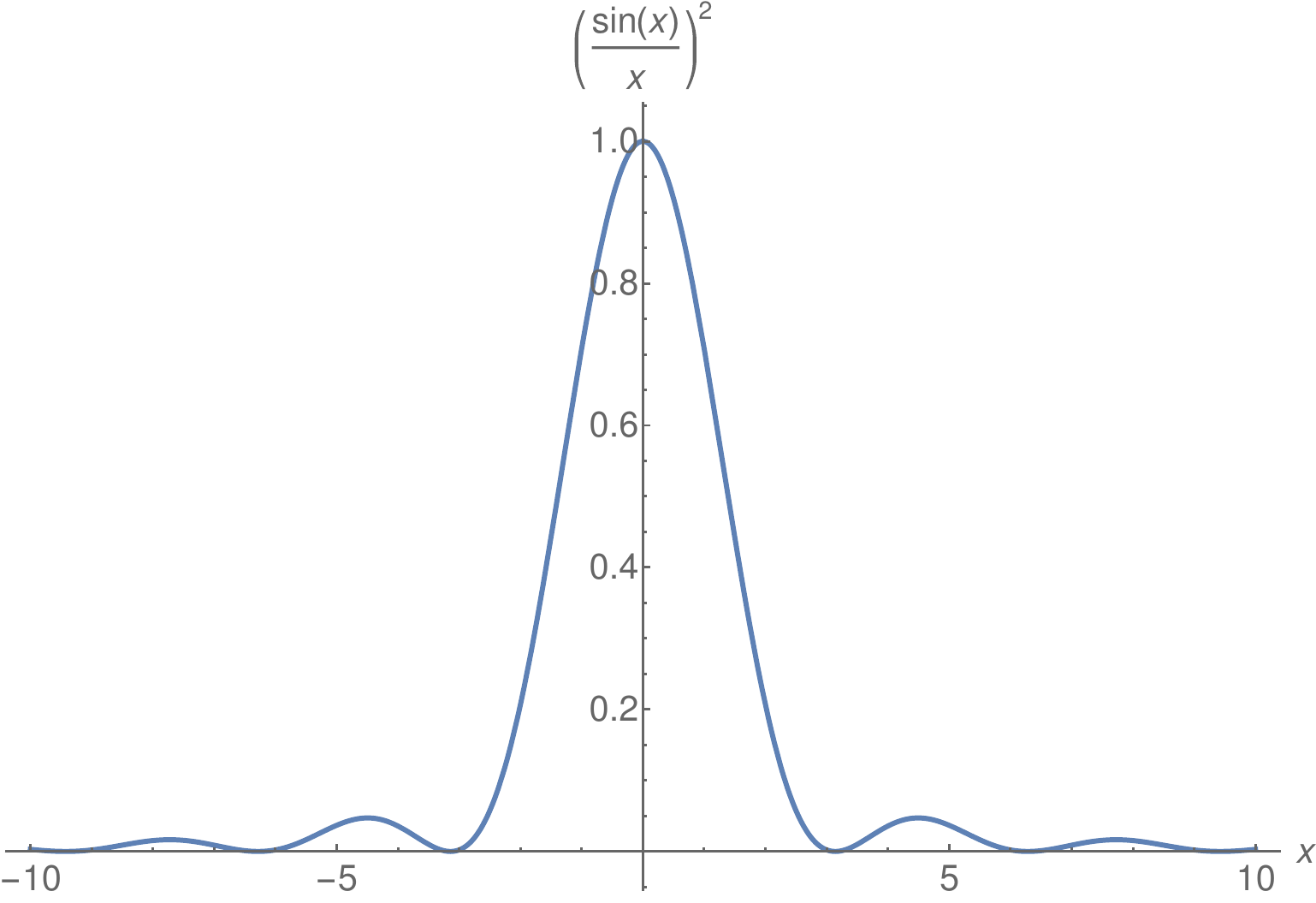}
	\caption{Plot of $\sin(x)^2/x^2$, which is related to $|\tilde{\phi}(k)|^2/|\tilde{\phi}(0)|^2$ in Eq.~\eqref{eq:remainingmodes} for $x = V_o^\frac{1}{3} k_\xi/2$. As discussed, the modes $k_\xi\in \frac{2\pi}{V_o^{1/3}} \mathbb{Z}$ are exactly removed, larger momenta are largely suppressed, but small momenta are still present.}
	\label{fig:sinx}
\end{figure}

\noindent
A plot of this momentum profile is shown in Fig.~\ref{fig:sinx} and we conclude that\footnote{This conclusion does not apply for compact spaces in which case the continuous and discrete Fourier decomposition are equivalent (details depend on the topology of the compact space) and thus homogeneity is implemented all over the space.}:
\begin{itemize}
	\item The integer modes $k_\xi \in (2\pi/V_o^{\frac{1}{3}})\cdot \mathbb{Z}$ are removed.
	\item The dominant mode is $\vec k =\mathbf  0$ with $|\tilde{\phi}(\mathbf 0)|^2 = V_o^2 C/(2\pi)^3$.
	\item There are still infinitely many modes left, even modes which have a shorter wavelength than the edge length of the fiducial cell.
	However these modes are suppressed of order $\mathcal{O}\left(1/V_o^{\frac{2}{3}}k_\xi^2\right)$. Interestingly, both $k_\xi$ and $V_o$ appear in the order of suppression. As such, small enough homogeneity region $V_o$ might yield non negligible higher modes contributions. This in turn might have sensible consequences for instance when dealing with small volumes in early-universe cosmology where inhomogeneities might then become non-negligible, thus breaking the homogeneous approximation. We will come back on this point later in the second part of the paper.
	\item We still work with a field theory, only the momentum profile of the field is very restricted as it has to have the form Eq.~\eqref{eq:remainingmodes}. 
	However, the dynamics of the function $f(\vec{k})$ is still undetermined and could be fixed by studying the field dynamics and initial conditions.
	This allows in principle for another approach to impose (approximate) homogeneity by fining different ansatze for the momentum profile to better control which momenta are kept and which not in the truncated theory. A possibility would be for example to enforce a Gaussian momentum profile.
\end{itemize}

Coming back now to the Hamiltonian \eqref{eq:scalarfieldHamiltonian2} above, one possibility is to truncate the sum over a finite number of boxes and simply ignore boundary terms.~This leads to the reduced Hamiltonian usually used to describe a spatially homogeneous scalar field
\begin{align}\label{eq:truncatedH}
	H = \frac{d L^3}{2} \left(\pi^2 + m^2 \phi^2\right) \;,
\end{align}

\noindent
and is truncated now as we simply ignore everything that happens outside the region $V_o$.
In particular, due to the above analysis, this means quantitatively that all modes $\tilde{\phi}(\vec k\neq\mathbf 0)$ are simply set to zero.
Due to Eq.~\eqref{eq:remainingmodes}, we see that this has a larger effect for small modes, but becomes negligible for modes $V_o^{1/3} k \gg 1$.
Therefore, this truncation seems to be reasonable for large fiducial cell volumes $V_o$.
This is also intuitively plausible as, for large cells, the boundary terms are negligible compared to the physics in the bulk.
For small boxes this will not be true any more and the truncation neglects physically relevant boundary terms and modes. It should be stressed however that this observation is purely kinematic in the sense that we know which momenta are truncated and how strong such a truncation is, but no statement is made about how relevant these momenta are for the dynamics of the smeared observables $\phi(V)$ and $\pi(V)$.
In other words, at this stage it is impossible to tell how good the truncation is from a dynamical standpoint, the latter requiring us to study the full theory.
We can look for instance at the dynamics of the averaged field $\phi(V)$ with $V\subset V_o$ in the full field theory. Due to the constraints, we find $\phi(x) = \phi(V) = \phi(V') = \phi(V_o)$ and the equations of motion written now w.r.t. the Dirac bracket leads to 
\be\label{eq:EoM}
	\begin{aligned}
		\frac{\dd}{\dd t}\phi(V) &= \Poisson{\phi(V)}{H}_D = \frac{\pi(V)}{\text{vol}(V)} \\
		\frac{\dd}{\dd t}\pi(V) &= \Poisson{\pi(V)}{H}_D = - \text{vol}(V) m^2 \phi(V)
	\end{aligned}
	\qquad \Rightarrow \qquad \frac{\dd^2}{\dd t^2} \phi(V) = -m^2 \phi(V) \;.
\ee

\noindent
Comparing Eq.~\eqref{eq:EoM} with the full theory result \eqref{eq:smeareddynamics}, the dynamics of the averaged quantity is well reproduced up to the fact that boundary terms are neglected.
In fact, if $\partial V \in V_o \setminus \partial V_o$, then the dynamics is exactly reproduced due to the homogeneity condition.
The argument can be further refined by considering the solution of the full theory dynamics as follows.
It is well known that the Klein-Gordon equation (resulting from Eqs.~\eqref{eq:FTEoM}) is solved by
\begin{equation}
	\phi(x,t) = \int_{\mathbb{R}^3} \frac{\dd^3k}{\sqrt{2 \omega(k)}}\,\left(a(k)e^{-i\omega(k)t + i \vec{k}\cdot \vec{x}} + a^*(k)e^{+i\omega(k)t - i \vec{k}\cdot \vec{x}}\right) \;,
\end{equation}

\noindent
with $\omega(k) = +\sqrt{k^2 + m^2}$.
Consequently, we find the full theory solution for the smeared field to be
\begin{equation}
	\phi(V,t) = \frac{1}{\text{vol}(V)}\int_{V}\dd^3 x \,\phi(x,t) = \int_{\mathbb{R}^3} \frac{\dd^3k}{\sqrt{2 \omega(k)}}\,\left(a(k)\,f_V(k)\,e^{-i\omega(k)t} + a^*(k)\,f^*_V(k)\,e^{+i\omega(k)t}\right) \;,
\end{equation}

\noindent
where we abbreviated
\begin{equation}
	f_V(k) := \int_V \frac{\dd^3x}{L^3} e^{+i\vec{k}\cdot \vec{x}} = \prod_{\xi = {x,y,z}} e^{i k_\xi \left(\xi_o + \frac{L}{2}\right)} \frac{2 \sin\left(\frac{k_\xi L}{2}\right)}{k_\xi L} \;,
\end{equation}

\noindent
and $L$ is the edge length of the volume $V$, i.e. $\text{vol}(V) = L^3$.
For momenta $k_\xi L \gg 1$, the function $f_V(k)$ suppresses all the contributions to the integral.
On the other hand, for small momenta, i.e. $k^2 \ll m^2$, we can approximate $\omega(k) \approx m$ and the exponential becomes just $k$-independent.
As long as $m \gg 1/L$, we are safe with this approximation and, when the momenta become of order of the mass $k_\xi \sim m \gg 1/L$, they are large enough for the function $f_V(k)$ to be safely approximated by zero and the integral can be cut off.
Finally, we arrive at
\begin{equation}
	\phi(V,t) \stackrel{m\gg 1/L}{\sim} A \cos\left(m\, t\right) + B\sin\left(m\,t\right) \;,
\end{equation}

\noindent
where $A$ and $B$ are constants obtained by the integration over $a(k)$ and $a^*(k)$ and combining the exponentials to trigonometric functions.
Obviously, this is the exact same solution we would obtain from Eq.~\eqref{eq:EoM} of the truncated theory.
Using therefore this full theory dynamical input, we can conclude that the homogeneous truncation works well as long as 
\begin{equation}\label{eq:truncationscale}
	m \gg \frac{1}{L} \;.
\end{equation}

\noindent
A more detailed analysis of this statement and its role in the quantum theory will presented later in Sec.~\ref{sec:symredscalarQFT} where the symmetry-reduction at the quantum field theory level is discussed.

In the following, we will work with the truncated Hamiltonian \eqref{eq:truncatedH} obtained by replacing the momentum profile \eqref{eq:remainingmodes} by simply $\propto\delta(\vec{k})$ and thus ignoring all other modes.
This is a direct consequence of neglecting all boundary terms.
Moreover, this means that, for a fixed fiducial cell $V_o$ over which spatial homogeneity is imposed, the smeared observables $\phi(V)$ and $\pi(V)$ are restricted to volumes $V \subset V_o$.
As the choice of $V_o$ enters explicitly the Hamiltonian and also the Dirac bracket, each choice corresponds in principle to a different homogeneously truncated theory.
Therefore, we actually deal with a full class of truncated theories, each of which is identified by the specification of the fiducial cell size, the latter interpreted as the region over which (spatial) homogeneity is imposed.
Different choices of $V_o$ are canonically independent from each other.
In fact, on the one hand, computing the Dirac bracket of the same observables in different truncated and reduced theories gives (cfr. Eq.~\eqref{eq:Diracsmeared})
\begin{equation}\label{eq:DBreducedtheory}
	\Poisson{\phi(V)}{\pi(V')}_D = \frac{\text{vol}(V')}{\text{vol}(V_o)} \;,
\end{equation}

\noindent
which changes its value for different $V_o$.
On the other hand, neither $\phi(x)$ nor $\pi(x)$ and consequently $\phi(V)$ and $\pi(V')$ depend (from a full theory point of view) on $V_o$, so that nothing could cancel the scaling of the Dirac bracket \eqref{eq:DBreducedtheory} with the inverse of $\text{vol}(V_o)$.
Consequently, changing $V_o$ changes the canonical structure of the symmetry-reduced theory and there is no canonical transformation relating the different symmetry reduced theories.
This is plausible as in fact each of these theories comes from the full theory after imposing different physical constraints.

Another difference in the theories with different $V_o$ is the set of independent observables.
In fact, the smeared observables $\phi(V)$ and $\pi(V)$, which are infinitely many in the full theory due to the infinitely many possible choices for $V$, come to be all equivalent in the truncated theory and can be replaced by the local quantity or the quantity smeared over the full fiducial cell, i.e.
\begin{equation}
	\forall\;V\,\subset V_o:\quad \phi(x) = \phi(V) = \phi(V_o) \qquad , \qquad \pi(x) = \frac{1}{\text{vol}(V)} \pi(V) = \frac{1}{\text{vol}(V_o)} \pi(V_o) \;.
\end{equation}

\noindent
Therefore, in the truncated theory there are only two independent observables $\phi(x) = \phi(V_o)$ and $\pi(x) = \pi(V_o)/\text{vol}(V_o)$.
Local fields become the same as the averaged fields over the fiducial cell and changing the fiducial cell is thus equivalent to picking a different subset of the full theory observables.
Summarising, if $V_o$ is different, the Hamiltonian, the set of observables, and the Poisson-algebra structure (given by the Dirac bracket as discussed above) are changed.\footnote{Again, on a compact space, we would be able to impose the homogeneity exactly, i.e. we do not need to throw away higher modes or argue about boundary terms in the Hamiltonian. Even though the symmetry reduction works then without any of these truncations, the theory remains dependent on the size of this compact background space. Choosing e.g. two tori $\mathbb{T}^3$ of different sizes $\text{vol}\left(\mathbb{T}\right)$, the resulting theories would be different in the exact same way as discussed above.}
However, at the classical level, it is possible to relate these different theories straight forwardly as follows. Let us consider the symmetry reduced theory with two different fiducial volumes, say $V_o^{(1)}$ and $V_o^{(2)}$, with the restriction $x \in V,V'\subset V_o^{(1)} \subset V_o^{(2)}$.
We can reconstruct all canonical data for the $V_o^{(1)}$-theory out of the $V_o^{(2)}$-one by relating their observables, Hamiltonians, and Dirac brackets according to
\begin{subequations}\label{eq:Vochange}
	\begin{align}
		\left.\phi(V)\right|_{V_o^{(1)}} = \left.\phi(V)\right|_{V_o^{(2)}} \qquad&,\qquad\left.\pi(V)\right|_{V_o^{(1)}} = \frac{\text{vol}(V_o^{(1)})}{\text{vol}(V_o^{(2)})}\left.\pi(V)\right|_{V_o^{(2)}} \;,\label{eq:Diracscalingb}
		\\
		\left. H\right|_{V_o^{(1)}} = \frac{\text{vol}\left(V_o^{(1)}\right)}{\text{vol}\left(V_o^{(2)}\right)} \left. H\right|_{V_o^{(2)}} \qquad&,\qquad
		\Poisson{\cdot}{\cdot}_D^{(1)} = \frac{\text{vol}\left(V_o^{(2)}\right)}{\text{vol}\left(V_o^{(1)}\right)} \Poisson{\cdot}{\cdot}_D^{(2)} \;.\label{eq:Diracscalingd}
	\end{align}
\end{subequations}

\noindent
Respecting this identification, we can change consistently between the symmetry reduced theories with different size of fiducial cells.
Caution is needed by carefully checking that the volume $V$ over which the field observables are integrated is smaller than the smallest fiducial cell.
Although, formally this does not matter, in principle this would contradict the assumptions to arrive at the symmetry reduced theory. It is however possible to show that the dynamics for all observables is actually independent of $V_o$.
Indeed, given any classical observable $\mathcal{O}$, its dynamics is given by
\begin{equation}
	\dot{\mathcal{O}} = \Poisson{\mathcal{O}}{\left.H\right|_{V_o^{(1)}}}^{(1)}_D \stackrel{\eqref{eq:Vochange}}{=} \frac{\text{vol}\left(V_o^{(2)}\right)}{\text{vol}\left(V_o^{(1)}\right)}\Poisson{\mathcal{O}}{\frac{\text{vol}\left(V_o^{(1)}\right)}{\text{vol}\left(V_o^{(2)}\right)}\left.H\right|_{V_o^{(2)}}}^{(2)}_D = \Poisson{\mathcal{O}}{\left.H\right|_{V_o^{(2)}}}^{(2)}_D\;,
\end{equation}

\noindent
and is thus independent of the fiducial cells as long as the support of $\mathcal{O}$ is smaller than both fiducial cells $V_o^{(1)}$ and $V_o^{(2)}$.
Explicitly, we have seen that the equations of motion \eqref{eq:EoM} are independent of $V_o$.\footnote{Note that this result does not rely on the specific form of the Hamiltonian and only its property to be extensive. Even if we would have added interaction terms, the equations of motion are independent of $V_o$.}
This is consistent with the fact that if the restriction to homogeneous modes would be performed at the level of the equations of motion, there would be no need for introducing a fiducial cell.
The fiducial cell only enters the canonical analysis off-shell at the level of the action, Hamiltonian, and symplectic structure to make the otherwise divergent integrals over the non-compact spacial slice regular by restricting them to the region $V_o$ and thus truncating everything outside of that region.
We can therefore conclude two important things:
First, the equations of motion and consequently all physical predictions (for $V_o$-independent observables), do not depend on the fiducial cell.
However, the choice of fiducial volume and thus the zero modes tells us which subset of full field theory observables are selected in the corresponding truncated reduced theory.
Moreover, even observables as $\phi(V)$ implicitly depend on the fiducial cell, and therefore the zero modes, due to the identification $\phi(V) \approx \phi(V_o)$.
This is plausible as the particular choice of zero modes determines how the physical requirement of homogeneity is implemented.
Different choices of the size of $V_o$ lead to different physical requirements for the scale of homogeneity.
Second, at the level of the equations of motion, we can send $\text{vol}(V_o) \rightarrow \infty$ and still have a regular theory with valid physical predictions. Note that this is only possible as classical predictions depends only on on-shell quantities and local equations of motion. This is however not the case for a quantum theory, which contains more physical output than just the dynamics of quantities.
In the quantum theory, quantum fluctuations and correlations are also physically relevant and are determined by exactly the above-mentioned otherwise ill-defined off-shell structures such as Hamiltonian, canonical brackets, and action.

Moreover, it is important to note that there is a very clear difference between $V_o$ and a region $V$.
As already discussed, $V_o$ is the scale on which homogeneity is imposed.
This is obviously a physical requirement as it makes a statement about the system's state.
However, this is not necessarily related to an observable $\phi(V)$.
Here, $V$ is simply the region whose physics we would like to track.
It is in fact conceivable to prepare in a laboratory an experiment, which is homogeneous on a scale $V_o$ much larger than the actual detector of size $V$, the latter operationally defining then our observables.
Classical physics depends only on local equations of motion and there is neither correlations nor entanglement.
Consequently, the physics of the observable $\phi(V)$ depends on $V$, but not on $V_o$.
Further, we should not forget that we are still dealing with a field theory after all.
This can be easily overlooked in the homogeneous setting as the Hamiltonian and phase space formally look as those of a particle mechanical system (no integrals and being finite dimensional).
There are still observables which are smeared against test functions, i.e. instead of treating $\phi(x)$ and $\pi(x)$, which are the local phase space points and have the same value at each $x \in V_o$, we should use
$$
\phi[f] = \int_{\Sigma_t} \dd^3 x \phi(x) f(x) \approx \text{vol}(V) \bar{f} \phi(V) = \text{vol}(V) \bar{f} \phi(V_o)\qquad , \qquad \pi[f] = \int_{\Sigma_t} \dd^3 x \pi(x) f(x) \approx \bar{f} \pi(V)\;,
$$

\noindent
with $\bar{f} = \frac{1}{\text{vol}(V)} \int_V\dd^3x f(x)$ and $V = \text{supp}(f) \subset V_o$.
Obviously, due to the homogeneity constraints, it is sufficient to have knowledge about the volume averaged observables $\phi(V)$ and $\pi(V)$ as any other field theoretic observable can be constructed out of these.
Again, now $V = \text{supp}(f)$ is dependent on the observable one is interested in, i.e. the volume one would like to track.
This is a physical input, which in a laboratory would be related to the detector.
This different from $V_o$, which is instead the region on which homogeneity is imposed.
The latter can be though of as the size of the total experiment, assuming it is possible to prepare the system in a homogeneous manner. 

\section{Quantisation of the Symmetry-Reduced Theory}\label{Sec:homogscalarfieldquantisation}

Let us now study the quantisation of the classically constrained spatially homogeneous scalar field theory discussed in the previous section.~The symmetry-reduction has been thus performed at the classical level and the resulting minisuperspace model is then quantised.~The discussion of a ``quantising first, then symmetry-reducing'' prescription for a scalar field theory is postponed to Sec.~\ref{sec:symredscalarQFT}.~Special focus in the following will lie on the analysis of the $V_o$-dependence and its consequences.~Specifically, in Sec.~\ref{sec:phiquantisation} we first study the quantisation of the non-interacting theory presented in the previous sub-sections.~The symmetry-reduced Hamiltonian is then modified in Sec.~\ref{sec:quantuminteractions} to include a $\phi^4$-interaction term and we analyse how the $V_o$-dependence affects the interacting (quantum) theory.

\subsection{Quantum Minisuperspace Model}\label{sec:phiquantisation}

Let us recall from the previous section that the classical spatially homogeneous minisuperspace model results from a twofold procedure consisting of setting to zero the inhomogeneous modes with wavelength larger than the cell size and truncating those remaining inhomogeneous modes inside the cell so that one is left only with the spatially homogeneous zero mode. As discussed in Sec. \ref{sec:truncation} this leads us not to a single symmetry-reduced theory, but rather infinitely many, depending on the choice of the spatial region $V_o$ over which homogeneity is imposed and the truncation of the corresponding Hamiltonian over such a region. It is therefore natural to ask how the quantum theory changes if we use a different starting classically symmetry-reduced model, i.e. different $V_o$.

Let us start by considering a given fiducial cell $V_o$ with volume $\text{vol}(V_o) = d\cdot L^3 =: V_o$, where from now on for simplicity of notation we will denote both the region and its volume by $V_o$.
The Hamiltonian \eqref{eq:truncatedH} then reads as
\begin{align}\label{eq:fixedVoHamiltonian}
	H = \frac{V_o}{2} \left(\pi^2 + m^2 \phi^2\right) \;,
\end{align}

\noindent
with the two canonically conjugate homogeneous fields $\phi=\phi(x)$ and $\pi = \pi(x)$ satisfying the Dirac bracket relations (cfr. Eq.~\eqref{eq:Dbracketphixpiy})
\begin{equation}\label{phipiDBhomogeneous}
	\Poisson{\phi(x)}{\pi(y)}_D = \frac{1}{V_o} \;,
\end{equation}

\noindent
for any $x,y\in V_o$. As field-theoretic observables, we use the volume averaged fields (cfr. Eq.~\eqref{eq:FTobssymred})
\begin{equation}
	\phi(V) = \frac{1}{\text{vol}(V)} \int_V\dd^3 x\, \phi(x) = \phi(x) \qquad , \qquad \pi(V) = \int_V \dd^3\, x \pi(x) = \text{vol}(V) \pi(x) \;,
\end{equation}
\noindent
where we recall that, in the second equality, spatial homogeneity can be imposed \textit{strongly} in the sense of Dirac's theory of constrained systems once the Dirac bracket has been constructed.

We shall now proceed to quantise the homogeneous scalar field theory corresponding to a given value of $V_o$ by means of Dirac quantisation. The symmetry reduced classical theory is thus quantised using the Dirac bracket as starting point, rather than symmetry reducing at the quantum field theory level. The latter case will be discussed in Sec. \ref{sec:symredscalarQFT}.
To this aim, we assign operators $\hat{\phi}$, $\hat{\pi}$ to the canonical homogeneous fields $\phi$, $\pi$ and seek for a representation of the canonical commutation relations associated with their Dirac bracket \eqref{phipiDBhomogeneous}, namely ($\hbar=1$)
\begin{equation}\label{eq:quantisationrule}
	[\hat{\phi},\hat{\pi}] = i \widehat{\Poisson{\phi}{\pi}}_D = \frac{i}{V_o} \;.
\end{equation}

\noindent
Following standard arguments based on the Stone-von Neumann theorem, the carrier Hilbert space is realised as $\mathscr{H} = L^2\left(\mathbb{R},\dd \phi\right)$ on which canonical operators are represented as\footnote{As will be discussed later, the appearance of the $V_o$ factor in \eqref{eq:canonicaloprepr} plays an important role.~The precise way in which such a factor enters the representation of the canonical field operators is in principle arbitrary.~In \eqref{eq:canonicaloprepr} we assigned it entirely to $\hat{\pi}$.~Restoring $\hbar$ in the numerator of the r.h.s. of the canonical commutation relations \eqref{eq:quantisationrule}, this is a somewhat natural choice if one think of the usual association $\{\cdot,\cdot\}\to -\frac{i}{\hbar}[\cdot,\cdot]$ between classical Poisson brackets and quantum commutators with an \textit{effective} reduced Planck constant $\hbar_{\text{eff}}=\hbar/V_0$ so that $\pi\to\hat{\pi}=-i\hbar_{\text{eff}}\partial_\phi$ as in \eqref{eq:canonicaloprepr}.~Note that $\hbar_{\text{eff}}\to0$ as $V_o\to\infty$.~Alternatively, however, the $V_o$ factors could appear in the action of $\hat{\phi}$ or distributed over the representations of both canonical operators. Indeed, more generically, we could include $V_o^\gamma$ and $V_o^\epsilon$ factors at the denominator of the operator representations for $\hat{\pi}$ and $\hat{\phi}$, respectively, with arbitrary powers $\gamma$ and $\epsilon$ ($\gamma+\epsilon=1$) parametrising the freedom in incorporating the $V_o$-factors in the representation of the quantum operators to ensure the canonical brackets \eqref{phipiDBhomogeneous} to be correctly represented as commutation relations. All these choices are unitarily equivalent.}
\be\label{eq:canonicaloprepr}
\hat{\phi} \Psi(\phi) = \phi \Psi(\phi)\qquad,\qquad \hat{\pi} \Psi(\phi) = -\frac{i}{V_o} \partdif{}{\phi} \Psi(\phi) \;,
\ee
	
\noindent
with $\Psi \in \mathscr{H}$.
This then also defines the Hamiltonian operator $\hat{H}$ associated with the classical Hamiltonian \eqref{eq:fixedVoHamiltonian}, which is simply like that of an harmonic oscillator, say
\begin{equation}
	\hat{H} \Psi(\phi)= \frac{1}{2} \left(-\frac{1}{V_o} \frac{\partial^2}{\partial \phi^2} + V_o m^2 \phi^2\right) \Psi(\phi) \;. 
\end{equation}

\noindent
Dynamics is then defined either in a Heisenberg or Schr\"odinger picture according to 
\be
	i\partdif{}{t} \Psi(\phi) = \hat{H} \Psi(\phi)
	\qquad\text{or equivalently}\qquad
	\partdif{}{t}\hat{\mathcal{O}} = -i[\hat{\mathcal{O}}, \hat{H}] \;.
\ee

\noindent
This fully defines the canonical quantisation of our homogeneous reduced theory, for a given value of $V_o$.

At this point, we can study how the quantisation changes if we would have chosen a different $V_o$.
According to the above steps, which explicitly show where $V_o$ enters the quantisation of the spatially homogeneous classical theory identified by that given value of $V_o$, the Hilbert space is still realised as $\mathscr{H} = L^2\left(\mathbb{R},\dd \phi\right)$, but the representation \eqref{eq:canonicaloprepr} of the field operators changes according to
\be\label{eq:Vochangeoperatorrepr}
	\left. \hat{\phi} \right|_{V_o^{(1)}} = \left. \hat{\phi} \right|_{V_o^{(2)}}\qquad,\qquad
	\left. \hat{\pi} \right|_{V_o^{(1)}} = \frac{V_o^{(2)}}{V_o^{(1)}}\left. \hat{\pi} \right|_{V_o^{(2)}}
\ee

\noindent
Remarkably, unlike the classical case where $\pi\mapsto\pi$ if the fiducial volume is changed (cfr. Eq. \eqref{eq:Diracscalingb}), this is not the case anymore at the quantum level and the operators have to pick up a scaling behaviour.
This is actually plausible keeping in mind the quantisation map given by Eq.~\eqref{eq:quantisationrule}.
In fact, while the Dirac bracket is proportional to $1/V_o$ and scales with the changing of $V_o$ (see Eq.~\eqref{eq:Diracscalingd}), the commutator is just the composition of linear operators, which cannot intrinsically scale by itself.
The scaling of the Dirac bracket is thus shifted to the operators representation itself.
Given the prescription for representing the elementary operators, we find the corresponding representation for the Hamiltonian operator, which has however no definite scaling behaviour.
Changing $V_o$ weights the kinetic and potential energy part differently in the one or the other direction.
Large $V_o$ makes the potential contribution larger, while small $V_o$ weights the kinetic contribution more.
In fact, $V_o$ plays the role of the mass in a standard quantum harmonic oscillator.
By analogy, as the harmonic oscillator is independent of its mass, one might expect that the Hamiltonian is actually independent of $V_o$.\footnote{This can trivially be seen by distributing $V_o$ symmetrically across $\hat{\phi} \Psi = \phi/\sqrt{V_o} \Psi$ and $\hat{\pi} \Psi = i/\sqrt{V_o} \Psi'$. The Hamiltonian becomes then simply $\hat{H} = -i\partial_\phi^2 + m^2 \phi^2$, which is independent of $V_o$.}
This can be checked by computing its eigenstates.
To this aim, we define the ladder operators
\begin{equation}
	\hat{a} = \sqrt{\frac{V_o m}{2}} \left(\hat{\phi} + \frac{i}{m} \hat{\pi}\right) \qquad , \qquad \hat{a}^\dagger = \sqrt{\frac{V_o m}{2}} \left(\hat{\phi} - \frac{i}{m} \hat{\pi}\right)\;,
\end{equation}

\noindent
with commutation relation $[\hat{a},\hat{a}^\dagger] = 1$, and the Hamiltonian reads as
\begin{equation}
	\hat{H} = m\left(\hat{a}^\dagger \hat{a} + \frac{1}{2}\right) \;.
\end{equation}

\noindent
It follows immediately that the spectrum of $\hat{H}$ is simply given by $E_n = m(n+\frac{1}{2})$ for $n \in \mathbb{N}$, and is thus independent of $V_o$.
However, as $\hat{a}$ and $\hat a^\dagger$ explicitly depend on $V_o$, the eigenstates $\Psi_n(\phi)$ of the Hamiltonian depend on $V_o$ as well.
We can easily construct these by defining $\zeta = \sqrt{V_o\,m}\, \phi$ and noticing that
\begin{equation}
	\hat{a} \Psi(\phi) = \sqrt{\frac{m}{2}} \left(\sqrt{V_o} \phi + \frac{1}{m \sqrt{V_o}} \partdif{}{\phi}\right) \Psi(\phi) = \frac{1}{\sqrt{2}} \left(\zeta + \partdif{}{\zeta}\right) \Psi(\zeta)
\end{equation}

\noindent
depends on $V_o$ only through $\zeta$. Therefore, the standard construction of the energy eigenstates of the harmonic oscillator can be directly performed for any $V_o$.
They only differ in their dependence on $\zeta$.
Let us denote the Hermite functions as 
\begin{equation}
	\Psi_n\left(\zeta\right) := \frac{1}{\sqrt{2^n \cdot n!}} \left(\frac{1}{\pi}\right)^{\frac{1}{4}} e^{-\frac{\zeta^2}{2}} H_n\left(\zeta\right) \;,
\end{equation}

\noindent
where $H_n$ are the Hermite polynomials.
As these states are then normalised with respect to the scalar product on $L^2\left(\mathbb{R},\dd \zeta\right)$, we have to rescale the wave function by a factor of $(V_o\,m)^{1/4}$.
The energy eigenfunctions $\Psi^{(V_o)}_n(\phi)$ for the given value $V_o$ then simply read as
\begin{equation}
	\Psi^{(V_o)}_n(\phi) = \left(V_o\,m\right)^{\frac{1}{4}} \Psi_n\left(\sqrt{V_o\,m} \phi\right) \;.
\end{equation}

\noindent
Consequently, we can relate the energy eigenstates (with $m$ kept constant) in the quantisation with $V_o^{(1)}$ to the ones in the $V_o^{(2)}$-quantisation according to
\begin{equation}\label{eq:Vochangeenergyeigenstates}
	\Psi_n^{(1)}(\phi) = \left(\frac{V_o^{(1)}}{V_o^{(2)}}\right)^{\frac
	{1}{4}}\Psi_n^{(2)}\left(\sqrt{\frac{V_o^{(1)}}{V_o^{(2)}}} \,\phi\right) \;,
\end{equation}

\noindent
where the superscripts $(1)$ and $(2)$ for the states refers to the theory with $V_o^{(1)}$ and $V_o^{(2)}$, respectively. These two sets of eigenfunctions must have the same dynamics.
In this respect let us notice that, having the eigenstates explicitly constructed and knowing their relation, the Hamiltonians of the two quantised theories can be written as
\begin{equation}\label{eq:minisupHspectrum}
	\hat{H}\ket{\cdot}= \sum_{n \in \mathbb{N}} m\left(n+\frac{1}{2}\right) \braket{\Psi_n^{(1)}}{\,\cdot \,} \ket{\Psi_n^{(1)}} \;,
\end{equation}

\noindent
and the transformation behaviour of $\hat{H}$ follows directly from the relation \eqref{eq:Vochangeenergyeigenstates} between $\Psi_n^{(1)}$ and $\Psi_n^{(2)}$.
It is then easy to check that the dynamics of two states in the $V_o^{(1)}$- and the $V_o^{(2)}$-quantisation are equivalent if and only if their expansion coefficients $\langle\Psi_n^{(1)}|\Psi\rangle$ are equal.
This leads to an isomorphism between the two $V_o$-labeled Hilbert spaces $\mathscr{H}^{(1)}$ and $\mathscr{H}^{(2)}$ carrying the quantum representation \eqref{eq:quantisationrule}, \eqref{eq:canonicaloprepr} for $V_o^{(1)}$ and $V_o^{(2)}$, respectively.
Indeed, although the Hilbert spaces are both realised as $\mathscr{H}^{(1)} = L^2\left(\mathbb{R},\dd \phi\right)$ and $\mathscr{H}^{(2)} = L^2\left(\mathbb{R},\dd \phi\right)$, in order for the dynamics of both quantisation schemes to be the same we have to identify the states according to the mapping
\be\label{VoHilbertspaceisomorphism}
	\mathscr{I}:\; \mathscr{H}^{(1)} \longrightarrow \mathscr{H}^{(2)}\qquad\text{by}\qquad\; \Psi^{(1)}(\phi) \longmapsto \Psi^{(2)}(\phi)= \mathscr{I}\bigl(\Psi^{(1)}\bigr)(\phi):= \left(\frac{V_o^{(2)}}{V_o^{(1)}}\right)^{\frac
		{1}{4}}\Psi^{(1)}\left(\sqrt{\frac{V_o^{(2)}}{V_o^{(1)}}} \,\phi\right).
\ee

\noindent
It is easy to verify that the map \eqref{VoHilbertspaceisomorphism} defines a unitary isomorphism, i.e. that the scalar product remains preserved
\begin{equation}
	\braket{\mathscr{I}\bigl(\Psi^{(1)}\bigr)}{\mathscr{I}\bigl(\Phi^{(1)}\bigr)}^{(2)} = \braket{\Psi^{(1)}}{\Phi^{(1)}}^{(1)}\qquad\forall\;\Psi^{(1)},\Phi^{(1)}\in\mathscr H^{(1)}\;.
\end{equation}

\noindent
Such an isomorphism is a map from the Hilbert space obtained by the $V_o^{(1)}$-quantisation and the \textit{different} Hilbert space obtained from $V_o^{(2)}$-quantisation.
It only happens that both representations have the ``same looking'' carrier Hilbert space, although they are a priory independent as resulting from quantisation of the canonically inequivalent classical symmetry-reduced theories labeled by the different volumes of the region over which homogeneity has been imposed. This is reminiscent of some Hamiltonian renormalisation in theory space (or at least in the homogeneous subspace of theory space) by relating Hilbert spaces and Hamiltonians corresponding to different $V_o$ values. Although the task of establishing an explicit relation with renormalisation goes beyond the purpose of the present work and is left for future investigations, let us emphasise here that the discussion of the mode truncation and the subsequent interpretation of $V_o$ as setting the scale of homogeneity rather than playing the role of a mere regulator devoid of physical meaning are in line with some results available in the LQC-literature based on the comparison with effective QFT techniques \cite{BojowaldTheBKLscenario, BojowaldEffectiveFieldTheory, BojowaldCanonicalderivationof, BojowaldEffectiveEquationsof,BojowaldMinisuperspacemodelsas, BojowaldEffectiveequationsforiso}. We will further comment on this point in Sec. \ref{sec:quantuminteractions} where the interacting scalar field theory is briefly discussed, as well as in Sec. \ref{Sec:Comsology} and \ref{Sec:QuantumCosmology} where a similar strategy to the one developed in this section is applied to cosmology.

To sum up, we find that quantising with a given fiducial cell $V_o^{(1)}$ leads to different representations for the field operators than quantising the theory with a different choice of the fiducial cell $V_o^{(2)}$.
This is plausible as already classically the canonical structure explicitly depends on $V_o$ after imposing homogeneity and different choices of the fiducial cell result into inequivalent theories.
However, as discussed in Sec. \ref{sec:truncation}, the classical dynamics was actually independent of this choice.
A similar independence of the dynamics at the quantum level can be achieved by identifying states in the Hilbert spaces carrying the $V_o$-dependent representation of the quantised theories according to the map $\mathscr{I}$ defined in Eq. \eqref{VoHilbertspaceisomorphism}.
Two states related by this map will have the same dynamical behaviour independently of it being evaluated in the $V_o^{(1)}$ or $V_o^{(2)}$ theory.
Therefore, the classical observation that the dynamics is independent of the fiducial cell can be reproduced at the quantum level.\footnote{Here, this was shown to hold true for the dynamics of states, i.e. in the Schr\"odinger picture. As it is well known, Schr\"odinger and Heisenberg pictures are equivalent and our results can easily be transferred to the Heisenberg picture. It should be pointed out however that, at this stage, the quantum dynamics to be the same for the different $V_o$-labeled theories was the input requirement leading us to the mapping \eqref{VoHilbertspaceisomorphism}. In principle, it is possible to find different arguments leading to different mappings between the Hilbert spaces. The physical viability of the results needs of course to be investigated afterwards.} But the quantum theory is more than just dynamics of states and observables as it also describes their correlations and quantum fluctuations. What remains to be studied is then the relation of expectation values, higher moments, and uncertainty relations in the different $V_o$-quantised theories. As an example, it is easy to show the following relation for the expectation value of the field operator $\hat\phi$
\be\label{eq:scalingphiexpval}
	\left\langle\hat{\phi}\bigl|_{V_o^{(1)}}\right\rangle_{\Psi^{(1)}}= \braket{\Psi^{(1)}}{\hat{\phi}\bigl|_{V_o^{(1)}} \Psi^{(1)}}=\sqrt{\frac{V_o^{(2)}}{V_o^{(1)}}} \braket{\Psi^{(2)}}{\hat{\phi}\bigl|_{V_o^{(2)}} \Psi^{(2)}}=\sqrt{\frac{V_o^{(2)}}{V_o^{(1)}}} \left\langle\hat{\phi}\bigl|_{V_o^{(2)}}\right\rangle_{\Psi^{(2)}}\;,
\ee

\noindent
where in the second equality we used the expression $\langle\hat{\phi}\bigl|_{V_o^{(1)}}\rangle_{\Psi^{(1)}}=\int_{-\infty}^{\infty} \dd\phi\,\Psi^{(1)}(\phi)^*\,\phi\,\Psi^{(1)}(\phi)$ for the expectation value of the multiplicative operator $\hat\phi$ over the state $\Psi^{(1)}\in\mathscr H^{(1)}$, Eq. \eqref{eq:Vochangeenergyeigenstates}, and redefined the integration variable as $\phi'=\sqrt{V_o^{(1)}/V_o^{(2)}}\,\phi$. Similar computations lead to the results
\be\label{eq:scalingmoments}
\begin{aligned}
		\left<\hat{\phi}\bigl|_{V_o^{(1)}}^n\right>_{\Psi^{(1)}} =\; \sqrt{\frac{V_o^{(2)}}{V_o^{(1)}}}^{\,n} \left<\hat{\phi}\bigl|_{V_o^{(2)}}^n\right>_{\Psi^{(2)}}\qquad&,\qquad\left<\hat{\pi}\bigl|_{V_o^{(1)}}^n\right>_{\Psi^{(1)}} =\; \sqrt{\frac{V_o^{(2)}}{V_o^{(1)}}}^{\,n} \left<\hat{\pi}\bigl|_{V_o^{(2)}}^n\right>_{\Psi^{(2)}}\;,\\
		\Delta_{\Psi^{(1)}} \hat{\phi}\bigl|_{V_o^{(1)}} =\; \sqrt{\frac{V_o^{(2)}}{V_o^{(1)}}} \Delta_{\Psi^{(2)}} \hat{\phi}\bigl|_{V_o^{(2)}}\qquad &,\qquad\Delta_{\Psi^{(1)}} \hat{\pi}\bigl|_{V_o^{(1)}} =\; \sqrt{\frac{V_o^{(2)}}{V_o^{(1)}}} \Delta_{\Psi^{(2)}} \hat{\pi}\bigl|_{V_o^{(2)}} \;.
\end{aligned}
\ee

\noindent
The above list could be continued for any statistical cumulant as they follow from the moments. Of particular interest are the variances as they enter the Heisenberg uncertainty principle.~Given the canonical commutation relations \eqref{eq:quantisationrule}, this reads
\begin{equation}
	\Delta_{\Psi^{(1)}} \hat{\phi}\bigl|_{V_o^{(1)}} \Delta_{\Psi^{(1)}} \hat{\pi}\bigl|_{V_o^{(1)}} \ge \frac{1}{2} \left|\left<\left[\hat{\phi}\left|_{V_o^{(1)}},\hat{\pi}\right|_{V_o^{(1)}}\right]\right>_{\Psi^{(1)}}\right| = \frac{1}{2V_o^{(1)}} \;,
\end{equation}

\noindent
and, according to Eqs.~\eqref{eq:scalingmoments}, we find consistently
\begin{equation}\label{scalarfieldUR}
	\frac{1}{2V_o^{(1)}} \le \Delta_{\Psi^{(1)}} \hat{\phi}\bigl|_{V_o^{(1)}} \Delta_{\Psi^{(1)}} \hat{\pi}\bigl|_{V_o^{(1)}} = \frac{V_o^{(2)}}{V_o^{(1)}} \Delta_{\Psi^{(2)}} \hat{\phi}\bigl|_{V_o^{(2)}} \Delta_{\Psi^{(2)}} \hat{\pi}\bigl|_{V_o^{(2)}} \ge \frac{V_o^{(2)}}{V_o^{(1)}} \frac{1}{2 V_o^{(2)}} \;.
\end{equation}

\noindent
This tells us that the uncertainty relation set up for any value of the fiducial cell volume is consistent with any other fiducial cell choice.
Moreover, a state $\Psi^{(1)}$, which saturates the bound in the $V_o^{(1)}$ quantum theory, does the same in the $V_o^{(2)}$ theory.
The absolute value of such a lower bound does instead depend on the fiducial cell choice.
The same holds true for expectation values, which change their values.
Using instead the smeared momentum $\widehat{\pi(V_o)} = V_o \hat{\pi}$ the uncertainty relation with $\widehat{\phi(V_o)} = \hat{\phi}$ is indeed $V_o$ independent as the additional $V_o$ factor cancels the scaling behaviour.
However, measuring the same smeared observable $\widehat{\pi(V)} = \text{vol}(V) \hat{\pi}$ over $V\subset V_o$ in two different $V_o$-quantum theories leads to different predictions about the amount of quantum fluctuations. This is physically plausible as, even though the volume $V$ is the same in both quantum prescriptions, i.e. the size of the detector is the same, the scales of homogeneity $V_o^{(1)}$ and $V_o^{(2)}$ are different.
Non-local features of the quantum theory such as entanglement and correlations have in principle knowledge about the global system's size and as such are affected by the different values of $V_o$. Thus, as already discussed in the classical case, the choice of $V_o$ is physical and one has to think carefully about which value is physically reasonable.~For concreteness, let us make the above considerations explicit for Gaussian states.~In the $V_o^{(1)}$-quantum theory, a Gaussian state centered around the point $(\phi_o,\pi_o)$ reads
\begin{equation}\label{eq:gaussianstate}
	\Psi^{(1)}\left(\phi\right) = \frac{1}{\left(\pi \sigma^{(1)}\right)^\frac{1}{4}} \exp\left(-\frac{1}{2\sigma^{(1)}}\left(\phi-\phi_o\right)^2 + i V_o^{(1)} \pi_o \phi\right) \;,
\end{equation}

\noindent
where $\sigma^{(1)}=\frac{1}{mV_o^{(1)}}$ ($\hbar=1$)\footnote{Note that for $\sigma^{(1)} = \frac{1}{mV_0^{(1)}}$ this is a coherent state of the system. Similar considerations hold also for arbitrary Gaussian states with $\sigma^{(1)}$ as free parameter.}.
Then, according to the mapping \eqref{VoHilbertspaceisomorphism}, the corresponding $\mathscr{I}$-related state in the $V_o^{(2)}$-quantum theory is given by
\begin{equation}\label{eq:gaussianstateV2}
	\Psi^{(2)}\left(\phi\right) = \frac{1}{\left(\pi \sigma^{(2)}\right)^\frac{1}{4}} \exp\left(-\frac{1}{2\sigma^{(2)}}\left(\phi-\sqrt{\frac{V_o^{(1)}}{V_o^{(2)}}}\phi_o\right)^2 + i V_o^{(2)}\, \sqrt{\frac{V_o^{(1)}}{V_o^{(2)}}}\pi_o\,\phi\right) \;,
\end{equation}
with $\sigma^{(2)}=\frac{V_o^{(1)}}{V_o^{(2)}}\sigma^{(1)}$.~The relevant moments will thus also scale accordingly.~The results are collected in Tab.~\ref{table:comparisonexpvalues} from which we see that, as expected from \eqref{scalarfieldUR}, the uncertainty relations are saturated for Gaussian states, but the absolute value of the lower bound changes with the corresponding value of $V_o$.~Moreover, even though the dynamics of the above states is the same in the two quantum theories, the expectation values of the field operators $(\phi_o,\pi_o)\mapsto \sqrt{\frac{V_o^{(1)}}{V_o^{(2)}}}(\phi_o,\pi_o)$ and the width $\sigma^{(1)}\mapsto \frac{V_o^{(1)}}{V_o^{(2)}}\sigma^{(1)}$ do change with the value of $V_o$, thus affecting the interpretation of the states in the corresponding representations.~In particular, if we start with a very sharply peaked Gaussian state around the point $(\phi_o,\pi_o)$, say e.g. $\Psi^{(1)}$ in \eqref{eq:gaussianstate} for sufficiently large $V_{o}^{(1)}$ so that $\sigma^{(1)}$ is small enough, the corresponding Gaussian state $\Psi^{(2)}$ would be sharply peaked around the same point only if the ratio of the starting and rescaled values of $V_o$ is of order one.

\begin{table}
	\centering
	\begingroup 
	\renewcommand{\arraystretch}{1.2}
	\begin{tabular}[h!]{p{0.15\textwidth}|p{0.25\textwidth}|p{0.45\textwidth}}
		&$V_o^{(1)}$-quantisation& $V_o^{(2)}$-quantisation\\
		\hline
		\vspace{-1.35cm}
		\hspace{1cm}\text{state}
		&\vspace{-2cm}\vbox{\begin{align*}
				\Psi^{(1)}(\phi)
		\end{align*}}
		& \vbox{\begin{align*}
				\Psi^{(2)}\left(\phi\right) = \left(\frac{V_o^{(2)}}{V_o^{(1)}}\right)^{\frac{1}{4}} \Psi^{(1)}\left(\sqrt{\frac{V_o^{(2)}}{V_o^{(1)}}} \,\phi\right)
		\end{align*}}
		\vspace{-1cm}\\
		\hline
		\vbox{\begin{align*}
				\|\Psi\|^2
		\end{align*}}
		\vspace{-1cm}&\vbox{\begin{align*}
				1
		\end{align*}}
		\vspace{-1cm}& \vbox{\begin{align*}
				1
		\end{align*}}
		\vspace{-1cm}\\
		\hline
		\vbox{\begin{align*}
				\langle\hat{\phi}\rangle
		\end{align*}}
		\vspace{-1cm}&
		\vbox{\begin{align*}
				\phi_o
		\end{align*}}
		\vspace{-1cm}&
		\vbox{\begin{align*}
				\sqrt{\frac{V_o^{(1)}}{V_o^{(2)}}}\phi_o
		\end{align*}}
		\vspace{-1cm}\\
		\hline
		\vbox{\begin{align*}
				\left<\hat{\pi}\right>
		\end{align*}}
		\vspace{-1cm}&
		\vbox{\begin{align*}
				\pi_o
		\end{align*}}
		\vspace{-1cm}&
		\vbox{\begin{align*}
				\sqrt{\frac{V_o^{(1)}}{V_o^{(2)}}}\pi_o
		\end{align*}}
		\vspace{-1cm}\\
		\hline
		\vbox{\begin{align*}
				\sigma
		\end{align*}}
		\vspace{-1cm}&\vbox{\begin{align*}
				\sigma^{(1)}=\frac{1}{m V_o^{(1)}}
		\end{align*}}
		\vspace{-1cm}& \vbox{\begin{align*}
		 \sigma^{(2)}=\frac{V_o^{(1)}}{V_o^{(2)}}\,\sigma^{(1)}
		\end{align*}}
		\vspace{-1cm}\\
		\hline
		\vbox{\begin{align*}
				\Delta\hat{\phi}
		\end{align*}}
		\vspace{-1cm}&
		\vbox{\begin{align*}
				\sqrt{\frac{\sigma^{(1)}}{2}}
		\end{align*}}
		\vspace{-1cm}&
		\vbox{\begin{align*}
				\sqrt{\frac{V_o^{(1)}}{V_o^{(2)}}} \sqrt{\frac{\sigma^{(1)}}{2}}
		\end{align*}}
		\vspace{-1cm}\\
		\hline
		\vbox{\begin{align*}
				\Delta\hat{\pi}
		\end{align*}}
		\vspace{-1cm}&
		\vbox{\begin{align*}
				\frac{1}{\sqrt{2\sigma^{(1)}} V_o^{(1)}}
		\end{align*}}
		\vspace{-1cm}&
		\vbox{\begin{align*}
				\frac{1}{\sqrt{V_o^{(1)}\,V_o^{(2)}}} \frac{1}{\sqrt{2\sigma^{(1)}}}
		\end{align*}}
		\vspace{-1cm}\\
		\hline
		\vbox{\begin{align*}
				\Delta\hat{\phi}\Delta\hat{\pi}
		\end{align*}}
		\vspace{-1cm}&
		\vbox{\begin{align*}
				\frac{1}{2 V_o^{(1)}}
		\end{align*}}
		\vspace{-1cm}&
		\vbox{\begin{align*}
				\frac{1}{2V_o^{(2)}}
		\end{align*}}
		\vspace{-1cm}\\
	\end{tabular}
	\endgroup
	\caption{Comparison of the expectation values and variances for Gaussian states \eqref{eq:gaussianstate} and \eqref{eq:gaussianstateV2} in the theories associated with two different values $V_o^{(1)}$ and $V_o^{(2)}$ of the fiducial cell.}
	\label{table:comparisonexpvalues}
\end{table}
\newpage

\subsection{Excursus: Interacting Theory}\label{sec:quantuminteractions}

So far we have discussed only a free scalar field without any interaction.
As we have seen in the previous section, at the quantum level, it is possible to relate the quantisation of two different truncations of the full theory.
Even more such a relation leads to perfectly equivalent dynamics.
This is a property of the Hamiltonian itself and it worked as we were able to explicitly solve the corresponding Schr\"odinger equations.
In more complicated cases this might not be possible any more and we might see some non-trivial modification when going from one quantum theory to another.
To illustrate this, let us consider now a simple interacting (scalar) field theory, namely a $\phi^4$-theory whose Hamiltonian is given by
\begin{equation}
	H = \int_{\Sigma_t} \dd^3 x \left[ \frac{1}{2}\left(\pi(x)^2 + \partial_a \phi(x) \partial^a \phi(x) + m^2 \phi(x)^2\right) + \frac{\lambda}{4!} \phi(x)^4 \right]\;,
\end{equation}

\noindent
with $\lambda$ denoting the coupling constant. The implementation of homogeneity, the constraint analysis, and the corresponding Dirac bracket are purely kinematic and work exactly as in Sec.~\ref{sec:implementinghom} also for the interacting case.
Using then the arguments of Sec.~\ref{sec:truncation} to truncate the theory, we arrive at the class of $V_o$-dependent spatially homogeneous theories described by the following Hamiltonians and Dirac brackets 
\begin{equation}
	H = \frac{V_o}{2} \left(\pi^2 + m^2 \phi^2\right) + \frac{V_o \lambda}{4!} \phi^4\qquad , \qquad \Poisson{\phi}{\pi}_D = \frac{1}{V_o} \;.
\end{equation}

\noindent
Quantising now along the lines of Sec.~\ref{sec:phiquantisation}, we get the quantum representations
\be
		\bigl[\hat{\phi},\hat{\pi}\bigr] = \frac{i}{V_o} \qquad,
		\qquad
		\hat{\phi} \Psi(\phi) = \phi \Psi(\phi) \qquad , \qquad \hat{\pi} \Psi(\phi) = -\frac{i}{V_o} \partdif{}{\phi} \Psi(\phi) \;,
\ee

\noindent
with $\Psi(\phi)\in\mathscr{H} = L^2\left(\mathbb{R},\dd \phi\right)$, for each given value of $V_o$, and the corresponding Hamiltonian operators read
\begin{equation}
	\hat{H} \Psi(\phi) = \frac{1}{2} \left(-\frac{1}{V_o} \frac{\partial^2}{\partial \phi^2} + V_o m^2 \phi^2 + \frac{V_o \lambda}{4!} \phi^4\right) \Psi(\phi)\;.
\end{equation} 

\noindent
Due to the interaction term, this is analogous to the Hamiltonian of an anharmonic oscillator, whose eigenstates cannot be computed analytically.
However, we do not need to solve for them explicitly but rather only figure out how these solutions depend on $V_o$.
In this regard, let us again introduce the quantity $\zeta = \sqrt{V_o\,m} \phi$ so that the energy eigenfunctions satisfy the equation
\begin{equation}
	 \frac{m}{2} \left(-\frac{\partial^2}{\partial \zeta^2} + \zeta^2 + \frac{\lambda}{4!\, V_o\, m^3} \zeta^4\right) \Psi(\zeta) = m \epsilon_n\left(\lambda/(V_o\,m^3)\right) \Psi(\zeta)\;,
\end{equation}

\noindent
The energy eigenvalues are thus given by $E_n(\lambda/(V_o m^3)) = m \epsilon_n(\lambda/(V_o m^3))$, where the energy is measured in terms of $m$ and $\epsilon_n$ is dimensionless.
In the non-interacting case, the above rescaling of $\phi$ leads to a Hamiltonian independent of $V_o$.
This is not the case when interaction terms are present as the latter still depend on $V_o$.
We could then use our favourite solution strategy to solve this system and find the energy eigenstates.
Whatever strategy one decides to use to tackle this problem, one would arrive at solutions
\begin{equation}
	\hat{H}\Psi_n(\zeta; \bar{\lambda}) = m\epsilon_n(\bar{\lambda}) \Psi_n(\zeta; \bar{\lambda})\;,
\end{equation}

\noindent
which only depend on the \textit{effective} coupling $\bar{\lambda} = \lambda/(V_o\, m^3)$ and energy eigenvalues depending multiplicatively on $m$.
Normalising these states w.r.t. the scalar product on $L^2\left(\mathbb{R},\dd \zeta\right)$, we can relate them again to the $V_o$-quantisation by means of
\begin{equation}
	\Psi_n^{(V_o)}(\phi; m , \lambda) = \left(V_o\,m\right)^{\frac{1}{4}} \Psi_n\left(\sqrt{V_o\,m} \phi; \lambda/V_o\,m^3\right) \qquad , \qquad E_n^{(V_o)}(m,\lambda) = E_n(\lambda/V_o\,m^3) \;.
\end{equation}

\noindent
Already from such a preliminary discussion, we see that the scale of homogeneity $V_o$ comes to play an even more prominent role in the interacting theory in that:
\begin{itemize}
	\item The Hamiltonian and eigenvalues depend on $V_o$. The latter then enters directly the time evolution.
	Thus, unlike the non interacting case discussed in the previous subsection, in the interacting case it is not possible any more to identify states in a $V_o^{(1)}$-quantisation with states in a $V_o^{(2)}$-quantisation as their dynamics will be different.
	\item This is a pure quantum effect as classically the equations of motion do not depend on $V_o$.
	\item More specifically, $V_o$ enters the effective coupling constant $\bar{\lambda}$ and therefore determines the strength of interaction, the latter depending on the choice of $V_o$.
	It would be tempting to view $V_o$ as the energy scale in a RG-flow, but this would be purely speculative at this stage.
	The validity of perturbation theory depends then on $\bar\lambda$, i.e. it depends on the ratio of the ``bare'' coupling constant $\lambda$ and $V_o$. 
	\item Dynamics of two different $V_o$-quantised theories can still be made independent of $V_o$/equivalent if one changes the values of the coupling constants $(m,\lambda)$.
	In theory space, it would then be possible to make the dynamics $V_o$-independent again as in the classical theory, but $(m,\lambda)$ must change with the scale $V_o$.
\end{itemize}

\section{Quantum Symmetry Reduction of a Scalar Field}\label{sec:symredscalarQFT}

In the previous sections, we discussed in detail how symmetry reduction of a classical field theory can be performed within the canonical Hamiltonian framework to arrive at the classical mini-superspace model and how to quantise the resulting spatially homogeneous and isotropic theories.~Quantum mini-superspace models such as cosmology should however be viewed as the symmetry reduced sector of the full quantum gravity theory.~In other words, since the world is ultimately quantum at a fundamental level and in order to bridge between the full QG theory and large scale phenomenology, a question raising somewhat naturally is then:~How can the symmetry reduction be performed directly at the full quantum level?~This has been studied extensively and is notably very hard for full quantum gravity.~A sample of works in LQG include \cite{AlesciQuantum-ReducedLoop,EngleRelatingLoopQuantum,EngleEmbeddingLoopQuantum,BeetleDiffeomorphismInvariantCosmological,BeetleQuantumisotropyand,BodendorferAnEmbeddingOf,BodendorferQuantumReductionTo, BodendorferOntherelation,BojowaldSymmetryReductionFor,KoslowskiReductionofa,EngleQFTandits,AlesciImprovedregularizationfrom,AlesciQuantumReducedLoopGravitySemiclassicalLimit,AlesciCosmologicalsingularityresolution,AlesciPhenomenologyofQRLG,HanEffectivedynamicsfromcoherent,HanLQGondynamicallattice,DaporCosmologicaleffectiveHamiltonian,AlesciLoopQuantumCosmology,AlesciQuantumreducedloop,BodendorferCoarseGrainingAs,BodendorferRenormalisationwithsu11} for the case of cosmology or \cite{EngleQFTandits} for a systematic analysis of the scalar field case.~In light of the results of the previous sections, the aim of the present section is to gain some intuition by focusing on the simple case of a quantum scalar field theory, select plausible observables and states in the full QFT, and identify the conditions for the quantum homogeneous description to hold within the full theory.~A systematic symmetry reduction imposing the second class constraints \eqref{eq:hominbox}-\eqref{eq:homacrossbox} at the quantum level along the lines of \cite{EngleQFTandits} for cylindrical symmetry is shifted elsewhere.

Let us start then by reviewing the main ingredients and fix the notation for a free real quantum scalar field on Minkowski background.
The Hamiltonian of the classical theory is given by (cfr.~Eq.~\eqref{eq:scalarfieldHamiltonian})
\begin{equation}
	H = \int_{\Sigma_t} \dd^3 x \frac{1}{2}\left(\pi(x)^2 + \partial_a \phi(x) \partial^a \phi(x) + m^2 \phi(x)^2\right) = \int_{\mathbb{R}^3} \frac{\dd^3k}{\sqrt{2\pi}^3} \frac{1}{2} \left(\left|\tilde{\pi}(k)\right|^2+\omega(k)^2 \bigl|\tilde{\phi}\left(k\right)^2\bigr|\right)\;,
\end{equation}

\noindent
with $\omega(k) = +\sqrt{k^2 + m^2}$ and the Fourier transform on the spatital slice $\Sigma_t$ reads as
\begin{equation}
	\tilde{\phi}(k) = \int_{\Sigma_t} \frac{\dd^3x}{\sqrt{2\pi}^3}\, \phi(x) e^{-ik\cdot x} \qquad , \qquad 
	\tilde{\pi}(k) = \int_{\Sigma_t} \frac{\dd^3x}{\sqrt{2\pi}^3}\, \pi(x) e^{-ik\cdot x} \;.
\end{equation}

\noindent
The assumption of the scalar field to be real amounts to the following reality conditions
\begin{equation}
	\tilde{\phi}(k) = \tilde{\phi}^*(-k) \qquad , \qquad \tilde{\pi}(k) = \tilde{\pi}^*(-k) \;.
\end{equation}

\noindent
Following the standard procedure for quantisation, we can define the creation and annihilation operators
\be
		\hat{a}(k) = \sqrt{\frac{\omega(k)}{2}} \hat{\tilde{\phi}}(k) + i\frac{\hat{\tilde{\pi}}(k)}{\sqrt{2\omega(k)}} \qquad , \qquad \hat{a}^\dagger(k) = \sqrt{\frac{\omega(k)}{2}} \hat{\tilde{\phi}}(-k) -i \frac{\hat{\tilde{\pi}}(-k)}{\sqrt{2\omega(k)}}\;,
\ee

\noindent
so that 
\be\label{eq:fieldops}
\hat{\tilde{\phi}}(k) = \frac{\hat{a}(k) + \hat{a}^\dagger(-k)}{\sqrt{2\omega(k)}}\qquad , \qquad \hat{\tilde{\pi}}(k) = -i\sqrt{\frac{\omega(k)}{2}} \left(a(k) - a^\dagger(-k)\right)\;,
\ee

\noindent
with the commutation relations
\begin{equation}
	\left[\hat{a}(k),\hat{a}^\dagger(p)\right] = \delta(k-p) \qquad , \qquad \left[\hat{\tilde{\phi}}(k) , \hat{\tilde{\pi}}(p)\right] = i\delta\left(k+p\right)\;.
\end{equation}

\noindent
The normal-ordered Hamiltonian operator then becomes
\begin{equation}\label{HamoperatorfullQFT}
	\hat{H} = \int_{\mathbb{R}^3} \dd^3 k\, \omega(k) \hat{a}^\dagger(k) \hat{a}(k) \;,
\end{equation}

\noindent
and the time-dependent field operators are given by
\begin{align}
	\hat{\phi}(x,t) &= e^{-it\hat{H}} \hat{\phi}(x) e^{it \hat{H}} = \int_{\mathbb{R}^3}\frac{\dd^3 k}{\sqrt{2\pi}^3 \sqrt{2\omega(k)}} \left( \hat{a}(k) e^{-i\omega(k)t+ik\cdot x} + \hat{a}^\dagger(k) e^{i\omega(k)t-ik\cdot x}\right)  \;,\\
	\hat{\pi}(x,t) &= e^{-it\hat{H}} \hat{\pi}(x) e^{it \hat{H}} = \partial_t \hat{\phi}(x,t) \;,
\end{align}

\noindent
where $\hat{\phi}(x)$ is the inverse Fourier transform of $\hat{\tilde{\phi}}(k)$ and these operators satisfy the standard equal-time commutation relations 
\begin{equation}
	\left[\hat{\phi}(x,t),\hat{\pi}(y,t)\right] = \left[\hat{\phi}(x),\hat{\pi}(y)\right] = i \delta(x-y) \;.
\end{equation}

\noindent
The Hilbert space is the usual Fock space with vacuum state $\ket{0}$
  such that $\hat{a}(k)\ket0=0$ for all anihilation operators $\hat{a}(k)$ (see e.g.~\cite{weinberg_1995,Peskin:1995ev} for standard textbook references).

We shall now proceed to the homogeneous symmetry reduction of our real quantum scalar field theory.
As mentioned above, in our present discussion we shall take a more intuitive approach rather than a fully fledged systematic analysis, while leaving the latter to future study.
First of all, let us select the observables of interest.
On a Cauchy slice $\Sigma_t$, these are given by the spatially averaged fields over a region $V\subset\Sigma_t$\footnote{Here we are defining the initial time operators. They still have to evolve according to the equations of motion to obtain fully dynamical operators.} 
\begin{equation}
	\hat{\phi}\left(V\right) = \frac{1}{\text{vol}(V)} \int_V \dd^3 x \hat{\phi}(x) \qquad , \qquad 
	\hat{\pi}\left(V\right) = \int_V \dd^3 x \hat{\pi}(x) \;,
\end{equation}

\noindent
and carry only homogeneous information about the volume of the smearing region $V$.
All local degrees of freedom in the interior of $V$ are integrated out.
For a cubic box of edge length $L$ (as chosen in Sec.~\ref{Sec:Warmupscalarfield}), this can be explicitly written as
\begin{equation}\label{eq:phiV}
	\hat{\phi}\left(V\right) = \frac{1}{L^3} \int_V \dd^3 x \int_{\mathbb{R}^3} \frac{\dd^3 k}{\sqrt{2\pi}^3} \hat{\tilde{\phi}}(k) e^{+ik\cdot x}= \int_{\mathbb{R}^3} \frac{\dd^3 k}{\sqrt{2\pi}^3} \hat{\tilde{\phi}}(k) f_V(k) \;,
\end{equation}

\noindent
with
\begin{equation}\label{def:fVofk}
	f_V(k) := \int_V \frac{\dd^3x}{L^3} e^{+ik\cdot x} = \prod_{\xi = {x,y,z}} e^{i k_\xi \left(\xi_o + \frac{L}{2}\right)} \frac{2 \sin\left(\frac{k_\xi L}{2}\right)}{k_\xi L} \;,
\end{equation}

\noindent
and $\xi_o=\left(x_o, y_o, z_o\right)$ the unit vector indicating the lower left edge of the box (cfr.~Sec.~\ref{sec:truncation}).
Similarly,
\begin{equation}\label{eq:scalfieldpiV}
	\hat{\pi}(V) = \int_V \dd^3 x \,\hat{\pi}(x) = L^3 \int_{\mathbb{R}^3} \frac{\dd^3k}{\sqrt{2\pi}^3}\, f_V(k) \hat{\tilde{\pi}}(k) \;
\end{equation}

\noindent
and it is straight forward to check that these operators satisfy the commutation relations
\begin{equation}
	\left[\hat{\phi}(V),\hat{\pi}(V)\right] = i \;.
\end{equation}

\noindent
The operators \eqref{eq:phiV}, \eqref{eq:scalfieldpiV} can be identified with the ones in Sec.~\ref{sec:phiquantisation} with $V = V_o$.\footnote{To be precise, we have to identify $\hat{\pi}(V)$ with the operator $L^3 \hat{\pi}$ in Sec.~\ref{sec:phiquantisation}, consistently with $\pi$ being a density 1 object as discussed multiple times throughout the paper.}
Consequently, we can define the creation and annihilation operators 
\begin{equation}\label{eq:QFTsymredaV}
	\hat{a}(V) = \sqrt{\frac{L^3\,m}{2}} \left(\hat{\phi}(V)+\frac{i}{m\,L^3} \hat{\pi}(V)\right) \qquad , \qquad \hat{a}^\dagger(V) = \sqrt{\frac{L^3\,m}{2}} \left(\hat{\phi}(V)-\frac{i}{m\,L^3} \hat{\pi}(V)\right)\,,
\end{equation}

\noindent
with the commutator
\begin{equation}
	\left[\hat{a}(V),\hat{a}^\dagger(V)\right] = 1 \;.
\end{equation}

\noindent
The states obtained by acting with the creation operators $\hat a^{\dagger}(V)$ on the vacuum $\ket0$ of our QFT, namely
\begin{align}\label{eq:symmredstates}
		\ket{V,n} &= \frac{\left(\hat{a}^\dagger(V)\right)^n}{\sqrt{n!}}\ket{0} \;,
\end{align}
	
\noindent
are orthonormal $\braket{V,n}{V,m} = \delta_{nm}$ and form a complete basis spanning a separable Hilbert subspace 
\begin{equation}\label{def:HVsubspace}
	\mathscr{H}_V = \ket{0} \oplus \text{span}_{\mathbb{C}} \left\{\ket{V,n} \;\middle| \; n \in \mathbb{N}\right\} \;.
\end{equation}

\noindent
Such a Hilbert subspace is isomorphic to the Hilbert space of the classically symmetry reduced theory in Sec.~\ref{sec:phiquantisation}.
However, we need to check that this Hilbert subspace remains closed under the action of the observables and, in order to preserve dynamics, also under the action of the Hamiltonian operator.

Before discussing this point, let us provide some further interpretation for the states \eqref{eq:symmredstates} by rewriting them in terms of the full theory operators.
To this aim, it is straight forward to verify that
\begin{subequations}
	\begin{align}
		\hat{a}(V) &= \sqrt{\frac{L^3 m}{2}} \int_{\mathbb{R}^3} \frac{\dd^3 k}{\sqrt{2\pi}^3} \, f_V(k) \left(\hat{\tilde{\phi}}(k) + \frac{i}{m} \hat{\tilde{\pi}}(k)\right)
		\notag
		\\
		&= \sqrt{\frac{L^3 m}{2}} \int_{\mathbb{R}^3} \frac{\dd^3 k}{\sqrt{2\pi}^3}\,\left(\frac{f_V(k)}{\sqrt{2\omega(k)}}\left(1+\frac{\omega(k)}{m}\right)\hat{a}(k)+ \frac{f^*_V(k)}{\sqrt{2\omega(k)}}\left(1-\frac{\omega(k)}{m}\right)\hat{a}^\dagger(k)\right)\label{aofVinfullqft}
		\\
		\hat{a}^\dagger(V) &= \sqrt{\frac{L^3 m}{2}} \int_{\mathbb{R}^3} \frac{\dd^3 k}{\sqrt{2\pi}^3}\,\left(\frac{f_V(k)}{\sqrt{2\omega(k)}}\left(1-\frac{\omega(k)}{m}\right)\hat{a}(k)+ \frac{f^*_V(k)}{\sqrt{2\omega(k)}}\left(1+\frac{\omega(k)}{m}\right)\hat{a}^\dagger(k)\right)\label{adaggerofVinfullqft}
	\end{align}
\end{subequations}

\noindent
from which we see that the creation operator $\hat{a}^\dagger(V)$ creates a particle with momentum profile $f^*_V(k)\left(1+\frac{\omega(k)}{m}\right)$, while it annihilates a particle with momentum profile $f_V(k)\left(1-\frac{\omega(k)}{m}\right)$.
Consequently, for any given $n$, the states $\ket{V, n}\in\mathscr{H}_V$ contains contributions from any other particle state of the full theory obtained by acting with combinations of up to $n$ creation operators onto the vacuum.
In a large mass, small momenta expansion, the momentum profile can be written as 
\begin{align}
	\frac{f_V(k)}{\sqrt{2\omega(k)}} \left(1- \frac{\omega(k)}{m}\right) = -2\frac{f_V(k)}{\sqrt{2m}} \left(\frac{k^2}{m^2} + \mathcal{O}\left(\frac{k^4}{m^4}\right)\right) \propto \frac{k^2}{L\, k_x\,k_y\,k_z} \cdot \frac{1}{L^2 m^2} + \mathcal{O}\left(\frac{k^4}{m^4}\right) \;. 
\end{align}

\noindent
which for $m \gg \frac{1}{L}$ is approximately zero and can be neglected. Intuitively, in fact, $\omega(k) \approx m$ for large masses and small momenta.
However, as the above quantity enters in a momentum integral in \eqref{adaggerofVinfullqft}, this would not be a useful criterion for this approximation as at some point the momentum will be larger than the mass.
Nevertheless, the momentum distribution $f_V(k)$ suppresses momenta larger than $L$.
Therefore, as long as we can guarantee that the scale where $k \sim m$ is in the regime $k \gg 1/L$, these contributions are suppressed by the function $f_V(k)$.
As an explicit example, let us look at the vacuum contribution to the state $\ket{V,2}$ which turns out to be small.\footnote{Similar computation is expected to go through for higher states.}
This can be seen from
\begin{align}
	\braket{0}{V,2} = \frac{L^3 m}{(2\sqrt{2}\pi)^3} \int_{\mathbb{R}^3} \dd^3k\, \frac{\left|f_V(k)\right|^2}{2\omega(k)} \left(1-\frac{\omega(k)^2}{m^2}\right) \;,
\end{align}

\noindent
which can be split into an integral with momenta $k_\xi < k_o $ and $k_\xi > k_o$, with $1 \ll L k_o \ll Lm$.
As $k_o \ll m$, the integral with $k_\xi < k_o$ is bounded by the order of $k_o^2/m^2$, which is then negligible.
For the momenta $k_\xi > k_o$, on the other hand, $k_\xi L \gg 1$ and the contributions coming from $\left|f_V(k)\right|^2$ are small.
Thus, approximating the integrand for $|k| > |k_o|$, we have
\begin{align}
	\left|L^3 m\,\dd^3 k \, \frac{\left|f_V(k)\right|^2}{2\omega(k)} \left(1-\frac{\omega(k)^2}{m^2}\right) \right| \stackrel{|k| \gg L^{-1}}{<}& \left|\text{angular integral}\cdot L^3 m \,\dd |k| |k|^2 \frac{1}{L^6\,|k|^6} \frac{1}{2 |k|} \frac{|k|^2}{m^2} \right|
	\notag
	\\
	=&\left| \text{angular integral}\cdot \dd |k| \frac{1}{2 |k|^3 m L^3}\right| \;,
\end{align} 

\noindent
whose contribution is finite and suppressed by the order of $1/L^3\,m\, k_o^2$.~By similar arguments, we can approximate 
\begin{equation}
	\sqrt{\frac{L^3 m}{2}} \frac{f_V(k)}{\sqrt{2\omega(k)}} \left(1+ \frac{\omega(k)}{m}\right) \simeq \sqrt{L^3} f_V(k) \qquad \text{for}\quad m \gg \frac{1}{L}\;,
\end{equation}

\noindent
so that, within this approximation, the $V$-creation and annihilation operators \eqref{aofVinfullqft}, \eqref{adaggerofVinfullqft} can be written as
\begin{equation}\label{largemassladderop}
	\hat{a}(V) \simeq \sqrt{L^3} \int_{\mathbb{R}^3}\frac{\dd^3 k}{\sqrt{2\pi}^3}\,f_V(k) \hat{a}(k) \qquad ,\qquad \hat{a}^\dagger(V) \simeq \sqrt{L^3} \int_{\mathbb{R}^3}\frac{\dd^3 k}{\sqrt{2\pi}^3}\,f^*_V(k) \hat{a}^\dagger(k)\,.
\end{equation}

\noindent
This allows us to interpret the states $\ket{V,n}$ as follows.
Since $f_V(k)$ is the Fourier transform of the characteristic function of the region $V$ (cfr.~Eq.~\eqref{def:fVofk}), in the approximation $Lm\gg 1$, the state $\ket{V,1}$ has the simple interpretation of a single particle being excited in the spatially bounded region $V$ only.
According to the r.h.s. of \eqref{def:fVofk}, its momentum profile is the one discussed in Sec.~\ref{sec:truncation} (cfr. Fig. \ref{fig:sinx} and the surrounding discussion) with the dominant contribution being homogeneous ($k_i=0$) in this volume and thus not containing any local information besides of the knowledge of $V$ itself.
However, as already discussed in Sec.~\ref{sec:truncation}, these states are not fully homogeneous as some subleading inhomogeneous modes in principle do contribute to the momentum profile as needed to localise them on a finite volume. Similarly, the states $\ket{V,n}$ correspond approximately to states where $n$ homogeneous particles are excited in the box $V$.
Moreover, the fact that, as we argued in Secs.~\ref{sec:truncation} and \ref{sec:phiquantisation}, the spatially homogeneous theory results from both truncating the theory and neglecting all remaining momenta, which translates into ignoring boundary conditions at $\partial V$, is equivalent to our condition $m \gg 1/L$, which was in fact already found in Sec.~\ref{sec:truncation}. The latter can be provided with the following physical interpretation.
For large masses, large momenta include also small velocities so that the particle is moving only slowly out of the volume $V$.
Similarly, large $L$ means large volumes and the flow through the boundary surface of $V$ becomes negligible compared to the bulk physics.
In both scenarios, the boundary contributions become sub-dominant on large homogeneity scales.

Let us now come back to the question of whether the Hilbert subspace $\mathscr{H}_V$ defined in Eq.~\eqref{def:HVsubspace} is preserved under the action of our observables $\hat{\phi}(V)$, $\hat{\pi}(V)$ and most importantly under dynamics, i.e. under the action of the Hamiltonian operator.
By construction, the operators $\hat{\phi}(V)$ and $\hat{\pi}(V)$ can be written in terms of the creation and annihilation operators only (cfr.~Eq.~\eqref{eq:QFTsymredaV}).
These are ladder operators on the Hilbert subspace $\mathscr{H}_V$ and therefore preserve its structure.
It remains then to check only if the Hamiltonian preserves this subspace.
To this aim, let us use again the approximation $m \gg 1/L$ to write 
\begin{equation}\label{Vnstatesapprox}
	\ket{V,n} \simeq N_n \int_{\mathbb{R}^3} \dd^3 k_n\, f_V(k_1) \dots f_V(k_n) \ket{k_1,\dots,k_n} \;,
\end{equation}
 
\noindent
where we used the definition \eqref{eq:symmredstates} of the states $\ket{V,n}$ together with \eqref{largemassladderop}, and the normalisation constant $N_n$ is given by $N_n = \sqrt{\frac{L^{3n}}{\left(2\pi\right)^{3n}\,n!}}$.
The action of the Hamiltonian \eqref{HamoperatorfullQFT} on the states \eqref{Vnstatesapprox} can be then approximated as
\begin{align}
	\hat{H} \ket{V,n} &\simeq N_n \int_{\mathbb{R}^3}\dd^3 k_1 \dots \dd^3 k_n f_V(k_1) \dots f_V(k_n) \hat{H} \ket{k_1,\dots,k_n} 
	\nonumber
	\\
	&= N_n \int_{\mathbb{R}^3}\dd^3 k_1 \dots \dd^3 k_n f_V(k_1) \dots f_V(k_n) \left(\sum_{i=1}^{n}\omega(k_i)\right) \ket{k_1,\dots,k_n}\nonumber
	\\
	&\hspace{-0.35cm}\stackrel{m \gg 1/L}{\approx} m\cdot n \ket{V,n} \in \mathscr{H}_V \;,\label{eq:approxenergyspectrum}
\end{align}

\noindent
i.e., within this approximation, the Hilbert subspace $\mathscr{H}_V$ is preserved under the action of the Hamiltonian\footnote{Note that this check is only for the Hamiltonian operator, i.e.~infinitesimal time steps. In the time evolution operator, the error is exponentiated, and moreover depends on the time.~Analysing this further would give a time scale at which this approximation fails again, even if the error is very small in the computation above.}. Note that Eq. \eqref{eq:approxenergyspectrum} also shows that, in the above approximation, $\ket{V,n}$ are eigenstates of $\hat H$ with the energy spectrum given by $m\cdot n$. This coincides with the spectrum for the quantised minisuperspace Hamiltonian \eqref{eq:minisupHspectrum} of Sec.~\ref{sec:phiquantisation} up to the vacuum energy, the latter being removed by normal ordering.

Therefore, using the above approximation, we arrive at a sector of the full quantum scalar field theory, which is equivalent to the one obtained from classically symmetry reducing first and then quantising.
Let us further emphasise that this approximation is exactly the truncation we did in Sec.~\ref{sec:truncation}.
Leaving this regime, it is not possible any more to embed the quantisation of the classically symmetry reduced theory into the full theory.
Moreover, we can now judge from the full quantum dynamics perspective under which conditions the truncation of Sec.~\ref{sec:truncation} is reasonable.
The result of the symmetry-reduction of the classical field theory discussed in Sec.~\ref{sec:truncation} can be in fact reproduced and we find that the truncated quantum theory well approximates the symmetry-reduced/volume-averaged homogeneous sector of the full quantum theory as long as the condition $m \gg 1/L$ holds.
This has the two-fold explanation that neglecting large momenta is fine, as the smeared quantum observables $\hat{\phi}(V)$ and $\hat{\pi}(V)$ themselves do not depend much on momenta $k_\xi \gg 1/L$.
Consequently, the physics is not substantially changed when neglecting these momenta with wavelength smaller than the edge length of the volume of interest as this local information is not considered.
The choice of the observables $\hat{\phi}(V)$ and $\hat{\pi}(V)$ is in principle dictated by the symmetry-reduction constraints, although here we have chosen an intuitive rather than a systematic approach.~On the other hand, the truncation in Sec.~\ref{sec:truncation} also ignores momenta with $k_\xi L < 1$.
However, these momenta are not relevant for the physics as we can approximate $\omega(k) = \sqrt{k^2 + m^2} \approx m$ without troubles as long as $m \gg 1/L$, which comes now as a dynamical statement as the spectrum of the Hamiltonian operator enters the argument.
Moreover, as this leads to an approximation for the dynamics of the states and observables, there exists a maximal time scale when this error accumulates and becomes not negligible any more.
It would be interesting to extend such considerations to the case of cosmology as a large time scale is covered.
This would however require inputs from the full QFT, so that the derivation of similar conditions for the cosmological case will in general be way more involved (see also the discussion in Sec. \ref{outlook:beyondhom}).

As a last point, let us close the discussion of the quantum symmetry-reduction of our free scalar field theory by making contact with the scaling behaviour of expectation values discussed in Sec.~\ref{sec:phiquantisation}. There, we saw that expectation values in the quantum theory resulting from quantising the classically symmetry-reduced homogeneous theory are $V$ dependent and have a well-defined scaling behaviour. We can now study the effect of changing the volume $V$ and the resulting scaling properties from a full QFT perspective.
To this aim, let us consider two boxes $V^{(1)}$ and $V^{(2)}$ with edge lengths $L_1$ and $L_2$, respectively.
We can then define the states
\begin{equation}
	\ket{\Psi^{(1)}} = \sum_{n=0}^{\infty} \Psi_n \ket{V^{(1)}, n} \qquad , \qquad \ket{\Psi^{(2)}} = \sum_{n=0}^{\infty} \Psi_n \ket{V^{(2)}, n}
\end{equation}

\noindent
both of which obviously belong to the full theory Fock space.
However, $\ket{\Psi^{(1)}} \notin \mathscr{H}_{V^{(2)}}$ and vice versa.
Nevertheless, in the approximation $L_1 m, \, L_2 m \gg 1$, they have the exact same dynamical behaviour as then (cfr.~Eq.~\eqref{eq:approxenergyspectrum})
\be
\hat{H} \ket{V^{(1)}, n} \simeq m\cdot n\,\ket{V^{(1)}, n} \qquad , \qquad  \hat{H} \ket{V^{(2)}, n} \simeq m\cdot n\,\ket{V^{(2)}, n} \;.
\ee

\noindent
Specifically, since $\ket{\Psi^{(1)}}$ and $\ket{\Psi^{(2)}}$ given above have the same expansion coefficients in the bases $\ket{V^{(1)},n}$ and $\ket{V^{(2)},n}$, their approximate dynamics is equivalent and a mapping between states with equivalent dynamics can be defined as in Sec.~\ref{sec:phiquantisation}, Eq. \eqref{VoHilbertspaceisomorphism}.
Inverting then the expressions~\eqref{eq:QFTsymredaV} to write $\hat\phi(V)$ in terms of ladder operators, a straight forward computation yields
\begin{align}
	\braopket{\Psi^{(1)}}{\hat{\phi}(V^{(1)})}{\Psi^{(1)}} =&\; \sum_{n,l=0}^{\infty} \Psi_n^*\,\Psi_l \braopket{V^{(1)},n}{\frac{\hat{a}(V)+\hat{a}^\dagger(V)}{\sqrt{2mL_1^3}}}{V^{(1)},l} 
	\notag
	\\
	=&\; \sum_{n=0}^{\infty} \frac{1}{\sqrt{2mL_1^3}} \left(\sqrt{n+1}\,\Psi_n^*\,\Psi_{n+1}+\sqrt{n}\,\Psi_n^*\,\Psi_{n-1}\right)
	\notag
	\\
	=&\; \sqrt{\frac{L_2^3}{L_1^3}}\, \braopket{\Psi^{(2)}}{\hat{\phi}(V^{(2)})}{\Psi^{(2)}}\;.
\end{align} 

\noindent
which is in agreement with the result~\eqref{eq:scalingphiexpval} obtained in Sec.~\ref{sec:phiquantisation} for the classically symmetry-reduced and then quantised theory. Similar computations can be then performed for the expectation value of $\hat{\pi}(V^{(1)})$ as well as for higher moments. Thus, consistently with the results of Sec.~\ref{sec:phiquantisation} we find that, in the approximation in which the above scaling behaviours can be derived, the uncertainty relations of the smeared operators are independent of the specific region $V$, namely $\Delta\hat\phi(V^{(1)})\Delta\hat\pi(V^{(1)})=\Delta\hat\phi(V^{(2)})\Delta\hat\pi(V^{(2)})$. However, in light of the above discussion, we see that the size of the region considered cannot become arbitrarily small for the behaviour of the minisuperspace theory to be reproduced within a full theory setting as otherwise the approximation $Lm\gg1$ would break down for the corresponding $\mathscr H_{V}$ sector.

Before moving to cosmology in the rest of the paper, let us close this section by emphasising that, as already mentioned, the above discussion was meant just as a sketch of how the symmetry reduction can be worked out at the quantum level. In particular, unlike it was the case for the study of the classical theory in Sec.~\ref{Sec:Warmupscalarfield}, a systematic introduction and implementation of the quantum analogue of the symmetry-reduction constraints would still be needed.
In this respect, it is worth mentioning that research in similar direction exists (see e.g. \cite{EngleQFTandits} and reference therein), although in terms of different symmetries.
It would be interesting to make contact with such line of research and eventually adapt it to the to the case of homogeneity for the setting presented in this paper.
This would also give a tool to tackle more complicated situations where one does not restrict to homogeneity and eventually other maybe not so intuitive symmetries.
Moreover, it would be insightful to repeat the above construction for the interacting theory.
In fact, as we have seen in Sec.~\ref{sec:quantuminteractions}, for the case of an interacting homogeneous scalar field, the coupling constant depends on the fiducial volume which in turn seems to suggest that, in order for the dynamics to be $V_o$-independent as in the classical theory, the constants of the theory $(m,\lambda)$ would need to change with the scale over which homogeneity is imposed. Performing the symmetry-reduction at the quantum level for the interacting theory might then help us to gain insight into the physics of such a dependence and eventually make an interpretation in terms of a renormalisation group flow possible. We shall comment on this in the outlook discussion of Sec.~\ref{sec:conclusion}

\section{Symmetry Reduction for Classical Cosmology}\label{Sec:Comsology}

Let us now turn our discussion to gravity. In particular, in the light of what we have learned from the study of classical mini-superspace reduction for the simple example of a scalar field theory and its implications for the quantisation of the resulting symmetry-reduced theories, we would like to import now similar conclusions for symmetry-reduction to homogeneous and isotropic cosmology. This will eventually help us to draw useful lessons and insights for the quantisation of classically symmetry-reduced cosmological models. Specifically, this will help us to shed some light on the role and interpretation of the homogeneity region (usually referred to as \textit{fiducial cell} in the LQC literature \cite{BojowaldLoopQuantumCosmology, AshtekarLoopquantumcosmology:astatusreport, BojowaldTheBKLscenario,BojowaldEffectiveFieldTheory,BojowaldMinisuperspacemodelsas, AshtekarQuantumNatureOf}) and its main consequences for the study of quantum fluctuations.~We will thus follow a similar logic to the scalar field case for implementing spatial homogeneity at the classical level via constraints and constructing the associated Dirac bracket. In order to avoid repetitions as much as possible, we will try to focus on the main conceptual and technical steps, while omitting detailed and extensive calculations (e.g. in the constraint algorithm) whenever possible without loosing any crucial ingredient to our discussion. Particular attention will be instead devoted to those steps in which subtle differences arise compared to the scalar field case.~This includes the mode decomposition and the role of fiducial structures when geometry is dynamical.   

\subsection{Setup}\label{sec:setup}

We start with the ADM formulation of General Relativity \cite{ArnowittTheDynamicsOf,DiracHamiltonianFormField, DiracGravityHamiltonianForm} minimally coupled with a massless scalar field.\footnote{As usual in cosmological models, the massless scalar field plays the role of a relational matter clock field used to deparametrise the theory and construct gauge-invariant relational observables.
Being cosmology also the aim of our present discussion, we included such a scalar field into the canonical analysis from the very beginning.}
The gravitational phase space is thus coordinatised by the spatial metric $q_{ab}$ and its momentum $P^{ab}$, living on a (non-compact) spatial slice $\Sigma$ of $\mathbb R^3$ topology, with non-vanishing canonical Poisson brackets
\be\label{ADMpb}
\{q_{ab}(x),P^{cd}(y)\}=\delta^c_{(a}\delta^d_{b)}\delta(x-y)\;.
\ee

\noindent
Similarly, the matter d.o.f. given by the scalar field $\phi$ and its momentum ${P}_\phi$ have non-vanishing Poisson brackets\footnote{To uniform the notation of the gravitational and matter sectors, we denoted the conjugate momentum to $\phi$ by ${P}_\phi$ rather than by $\pi$ as done in the previous sections.}
\be\label{matterPB}
\{\phi(x),{P}_\phi(y)\}=\delta(x-y)\;.
\ee

\noindent
Recall that the conjugate momentum $P_\phi$ is a scalar density of weight 1.
Gravitational and matter d.o.f. are subject to the spatial diffeomorphism $\mathcal C_a$ and Hamiltonian $\mathcal H$ first-class constraints densities given by \cite{BodendorferAnEmbeddingOf} ($\kappa=8\pi G=1$)
\begin{align}\label{ADMconstraints}
\mathcal C_a&=-2\nabla_bP^b_a+{P}_\phi\partial_a\phi\approx0\;,\\
\mathcal H&=-\frac{P^2}{3\sqrt{q}}+\frac{4}{\sqrt{q}}\left(P_{ab}-\frac{1}{3}q_{ab}P\right)\left(P^{ab}-\frac{1}{3}q^{ab}P\right)-\frac{\sqrt{q}}{2}R+\frac{{P}_\phi^2}{2\sqrt{q}}+\frac{1}{2}\sqrt{q}q^{ab}\partial_a\phi\partial_b\phi\approx0\;,
\end{align}

\noindent
with $P=q_{ab}P^{ab}$ denoting the trace of $P^{ab}$, $q$ the determinant of $q_{ab}$, and $R$ the spatial Ricci scalar.

Following \cite{BodendorferAnEmbeddingOf,BodendorferQuantumReductionTo} (see also \cite{AlesciQuantumreducedloop}), we work in the diagonal metric gauge $q_{ab}=\text{diag}(q_{11},q_{22},q_{33})_{ab}$, thus partially gauge-fixing spatial diffeomorphisms to the subset of gauge-preserving transformations which remain first class. 
The off-diagonal components of $P^{ab}$ are solved for by the constraint $\mathcal C_a=0$ so that the reduced phase space is parametrised by the diagonal components of $q_{ab}$ and $P^{ab}$. 
Furthermore, assuming local isotropy\footnote{Here we follow \cite{BodendorferAnEmbeddingOf}, where a canonical transformation to the pairs $(\alpha, P_\alpha)$ as in the main text and $(\beta = P^{11} q_{11} - P^{22} q_{22}, P_\beta = \log\left(q_{22}/q_{11}\right)/2)$, $(\gamma = P^{11} q_{11} - P^{33} q_{3}, P_\gamma = \log\left(q_{33}/q_{11}\right)/2)$ was performed. The quantities $P_\beta$ and $P_\gamma$ measure the local deviations of the geometry from a cube. Thus, enforcing $\beta  =\gamma = P_\beta = P_\gamma = 0$ leads to the presented system, which in each tangent space is isotropic, i.e. no direction is special. However, this only holds locally as $\alpha$ depends still on the spacetime point.} so that $q_{ab}(x)=a^2(x)\delta_{ab}$ and $P^{ab}(x)=p(x)\delta^{ab}$ in a suitable coordinate system, we introduce the canonical variables for the gravitational sector
\be\label{diagisotropyvariables}
\alpha=\sqrt{q_{11}q_{22}q_{33}}=\sqrt{q}=a^3\qquad,\qquad P_\alpha=\frac{2P^{ab}q_{ab}}{3\sqrt{q_{11}q_{22}q_{33}}}=\frac{2p}{a}
\ee 

\noindent
with non-vanishing canonical Poisson brackets
\be\label{alphaPB}
\{\alpha(x),P_\alpha(y)\}=\delta(x-y)\;.
\ee

\noindent
Note that the metric is still neither homogeneous nor isotropic due to the spatial dependence of $a(x)$.
There are no global Killing fields for the system, and therefore the is no global symmetry present.~Only locally, in tangent space of a point, we have isotropy in the sense that each direction is equivalent.
The variables \eqref{diagisotropyvariables} are the isotropic version of the set of adapted variables introduced in \cite{BodendorferAnEmbeddingOf,BodendorferQuantumReductionTo} to implement symmetry-reduction constraints at the quantum level.~These were the starting point to construct an embedding of LQC $(b,v)$-variables into a full LQG context, the latter being formulated in terms of such adapted variables rather than standard Ashtekar's connection variables \cite{BodendorferAnEmbeddingOf}.~In \cite{BodendorferAnEmbeddingOf,BodendorferQuantumReductionTo}, however, a 3-torus compact spatial topology was assumed to avoid dealing with boundary terms or infinite volume.~As our main interest in the following will rather lie into the role of fiducial structures when implementing spatial homogeneity over some spatial region, we consider the spatial hypersurfaces to be non-compact and without boundary, while making a shortcut and assume isotropy to be already imposed. 

On the reduced phase space coordinatised by the canonical pairs $(\alpha,P_\alpha)$ and $(\phi,{P}_\phi)$, the spatial diffeomorphisms and Hamiltonian constraints \eqref{ADMconstraints} become
\begin{align}
\mathcal C_a&=-2\partial_a\left(a^2p\right)+{P}_\phi\partial_a\phi=-\partial_a\left(\alpha P_\alpha\right)+{P}_\phi\partial_a\phi\approx0\;,\label{isotropicvectorconstraint}\\
\mathcal H&=\frac{{P}_\phi^2}{2\alpha}+\frac{1}{2}\alpha^{\frac{1}{3}}\left(\nabla\phi\right)^2-\frac{1}{2}\alpha R(\alpha)-\frac{3}{4}\alpha P_\alpha^2\approx0\;,\label{isotropichamconstraint}
\end{align}

\noindent  
where the spatial Ricci scalar reads in terms of $\alpha$ and its derivatives as
\be\label{spatialRicci}
R(\alpha)= \frac{2}{\alpha^{\frac{8}{3}}} \sum_{a=1}^3\left(\frac{5}{9}\left(\partial_a\alpha\right)^2-\frac{2}{3}\alpha\partial_a^2\alpha\right)\;.
\ee

Finally, let us stress that the canonical variables $(\alpha(x),P_\alpha(x))$, $(\phi(x),{P}_\phi(x))$ still depend on the spatial points so that we are still dealing with a field theory at this stage. The spatial dependence will only be dropped after imposing homogeneity but, as it was the case also for the scalar field example discussed in the previous sections, the field-theoretic character of the system will be encoded into averaged smeared quantities over a region $V\subset\Sigma_t$. However, unlike the scalar field example of Sec. \ref{Sec:Warmupscalarfield}, where a fixed background geometry was available, the definitions of such observables and their Poisson brackets are now more subtle due to the geometry of the smearing region being itself dynamical. Indeed, for a given spatial region $V\subset\Sigma_t$, the averaged smeared fields over that region are given by
\begin{align}
\alpha(V)&=\int_{V}\dd^3x\,\alpha(x)=\int_{V}\dd^3x\sqrt{q}=\text{vol}(V)\;,\label{fullalphaobservable}\\
P_\alpha(V)&=\frac{1}{\text{vol}(V)}\int_{V}\dd^3x\sqrt{q}\,P_\alpha(x)=\frac{1}{\text{vol}(V)}\int_{V}\dd^3x\,\alpha(x)P_\alpha(x)\;,\label{fullPalphaobservable}
\end{align}

\noindent
for the gravitational sector, and
\begin{align}
\phi(V)&=\frac{1}{\text{vol}(V)}\int_{V}\dd^3x\,\sqrt{q}\,\phi(x)=\frac{1}{\text{vol}(V)}\int_{V}\dd^3x\,\alpha(x)\phi(x)\;,\label{fullphiobservable}\\
P_\phi(V)&=\int_{V}\dd^3x\,{P}_\phi(x)=\int_{V}\dd^3x\sqrt{q}\,\overline P_\phi(x)=\int_{V}\dd^3x\,\alpha(x)\overline P_\phi(x)\;,\label{fullPphiobservable}
\end{align}

\noindent
for the matter sector (cfr.~Eq.~\eqref{eq:smearedphipi}), respectively.
Here $\overline{P}_\phi = P_\phi/\alpha$ is a scalar constructed out of the two densities $P_\phi$ and $\alpha$.
Note that, unlike the scalar field case/sector where $\phi(x)$ is a scalar density and $P_\phi(x)$ has density weight 1, the density weights of the configuration field and its momentum are exchanged in the gravitational sector. As anticipated in Sec.~\ref{sec:scalarfieldsetup} (cfr.~below Eq.~\eqref{eq:smearedphipi}), this corresponds to the quantities $\alpha(V)$ and $P_\alpha(V)$ being extensive and intensive, respectively. The extensive nature of the former is expected from its physical interpretation \eqref{fullalphaobservable} as actually being the volume of the region $V$ so that $\alpha(V_1\cup V_2)=\text{vol}(V_1\cup V_2)=\text{vol}(V_1)+\text{vol}(V_2)=\alpha(V_1)+\alpha(V_2)$, for any non-intersecting $V_1,V_2\subset\Sigma_t$.
 
As can be checked by direct computation, the smeared quantities \eqref{fullalphaobservable}-\eqref{fullPphiobservable} satisfy the following Poisson bracket relations
\begin{align}
&\Poisson{\alpha(V)}{P_\alpha(V')}=-\Poisson{P_\phi(V)}{\phi(V')}=\frac{\text{vol}(V\cap V')}{\text{vol}(V')}\;,\label{smearedfieldsPB1}\\
&\Poisson{\alpha(V)}{\phi(V')}=\Poisson{\alpha(V)}{P_\phi(V')}=0\;,\label{smearedfieldsPB2}\\
&\Poisson{P_\alpha(V)}{\phi(V')}=\frac{1}{\text{vol}(V)}\int_{V}\dd^3x\,\alpha(x)\Biggl(-\frac{1}{\text{vol}(V')^2}\int_{V'}\dd^3z\Poisson{P_\alpha(x)}{\alpha(z)}\int_{V'}\dd^3y\,\alpha(y)\phi(y)\,+\nonumber\\
&\qquad\qquad\qquad\qquad\;\,+\frac{1}{\text{vol}(V')}\int_{V'}\dd^3y\Poisson{P_\alpha(x)}{\alpha(y)}\phi(y)\Biggr)\nonumber\\
&\qquad\qquad\qquad\quad\;=-\frac{1}{\text{vol}(V)\text{vol}(V')^2}\int_{V}\dd^3x\,\alpha(x)\chi_{V'}(x)\int_{V'}\dd^3y\,\alpha(y)\phi(y)\,+\nonumber\\
&\qquad\qquad\qquad\qquad\;\,+\frac{1}{\text{vol}(V)\text{vol}(V')}\int_{V}\dd^3x\,\alpha(x)\phi(x)\chi_{V'}(x)\nonumber\\
&\qquad\qquad\qquad\quad\;=\frac{\text{vol}(V\cap V')}{\text{vol}(V)\text{vol}(V')}\Bigl(\,\phi(V')-\phi(V\cap V')\Bigr)\;.\label{smearedfieldsPB3}
\end{align}

\noindent
Therefore, as it was also the case in Sec. \ref{sec:scalarfieldsetup} (cfr. Eq. \eqref{eq:smearedPB}), we see that the Poisson brackets \eqref{smearedfieldsPB1} yield a non-trivial result when the two smearing regions intersect.
Moreover, due to the spatial geometry being itself part of the canonical variables, the Poisson bracket \eqref{smearedfieldsPB3} is a priori not straightforwardly vanishing and this in turn poses restrictions on the smearing regions\footnote{This is also compatible with the fact that, by computing the Poisson brackets with the Hamiltonian constraint, the dynamics of the smeared quantities -- as e.g. $\dot\phi(V)$ -- will generically get contributions from the evolution of both matter ($\dot \phi(x)$) and geometry ($\dot \alpha(x)$) local d.o.f.}. 
Specifically, the r.h.s. of Eq. \eqref{smearedfieldsPB3} vanishes for finite volumes only if $V\cap V'=V'$, i.e. $V'\subseteq V$, in which case we also have that the r.h.s. of the bracket \eqref{smearedfieldsPB1} becomes simply 1 and $(\alpha(V),P_\alpha(V))$, $(\phi(V),P_\phi(V))$ are canonically conjugate pairs.
On the other hand, if $V\subset V'\to\infty$, all brackets trivialise compatibly with correlations being negligible when averaging over the whole spatial slice. 
Finally, we notice that the above-mentioned subtleties with spatial domains are smeared out once homogeneity over a region $V_o$ has been imposed in which case $V,V'\subseteq V_o$ and the spatially homogeneous averaged field-theoretic quantities reduce to the local field variables restricted to the homogeneity region $V_o$ (cfr. Eqs. \eqref{homogeneoussmearedphi}-\eqref{reducedandfullpi}).
We will come back on this point later in Sec.~\ref{Sec:gravityDBtruncatedtheory} after discussing the implementation of spatial homogeneity constraints in the next (sub)section.

\subsection{Implementing Homogeneity Constraints}\label{sec:gravityconstraintalgorithm}

As starting point to implement spatial homogeneity via constraints, we consider local Killing equations for the spacetime metric $g_{\mu\nu}$ of the kind $\mathcal L_K g_{\mu\nu}=0$, for any vector field $K=K^a(x)\partial_a$, or equivalently $K^a\partial_ag_{\mu\nu}+(\partial_\mu K^a)g_{a\nu}+(\partial_\nu K^a)g_{\mu a}=0$. 
In particular, for coordinate vector fields $\frac{\partial}{\partial x^a}$, this yields the conditions $\partial_a g_{\mu\nu}=0$ which in turn, using the ADM decomposition of the metric \cite{ArnowittTheDynamicsOf}
\begin{equation}\label{eq:ADMmetric}
	g_{\mu\nu}=\begin{pmatrix}-N^2+N^aN_a&N_a\\N_a&q_{ab}\end{pmatrix} \;,
\end{equation}

\noindent
amounts to the following constraints for the spatial metric
\be\label{metrichomogeneityconstraints}
\Psi_a^\alpha:=\partial_a\alpha(x)\approx0\;,
\ee

\noindent
as well as the homogeneity conditions for lapse and shift
\be\label{lapseshifthomogeneity}
\partial_a N(x)=0\qquad,\qquad \partial_a N^b(x)=0\;,
\ee

\noindent
restricting the diffeomorphism gauge symmetry to those diffeomorphisms compatible with spatial homogeneity. 
Similarly, for the matter sector, we demand spatial homogeneity of the configuration scalar field via the constraints\footnote{As discussed below, in principle we could have just considered Eqs.~\eqref{metrichomogeneityconstraints} and \eqref{lapseshifthomogeneity} as starting point for our constraint analysis, while homogeneity for the scalar field would follow from the spatial diffeomorphism vector constraint $\mathcal C_a\approx0$ which relates gravitational and matter degrees of freedom (cfr. Eq. \eqref{isotropicvectorconstraint}).}
\be\label{phihomogeneityconstraints}
\Psi_a^\phi:=\partial_a\phi(x)\approx0\;.
\ee

\noindent
Using then the total Hamiltonian
\begin{align}
H_T&=\int\dd^3y\,\left(N(y)\mathcal H(y)+N^b(y)\mathcal C_b(y)+f^b(y)\partial_b\alpha(y)+\mu^b(y)\partial_b\phi(y)\right)\nonumber\\
&=H[N]+C_b[N^b]+\Psi_b^\alpha[f^b]+\Psi_b^\phi[\mu^b]\;,
\end{align}

\noindent
with $\mathcal C_b$ and $\mathcal H$ given in Eqs.~\eqref{isotropicvectorconstraint} and \eqref{isotropichamconstraint}, together with the canonical Poisson brackets \eqref{matterPB}, \eqref{alphaPB}, demanding stability of the constraints \eqref{metrichomogeneityconstraints}, that is $\dot \Psi_a^\alpha[f^a]\approx0$, yields after integration by parts
\begin{align}
\dot \Psi_a^\alpha[f^a]&=\Poisson{\Psi_a^\alpha[f^a]}{H_T}\nonumber\\
&\approx\int\dd^3 x\,f^a(x)\left(\partial_a\Bigl(\alpha(x)\partial_bN^b(x)\Bigr)-\frac{3}{2}\partial_a\Bigl(N(x)\alpha(x)P_\alpha(x)\Bigr)\right)\approx0\qquad\forall\;f_a\;,
\end{align}

\noindent
i.e.
\be
\dot\Psi_a^\alpha\approx \partial_a\bigl(\alpha\partial_bN^b\bigr)-\frac{3}{2}\partial_a\left(N\alpha P_\alpha\right)\approx0\;,
\ee

\noindent
from which, using Eqs. \eqref{metrichomogeneityconstraints} and \eqref{lapseshifthomogeneity}, spatial homogeneity constraints for the conjugate momentum $P_\alpha$ follow, namely
\be\label{Palphahomogeneityconstriants}
\dot\Psi_a^\alpha\approx0\qquad\Rightarrow\qquad \Psi_a^{P_\alpha}=\partial_aP_\alpha(x)\approx0\;.
\ee

\noindent
Similarly, stability of the constraints \eqref{phihomogeneityconstraints} w.r.t. the total Hamiltonian
\be
H_T=\int\dd^3y\,\left(N(y)\mathcal H(y)+N^b(y)\mathcal C_b(y)+f^b(y)\partial_b\alpha(y)+\lambda^b(y)\partial_bP_\alpha(y)+\mu^b(y)\partial_b\phi(y)\right)\;,
\ee

\noindent
yields after integration by parts
\be
\dot\Psi_a^\phi[\mu^a]=\Poisson{\Psi_a^\phi[\mu^a]}{H_T}\approx\int\dd^3x\,\mu^a(x)N(x)\partial_aP_\phi(x)\approx0\qquad\forall\;\mu^a,N
\ee

\noindent
thus resulting into the homogeneity constraints for the conjugate momentum $P_\phi$, i.e.
\be\label{Pphihomogeneityconstriants}
\dot\Psi_a^\phi\approx0\qquad\Rightarrow\qquad \Psi_a^{P_\phi}=\partial_aP_\phi(x)\approx0\;.
\ee

\noindent
The constraints \eqref{metrichomogeneityconstraints}, \eqref{Palphahomogeneityconstriants} and \eqref{phihomogeneityconstraints}, \eqref{Pphihomogeneityconstriants} respectively form second class sets as can be easily seen by computing their Poisson brackets
\be\label{homogeneity2ndclassPB}
\Poisson{\Psi_a^\alpha(x)}{\Psi_b^{P_\alpha}(y)}=\Poisson{\Psi_a^\phi(x)}{\Psi_b^{P_\phi}(y)}=\frac{\partial}{\partial x^a}\frac{\partial}{\partial y^b}\delta(x-y)\neq0\;.
\ee

\noindent
Finally, collecting all the constraints in the new total Hamiltonian and recalling the Poisson brackets \eqref{matterPB}, \eqref{alphaPB}, and \eqref{homogeneity2ndclassPB}, after some lengthy but straightforward algebra, stability of $\Psi_a^{P_\alpha}$ and $\Psi_a^{P_\phi}$ just yield conditions on the Lagrange multipliers of the respective conjugate second class constraints
\begin{align}
\dot\Psi_a^{P_\alpha}[\lambda^a]\approx0\qquad&\Rightarrow\qquad\partial_a\partial_b f^b\approx0\;,\\
\dot\Psi_a^{P_\phi}[\omega^a]\approx0\qquad&\Rightarrow\qquad\partial_a\partial_b \mu^b\approx0\;,
\end{align}

\noindent
so that no further constraints arise and the algorithm terminates here.

Before moving to the discussion of the mode decomposition and the construction of the corresponding symmetry-reduced homogeneous theory resulting from the introduction of the Dirac bracket allowing us to implement the above homogeneity constraints strongly, let us close this section with few comments. 
First of all, we notice that the homogeneity constraints \eqref{metrichomogeneityconstraints}, \eqref{Palphahomogeneityconstriants}, and \eqref{phihomogeneityconstraints} imply the spatial diffeomorphism constraint $\mathcal C_a\approx0$ to be automatically satisfied (cfr. Eq.~\eqref{isotropicvectorconstraint}). 
Alternatively, as anticipated above, even though at the beginning of our constraint algorithm we do not impose homogeneity for the scalar field $\phi$, the latter will follow from the homogeneity constraints for $\alpha$ and $P_\alpha$ via the vector constraint $\mathcal C_a\approx0$. Moreover, looking at the EoM for $\alpha(x)$, that is
\begin{align}
&\dot\alpha(x)=\Poisson{\alpha(x)}{C_a[N^a]+H[N]}\nonumber\\
&\qquad=-N^a(x)\partial_a\alpha(x)-\frac{3}{2}N(x)\alpha(x)P_\alpha(x)\nonumber\\
&\quad\,\,\underset{\Psi_a^\alpha\approx0}{\approx}-\frac{3}{2}N(x)\alpha(x)P_\alpha(x)\;,\label{alphaEoM}
\end{align}

\noindent
we see that, upon imposing the homogeneity constraints, $P_\alpha\approx-\frac{2\dot\alpha}{3N\alpha}$ is related to the Hubble rate as expected also from the so-called $(b,v)$-variables for isotropic and homogeneous cosmology \cite{AshtekarLoopquantumcosmology:astatusreport,BodendorferAnElementaryIntroduction} where the on-shell value of the momentum $b$ conjugate to the volume $v$ is related to the Hubble rate $\dot v/v$.  

\subsection{Dirac Bracket and Truncated Theory}\label{Sec:gravityDBtruncatedtheory}

As for the scalar field case, our strategy to obtain a spatially homogeneous theory consists of implementing the second-class constraints strongly by
using the Dirac bracket. We will then proceed along similar steps to those of the mini-superspace reduction procedure outlined in Sec. \ref{sec:implementinghom}. As we will discuss in this section, however, a major difference compared to the scalar field case would reside in the way in which the mode decomposition of the fields variable is achieved when there is no background fixed geometry. In fact, as anticipated also in Sec.~\ref{sec:implementinghom}, unlike the analysis of the previous sections where the presence of a fixed background metric allowed us to canonically associate dual modes via $L^2$-pairing (see also Eq.~\eqref{L2pairingdualmodes} and the surrounding discussion in Appendix \ref{app:modedecomposition}), the dynamical nature of geometry in the case of gravity requires us to introduce a \textit{fiducial metric} $\fidmetric{a}{b}$ associated with local Cartesian coordinate axes along the edges of the boxes partitioning the region $V_o$ over which spatial homogeneity shall be imposed. Their physical volume will thus involve both the fiducial and the dynamical metric so that, physically speaking, such a fiducial metric will provide us with a measure of a wavelength cutoff discriminating between the spatially constant homogeneous $\vec k=\mathbf{0}$ and higher order modes within the boxes. In particular, the dependence of the $\vec k\neq\mathbf{0}$ modes and their truncation in the resulting homogeneous theory on such \textit{fiducial structures} suggests us that the latter are not playing the role of mere regulators for the canonical theory but rather setting the scale for the validity of the homogeneous approximation. Their physical role becomes even more prominent when the homogeneity region is shrinking dynamically with the Universe's volume as it is the case in cosmology, thus having physical implications also for the validity of the homogeneous description at early times and quantum fluctuations in the quantum theory resulting from quantisation of the symmetry-reduced truncated classical theory as discussed in Sec. \ref{Sec:QuantumCosmology}.

Details on the general strategy of the mode decomposition, especially when there is no fixed background geometry, can be found in Appendix~\ref{app:modedecomposition}.
Along similar lines to Sec. \ref{sec:constraintimplementation}, the starting point for this construction and the evaluation of the Dirac bracket is to partition the spatial slice $\Sigma_t$ of $\mathbb R^3$-topology into the disjoint union $\Sigma_t=\bigsqcup_{n=1}^\infty V_n$ of boxes $V_n$, which is a purely topological construction at this stage.~The fields (say $\alpha(x)$ for concreteness) can be thus decomposed without loss of generality as
\be\label{fieldpartitioning}
\alpha(x)=\sum_{n=1}^\infty\chi_{V_n}(x)\,\alpha^{n}(x)\;,
\ee

\noindent
and similarly for the other field variables. In each box $V_n\subset\Sigma_t$, the relevant information is thus encoded into the fields supported in the given box, say $\alpha^{n}(x)=\alpha(x)\bigl|_{x\in V_n}$, while the information carried by the fields $\alpha^{n}(x)$ outside $V_n$ is redundant so that we can set $\alpha^{n}(x)\bigl|_{x\in V_n^\mathrm{C}}=0$ and impose boundary conditions as e.g. periodic boundary conditions.~We can then proceed to decompose $\alpha^{n}(x)$ into countably many modes, for instance via a discrete Fourier decomposition, with the zero modes being independent of $x\in V_n$. For this to be possible, however, we need to define an $L^2$-pairing on each $V_n$\footnote{The $L^2$-pairing is only needed to reconstruct the local field $\alpha(x)$, while the expansion works without extra structure for the smeared field $\alpha[f]$. We refer to Appendix~\ref{app:modedecomposition} for details.}. To this aim, let us introduce an arbitrary spatially flat \textit{fiducial metric}
\be\label{eq:deffidmetric}
\fidmetric{a}{b}\dd x^a\dd x^b:=\beta^{\frac{2}{3}}\delta_{ab}\dd x^a\dd x^b\qquad,\qquad\beta\in\mathbb R\;.
\ee

\noindent
This gives the boxes $V_n$, introduced so far only at topological level, a cubic shape with the $x^a$-coordinate axes associated with the fiducial triad field along the edges of the box. 
Denoting by $L$ the coordinate edge length of each $V_n$, i.e. $V_n=\{\textsf{p}\in\Sigma_t\mid x_n(\textsf{p})=\{x_n^a + \xi_a\}_{a=1,\dots,3}\;,\; \xi_a\in[0,L)\}$ topologically specifying cubic neighbours around a point in space, the volumes of $V_n$ measured w.r.t. the fiducial and physical metric are given by
\be
\text{vol}_{\fidmetric{}{}}(V_n)=\int_{[0,L]^3}\dd^3x\,\sqrt{\fidmetric{}{}}=\beta\,L^3\qquad,\qquad \text{vol}_{q}(V_n)=\int_{[0,L]^3}\dd^3x\,\alpha(x)
\ee

\noindent
respectively. We can decompose then scalar quantities in terms of $\fidmetric{}{}$-Fourier modes, that is the module $\|\vec k\|$ of the wave vector $\vec k$ (and hence the wavelength) of the modes is measured w.r.t. $\fidmetric{}{}$. Specifically, by still referring to the density 1 field $\alpha(x)$ for concreteness, and constructing the scalar density $\overline{\alpha}(x)=\sqrt{q}/\sqrt{\fidmetric{}{}}$, similarly for the $\overline{\alpha}^{n}(x)$ in Eq. \eqref{fieldpartitioning}, we have
\be\label{alphamodedecomp}
\alpha^n(x) = \sqrt{\fidmetric{}{}}\,\overline\alpha^{n}(x)=\sqrt{\fidmetric{}{}}\,\sum_{\vec k\in\mathbb Z^3}\tilde\alpha^{n}_{\vec k}\frac{e^{-i\frac{2\pi}{L}\vec k\cdot\vec x}}{\beta L^3}\qquad,\qquad\tilde\alpha^{n}_{\vec k}=\int_{V_n}\dd^3x\,\alpha^{n}(x)e^{+i\frac{2\pi}{L}\vec k\cdot\vec x}
\ee 

\noindent
where $\vec k\cdot\vec x=\delta_{ab}k^ax^b$ is the standard contraction between vectors and covectors. As can be checked by direct computation, the modes satisfy the orthogonality and completeness properties (cfr. Eq.~\eqref{eq:orthogonalitycompleteness})
\be
\frac{1}{\beta L^3}\int_{V_n}\dd^3x\,\sqrt{\fidmetric{}{}}\,e^{-i\frac{2\pi}{L}(k_a-p_a)x^a}=\prod_{a=1}^3\delta_{k_a,p_a}\equiv\delta_{\vec k,\vec p}\quad,\quad\sum_{\vec k\in\mathbb Z^3}\braket{e^{-i\frac{2\pi}{L}\vec k\cdot\vec y}}{f(y)}e^{i\frac{2\pi}{L}\vec k\cdot\vec x}=f(x)\quad\forall f\;,
\ee

\noindent
with $L^2$-pairing between field modes and their dual modes defined w.r.t. $\fidmetric{a}{b}$, namely (cfr. Eq. \eqref{L2pairingdualmodes})
\be
\braket{f}{g}:=\frac{1}{\beta L^3}\int_{V_n}\dd^3x\,\sqrt{\fidmetric{}{}} f^*(x)g(x)\;.
\ee

\noindent
The momentum field $P_\alpha(x)$ is a scalar and can straight forwardly be decomposed in terms of the discrete Fourier modes as
\be\label{eq:Palphamodedecomp}	
		P_\alpha(x) = \sum_n \chi_{V_n}(x) P_\alpha^n(x) \quad , \quad P_\alpha^n(x) = \sum_{\vec{k} \in \mathbb{Z}^3} \tilde{P}_{\alpha,\vec{k}}^n e^{+i\frac{2\pi}{L} \vec{k} \cdot \vec{x}} \quad,\quad\tilde{P}_{\alpha,\vec{k}}^n= \frac{1}{\beta L^3} \int_{V_n} \dd^3 x \sqrt{\fidmetric{}{}}\, P_{\alpha}^n e^{-i\frac{2\pi}{L} \vec{k}\cdot \vec{x}} \;.
	\ee
Similar decomposition can be written for the remaining canonical fields $\phi(x)$, and $\overline{P}_\phi(x)=P_\phi(x)/\sqrt{\fidmetric{}{}}$, where we recall that $\phi$ is a density 0 object while $P_\phi$ is a density 1 object.
As such, the expansion of $\phi$ is analogue to Eqs.~\eqref{eq:Palphamodedecomp} and the expansion for $P_\phi$ similar to Eq.~\eqref{alphamodedecomp}.
Plugging the mode decompositions into the canonical Poisson brackets \eqref{matterPB} and \eqref{alphaPB}, the non-vanishing Poisson brackets for the corresponding modes in $x$- and $k$-space can be then readily computed to be
\begin{align}
&\Poisson{\overline{\alpha}^{n}(x)}{P_\alpha^{m}(y)}=\frac{\chi_{V_n\cap V_m}(x)}{\sqrt{\fidmetric{}{}}}\delta_{nm}\delta(x-y)=\frac{1}{\sqrt{\fidmetric{}{}}}\delta_{nm}\delta(x-y)=\Poisson{\phi^{n}(x)}{\overline{P}_\phi^{m}(y)}\;,\\
&\Poisson{\tilde{\alpha}^{n}_{\vec k}}{\tilde{P}_{\alpha,\vec p}^{n}}=\delta_{nm}\delta_{\vec k,\vec p}=\Poisson{\phi^{n}_{\vec k}}{\tilde{P}_{\phi,\vec p}^{m}}\;.\label{kmodesPB}
\end{align}

We are now in the position to translate the homogeneity constraints of Sec.~\ref{sec:gravityconstraintalgorithm} into constraints for the corresponding field modes as follows. Plugging Eq.~\eqref{fieldpartitioning} into the constraint \eqref{metrichomogeneityconstraints}, we get
\be\label{alphahomogconstraintdecomposition}
0\approx\Psi_a^\alpha(x)=\partial_a\alpha(x)=\sum_{n=1}^\infty\Bigl(\chi_{V_n}(x)\partial_a\alpha^{n}(x)+\alpha^{n}(x)\partial_a\chi_{V_n}(x)\Bigr)\;,
\ee

\noindent
where the first term refers only to the spatial dependence of the fields within each region $V_n$, while the second term $\partial_a\chi_{V_n}(x)\propto\delta_S(x-\partial V_n)$, which is a boundary term, refers to the changes of the fields across different boxes thus encoding the interaction through the boundary of adjacent regions. Therefore, the three constraints per point $\Psi_a^\alpha(x)\approx0$, $a=1,2,3$, decompose into two sets of constraints
\begin{align}
&\textsf{1)}\;\partial_a\alpha^{n}(x)\bigl|_{V_n}=0\qquad\forall\; n\label{homogconstrmodes1a}\\
&\textsf{2)}\;\lim_{x_n\to\partial V_n\cap\partial V_m}\alpha^{n}(x_n)=\lim_{x_m\to\partial V_n\cap\partial V_m}\alpha^{m}(x_m)\;,\label{homogconstrmodes1b}
\end{align}

\noindent
respectively demanding the fields to be homogeneous within each box, and to not change across one box and the other. Recalling now that $\alpha(x)=\sqrt{\fidmetric{}{}}\,\overline\alpha(x)$ and using the decomposition \eqref{alphamodedecomp}, the constraints \eqref{homogconstrmodes1a}, \eqref{homogconstrmodes1b} lead to the following constraints for the $\vec k$-modes
\be\label{homogconstraintsmodesalpha}
\textsf{1)}\quad\,\;\Psi_{\vec k}^{\alpha, n}:=\tilde{\alpha}^{n}_{\vec k}\approx0\qquad\vec k\neq\mathbf{0}\qquad\;,\;\qquad\textsf{2)}\;\;\;\Psi_{\alpha}^{n}:=\tilde{\alpha}^{n}_\mathbf{0}-\tilde{\alpha}^{1}_\mathbf{0}\approx0\qquad\;\;\, n\neq1\;.
\ee

\noindent
Similar sets of constraints follow from the mode decomposition of the other canonical fields by means of analogous steps, namely
\begin{align}
\textsf{1)}\;\;\Psi_{\vec k}^{P_\alpha,n}:=\tilde P_{\alpha,\vec k}^{n}\approx0\qquad\vec k\neq\mathbf{0}\;\qquad&,\qquad\textsf{2)}\,\,\,\,\Psi_{P_\alpha}^{n}:=\tilde P_{\alpha,\mathbf{0}}^{n}-\tilde P_{\alpha,\mathbf{0}}^{1}\approx0\qquad\; n\neq1\;,\label{homogconstraintsmodesPalpha}\\
\Psi_{\vec k}^{\phi,n}:=\tilde\phi_{\vec k}^{n}\approx0\qquad\vec k\neq\mathbf{0}\;\qquad&,\qquad\quad\,\,\;\quad\,\Psi_{\phi}^{n}:=\tilde\phi_{\mathbf{0}}^{n}-\tilde\phi_{\mathbf{0}}^{1}\approx0\qquad\;\; n\neq1\;,\label{homogconstraintsmodesphi}\\
\Psi_{\vec k}^{P_\phi,n}:=\tilde{P}_{\phi,\vec k}^{n}\approx0\qquad\vec k\neq\mathbf{0}\;\qquad&,\qquad\quad\,\,\;\Psi_{P_\phi}^{n}:=\tilde{P}_{\phi,\mathbf{0}}^{n}-\tilde{P}_{\phi,\mathbf{0}}^{1}\approx0\qquad\;\, n\neq1\;.\label{homogconstraintsmodesPphi}
\end{align}

\noindent
Moreover, using the Poisson brackets \eqref{kmodesPB}, we find the following non-vanishing Poisson brackets between the constraints \eqref{homogconstraintsmodesalpha}-\eqref{homogconstraintsmodesPphi}
\be\label{PBconstraints1)}
\Poisson{\Psi_{\vec k}^{\alpha,n}}{\Psi_{\vec p}^{P_\alpha,m}}=\delta_{nm}\delta_{\vec k,\vec p}=\Poisson{\Psi_{\vec k}^{\phi,n}}{\Psi_{\vec p}^{P_\phi,n}}\qquad,\qquad\Poisson{\Psi_\alpha^{n}}{\Psi_{P_\alpha}^{m}}=\delta_{nm}+1_{nm}=\Poisson{\Psi_\phi^{n}}{\Psi_{P_\phi}^{m}}
\ee

\noindent
with $1_{nm}$ being the matrix whose elements are all equal to $1$.
The constraints in fact form two distinct (mutually commuting) second class sets for the gravitational and matter sectors, respectively.

The construction of the Dirac brackets associated with the above second class constraints proceeds along similar steps as in the second part of Sec. \ref{sec:constraintimplementation}. The idea is then to impose the constraints of the kind \textsf{1)} and \textsf{2)} in two steps by constructing the associated Dirac bracket w.r.t. the two kinds of second class constraints. Specifically, denoting type \textsf{1)} constraints collectively by $\Psi_{\textsf{A}}^{\textsf{1)}}$, with $\textsf{A}=\alpha,P_\alpha,\phi,P_\phi$, and by $\Poisson{\cdot}{\cdot}_{D,\textsf{1)}}$ the Dirac bracket associated with them, we have
\be\label{type1DB}
\Poisson{\cdot}{\cdot}_{D,\textsf{1)}}=\Poisson{\cdot}{\cdot}-\Poisson{\cdot}{\Psi_{\textsf{A}}^{\textsf{1)}}}C^{\textsf{AB}}\Poisson{\Psi_{\textsf{B}}^{\textsf{1)}}}{\cdot}\;,
\ee

\noindent
where $C^{\textsf{AB}}$ denotes the inverse matrix of $C_{\textsf{AB}}:=\Poisson{\Psi_{\textsf{A}}^{\textsf{1)}}}{\Psi_{\textsf{B}}^{\textsf{1)}}}$ which, taking into account the Poisson brackets \eqref{PBconstraints1)}, read as
\be
	C_{\textsf{AB}} = \delta_{nm}\delta_{\vec k,\vec p}\left(\begin{array}{cc|cc}
		0 & 1 & 0 & 0 \\
		-1 & 0 & 0 & 0\\
		\hline
		0 & 0 & 0 & 1\\
		0 & 0 & -1 & 0
	\end{array}\right)\quad,\quad C^{\textsf{AB}} = \delta_{nm}\delta_{\vec k,\vec p}\left(\begin{array}{cc|cc}
		0 & -1 & 0 & 0 \\
		1 & 0 & 0 & 0\\
		\hline
		0 & 0 & 0 & -1\\
		0 & 0 & 1 & 0
	\end{array}\right)\;.
\ee

\noindent
Evaluating then the Dirac bracket \eqref{type1DB} for the type $\textsf{1)}$ field modes \eqref{alphamodedecomp} in $\vec k$-space, we get
\be\label{modesDB1)}
\Poisson{\tilde{\alpha}^{n}_{\vec k}}{\tilde P_{\alpha,\vec p}^{m}}_{D,\textsf{1)}}=\delta_{nm}\delta_{\vec k,\mathbf{0}}\delta_{\vec p,\mathbf{0}}=\Poisson{\tilde\phi^{n}_{\vec k}}{\tilde{P}_{\phi,\vec p}^{m}}_{D,\textsf{1)}}\;,
\ee

\noindent
and all the other brackets vanish. 
As expected, only the spatially constant $\vec k=\mathbf{0}$ modes $\tilde{\alpha}^{n}_{\mathbf{0}}$, $\tilde P^{n}_{\alpha,\mathbf{0}}$, $\tilde \phi^{n}_{\mathbf{0}}$, $\tilde{P}^{n}_{\phi,\mathbf{0}}$ have non-trivial Dirac brackets and we can set the inhomogeneous  higher momentum modes to zero strongly by using the above bracket.

As for type \textsf{2)} constraints, denoting them collectively by $\Psi_{\textsf{A}}^{\textsf{2)}}$, we define the full Dirac bracket $\Poisson{\cdot}{\cdot}_{D}$ as
\be\label{type2DB}
\Poisson{\cdot}{\cdot}_{D}=\Poisson{\cdot}{\cdot}_{D,\textsf{1)}}-\Poisson{\cdot}{\Psi_{\textsf{A}}^{\textsf{2)}}}_{D,\textsf{1)}}M^{\textsf{AB}}\Poisson{\Psi_{\textsf{B}}^{\textsf{2)}}}{\cdot}_{D,\textsf{1)}}\;,
\ee

\noindent
with $M^{\textsf{AB}}$ denoting the inverse of the matrix $M_{\textsf{AB}}=\Poisson{\Psi_{\textsf{A}}^{\textsf{2)}}}{\Psi_{\textsf{B}}^{\textsf{2)}}}_{D,\textsf{1)}}$ and, using the brackets \eqref{modesDB1)}, they are given by
\be
	M_{\textsf{AB}} = \left(\begin{array}{c|c}
		Q_{nm} & 0 \\
		\hline
		0 & Q_{nm}
	\end{array}\right)\qquad,\qquad M^{\textsf{AB}} = \left(\begin{array}{c|c}
		Q_{nm}^{-1} & 0 \\
		\hline
		0 & Q_{nm}^{-1}
	\end{array}\right)\;.
\ee

\noindent
where, for a finite spatial region partitioned into $d$ boxes $V_n$, $Q_{nm}=\delta_{nm}+1_{nm}$ and $Q_{nm}^{-1}=\delta_{nm}-\frac{1}{d}1_{nm}$ with $n,m=2,\dots, d$ as $n,m\neq1$ for the type \textsf{2)} constraints in Eqs. \eqref{homogconstraintsmodesalpha}-\eqref{homogconstraintsmodesPphi}, $1_{nm}$ being the matrix whose elements are all equal to $1$ (cfr. discussion surrounding Eqs. \eqref{eq:Minverse}-\eqref{eq:Qinverse}). Evaluating the Dirac bracket \eqref{type2DB} for the $\vec k=\mathbf{0}$ modes and using the identities $\sum_{n=2}^d Q^{-1}_{nm} = 1/d$ and $\sum_{n,m=2}^d Q^{-1}_{nm} = (d-1)/d$, yields the following non-vanishing brackets (cfr. Eq. \eqref{scalarfieldmodesfullDB} for the scalar field case)
\be\label{gravityDB}
\Poisson{\tilde{\alpha}_{\mathbf{0}}^{n}}{\tilde P_{\alpha,\mathbf{0}}^{m}}_{D}=\frac{1}{d}=\Poisson{\tilde\phi_{\mathbf{0}}^{n}}{\tilde{P}_{\phi,\mathbf{0}}^{m}}_{D}\;,
\ee

\noindent
or going back to position space representation
\be\label{gravityDBxspace}
\Poisson{\overline\alpha(x)}{P_\alpha(y)}_D=\frac{1}{d\beta L^3}=\Poisson{\phi(x)}{\overline{P}_\phi(y)}_D\;,
\ee

\noindent
with $x,y\in\bigsqcup_{n=1}^dV_n$. Therefore, as it was the case also for the scalar field discussed in Sec. \ref{sec:constraintimplementation}, the effect of implementing the type \textsf{2)} sets of constraints is again that all $V_n$-boxes have non-trivial Dirac bracket with each other as expected from all the equal boxes carrying the same information, and knowing about their neighbours to ensure homogeneity across the different boxes. Moreover, the brackets \eqref{gravityDBxspace} depend of the \textit{fiducial volume} $d\cdot\beta L^3=d\cdot\text{vol}_{\fidmetric{}{}}(V_n)=\text{vol}_{\fidmetric{}{}}(V_o)$ of the spatial region $\bigsqcup_{n=1}^dV_n=V_o$ over which homogeneity is imposed. The latter region is thus not only a regulator for the otherwise divergent integrals of the canonical theory with non-compact spatial topology, but acquires physical meaning by setting the scale over which the homogeneous approximation is implemented. Indeed, according to the constraints \eqref{homogconstraintsmodesalpha}-\eqref{homogconstraintsmodesPphi}, spatial homogeneity is first implemented within each given elementary cell $V_n$ by setting to zero all the $\vec k\neq\mathbf 0$ modes, which carry an $\vec x$-dependence, via type \textsf{1)} constraints, and then imposing continuity of the homogeneous $\vec k=\mathbf{0}$ modes across the cells, via type \textsf{2)} constraints, to implement homogeneity all over the spatial region $V_o$. The latter region must be finite as otherwise the canonical brackets \eqref{gravityDB}-\eqref{gravityDBxspace} would triavialise in the $\text{vol}_{\fidmetric{}{}}(V_o)\to\infty$ limit, compatibly with the discussion of Poisson brackets \eqref{smearedfieldsPB1}-\eqref{smearedfieldsPB3} for the full theory smeared quantities (cfr. discussion below Eq. \eqref{smearedfieldsPB3}). Similarly to what we have discussed at the end of Sec. \ref{sec:constraintimplementation}, we can actually make a comparison with the smeared fields and their Poisson brackets as follows.

Using the expression \eqref{fieldpartitioning}, restricted to $V_o=\bigsqcup_{n=1}^dV_n$, together with the expansion \eqref{alphamodedecomp}, we have
\begin{align}\label{homogandsmearedalpha}
	\overline{\alpha}(x)\bigl|_{x \in V_o} &\;\;\;\,=\,\sum_{n=1}^d \chi_{V_n}(x) \sum_{\vec k \in \mathbb{Z}^3} \tilde{\alpha}_{\vec k}^{n}\frac{e^{-i\frac{2\pi}{L} \vec k\cdot\vec x}}{\beta L^3}
	\notag
	\\
	&\stackrel{\Psi_{\vec k}^{\alpha, n}=0}{=}\,\frac{1}{\beta\,L^3}\,\sum_{n=1}^d \chi_{V_n}(x)\tilde{\alpha}_{\mathbf{0}}^{n}
	\notag
	\\
	&\;\stackrel{\Psi^{n}_\alpha=0}{=}\,\frac{1}{\beta L^3}\,\tilde{\alpha}_{\mathbf{0}}^{1}=\frac{1}{d\beta L^3} \sum_{n=1}^d \int_{V_n} \dd^3 x\,{\alpha}(x)
	\notag
	\\
	&\;\;\;=\,\frac{1}{d\beta L^3}\int_{V_o}\dd^3x\,\alpha(x)
	\notag
	\\
	&\;\;\;=\, \frac{\text{vol}_q(V_o)}{\text{vol}_{\fidmetric{}{}}(V_o)}=\frac{{\alpha}(V_o)}{\text{vol}_{\fidmetric{}{}}(V_o)}\quad\quad\text{i.e.}\quad\quad\alpha=\alpha(x)\bigl|_{x \in V_o}=\sqrt{\fidmetric{}{}}\,\overline{\alpha}(x)\bigl|_{x \in V_o}=\frac{{\alpha}(V_o)}{dL^3}\;.
\end{align}

\noindent
and
\be\label{homogandsmearedPalpha}
P_\alpha=P_\alpha(x)\bigl|_{x \in V_o}=\frac{1}{\text{vol}_{\fidmetric{}{}}(V_o)} \int_{V_o} \dd^3x\, \sqrt{\mathring{q}}\, P_\alpha(x) = \frac{1}{\text{vol}_{q}(V_o)} \int_{V_o} \dd^3x\, \alpha\, P_\alpha(x) = P_\alpha(V_o)\;.
\ee

\noindent
Similar results hold for the matter fields sector (cfr. Eqs. \eqref{homogeneoussmearedphi} and \eqref{reducedandfullpi}). Therefore, imposing the constraints $\Psi_{\vec k}^{\alpha, n}=0$, $\Psi^{n}_\alpha=0$ strongly after introducing the Dirac bracket, the homogeneous local fields restricted to the region $V_o$ are equivalent to their $V_o$-averaged counterparts. On the other hand, due to the homogeneity constraints, the full theory fields \eqref{fullalphaobservable}-\eqref{fullPphiobservable} smeared over a region $V\subset V_o$ read
\begin{align}
\alpha(V) &= \text{vol}_q(V)=\int_{V} \dd^3 x\,\alpha(x) =\text{vol}_{\fidmetric{}{}}(V)\,\overline{\alpha}=(d\, L^3)\,\alpha\;,\label{homogfullsmearedalpha}\\
{P}_\phi(V)&= \int_{V} \dd^3x\,\alpha(x)P_\phi(x) =\text{vol}_{\fidmetric{}{}}(V)\overline{P}_\phi=(d\, L^3)\,P_\phi\;,\label{homogfullsmearedPphi}
\end{align}

\noindent
for the smeared densities, and
\begin{align}
{P}_\alpha(V)&=\frac{1}{\text{vol}(V)}\int_V\dd^3x\,\alpha(x)P_\alpha(x)=\frac{1}{\text{vol}_{\mathring{q}}(V)} \int_V \dd^3x \sqrt{\mathring{q}}\, P_\alpha(x)=P_\alpha\;,\label{homogfullsmearedPalpha}\\
{\phi}(V)&=\frac{1}{\text{vol}(V)}\int_V\dd^3x\,\alpha(x)\phi(x)= \frac{1}{\text{vol}_{\mathring{q}}(V)} \int_V \dd^3x \sqrt{\mathring{q}}\, \phi(x)=\phi \;.\label{homogfullsmearedphi}
\end{align}

\noindent
for the scalar quantities. Note that, under a rescaling $\text{vol}_{\fidmetric{}{}}(V)\mapsto\gamma\,\text{vol}_{\fidmetric{}{}}(V)$, the extensive quantities \eqref{homogfullsmearedalpha} and \eqref{homogfullsmearedPphi} scale as ${\alpha}(V)\mapsto\gamma\,{\alpha}(V)$ and ${P}_\phi(V)\mapsto\gamma\,{P}_\phi(V)$, while the intensive quantities \eqref{homogfullsmearedPalpha} and \eqref{homogfullsmearedphi} do not scale. 
As already anticipated in Sec. \ref{sec:scalarfieldsetup} (cfr. below Eq. \eqref{eq:smearedphipi} and the footnote there), these match with the scaling behaviours expected from LQC literature \cite{BojowaldLoopQuantumCosmology,AshtekarLoopquantumcosmology:astatusreport,BodendorferAnElementaryIntroduction}. Moreover, using the above expressions \eqref{homogfullsmearedalpha}-\eqref{homogfullsmearedphi} for the smeared quantities over the regions $V,V'\subseteq V_o$ together with the Dirac brackets \eqref{gravityDBxspace}, we find
\begin{align}
&\Poisson{{\alpha}(V)}{{P}_\alpha(V')}_D=-\Poisson{{P}_\phi(V)}{{\phi}(V')}_D=\frac{\text{vol}_{\fidmetric{}{}}(V)}{\text{vol}_{\fidmetric{}{}}(V_o)}=\frac{\text{vol}_q(V)}{\text{vol}_q(V_o)}\stackrel{V\subseteq V_o}{=}\frac{\text{vol}_q(V\cap V_o)}{\text{vol}_q(V_o)}\;,\label{homogsmearedDB1}\\
&\Poisson{\alpha(V)}{{\phi}(V')}_D=\Poisson{\alpha(V)}{{P}_\phi(V')}_D=\Poisson{{P}_\alpha(V)}{\phi(V')}_D=0\;,\label{homogsmearedDB2}
\end{align}

\noindent
and, consistently with the full theory brackets in Eqs. \eqref{smearedfieldsPB1}-\eqref{smearedfieldsPB3} and the surrounding discussion, we see that the brackets \eqref{homogsmearedDB1} do not depend on $V'$ as the averaging \eqref{homogfullsmearedPalpha} and \eqref{homogfullsmearedphi} of $P_\alpha$ and $\phi$ is the same for all volumes within the homogeneity region. Therefore, averaging over $V'$ is the same as averaging over $V_o$ and the result is then simply the same as the full theory result with the largest equivalent averaging volume, i.e. the volume of the whole region $V_o$ over which homogeneity is imposed. In particular, this implies that the last bracket in \eqref{homogsmearedDB2} between ${P}_\alpha(V)$ and $\phi(V')$ vanishes (cfr. Eq. \eqref{smearedfieldsPB3}). On the other hand, the extensive quantities \eqref{homogfullsmearedalpha} and \eqref{homogfullsmearedPphi} keep track of the smearing volume under consideration.

Finally, going back to the diffeomorphism constraints \eqref{isotropicvectorconstraint} and \eqref{isotropichamconstraint}, imposing spatial homogeneity over the finite region $V_o=\bigsqcup_{n=1}^dV_n$, and neglecting the boundary terms originating from interactions across the codimension 2 surface separating $V_o$ and its complement (cfr. Eq. \eqref{alphahomogconstraintdecomposition})\footnote{As already discussed for the scalar field case in Sec. \ref{sec:truncation}, this amounts to a truncation of the theory by restricting it to the region $V_o$, where all $\vec k\neq\mathbf{0}$ modes are set to zero, thus neglecting the backreaction of the inhomogeneities outside of that region encoded into their effect on the homogeneous modes through the boundary.}, we find that the vector constraint \eqref{isotropicvectorconstraint} is automatically satisfied as anticipated in Sec. \ref{sec:gravityconstraintalgorithm}, while the symmetry-reduced homogeneous truncated Hamiltonian constraint reads as\\
\be\label{homogeneoustruncatedhamconstraint}
H = N \mathcal{H} \qquad , \qquad \mathcal{H} = V_o \frac{P_\phi^2}{2\alpha} -\frac{3 \kappa}{4} V_o \alpha P_\alpha^2 \approx 0 \;,
\ee

\noindent
where the dynamical quantities are all local, i.e. evaluated at a point $x\in V_o$, we reintroduced units such that $\kappa = 8 \pi G$ to allow for a dimensional analysis in the coming sections\footnote{To be precise, we work now in units where $\hbar = c = 1$ and $[G] = [\kappa] = \text{\textit{length}}^2$.}, and with a slight abuse of notation we denoted both the region and its coordinate volume $d\,L^3$ simply by $V_o$ to ease the comparison with the literature.

Similarly to the scalar field case discussed in Sec.~\ref{sec:truncation}, we can study how good the homogeneous approximation is and whether $V_o$ plays any role in the error originating from the truncation of the theory. To this aim, we look at the momentum profile obtained via a continuous Fourier transformation of the symmetry-reduced homogeneous field $\alpha(x)\bigl|_{V_o^{(\vec n)}}$ obtained in~Eq.~\eqref{homogandsmearedalpha}, where we recall from Sec.~\ref{sec:truncation} that $\vec n\in\mathbb Z^3$ denotes the three-dimensional vector attached to a corner of the cells $V_o^{(\vec n)}$ with components $n_\xi$ pointing along the $\xi=x,y,z$ directions. Explicitly, we have
\begin{align}
	\tilde{\alpha}(k) :&= \frac{1}{\sqrt{2\pi}^3} \int_{-\infty}^\infty \dd^3 x \, \alpha(x) e^{-i \vec k \cdot \vec x}\notag\\
	&= \frac{1}{\beta L^3\sqrt{2\pi}^3} \int_{-\infty}^\infty \dd^3 x \, \sqrt{\fidmetric{}{}} e^{-i \vec k \cdot \vec x}\sum_{\vec{n} \in \mathbb{Z}^3}\chi_{V_{\vec{n}}}\,\overline{\alpha}_{\mathbf{0}}^{(\vec{n})}
	\notag
	\\
	&= \frac{1}{(\sqrt{2\pi}L)^3} \sum_{\vec{n} \in \mathbb{Z}^3}\overline{\alpha}_{\mathbf{0}}^{(\vec{n})}\int_{V_o^{(\vec n)}}\dd^3x\,e^{-i\vec k\cdot\vec x}
	\notag
	\\
	&= \frac{1}{(\sqrt{2\pi}L)^3} \sum_{\vec{n} \in \mathbb{Z}^3}\overline{\alpha}_{\mathbf{0}}^{(\vec{n})} \prod_{\xi = {x,y,z}}\int_{V_o^{1/3} n_\xi}^{V_o^{1/3}( n_\xi+1)} \dd \xi \, e^{-i k_\xi \cdot \xi}
	\notag
	\\
        &= \frac{1}{\sqrt{2\pi}^3}\underbrace{\left(\sum_{\vec{n} \in \mathbb{Z}^3} \bar{\alpha}_{\mathbf{0}}^{(\vec{n})} e^{-iV_o^{1/3} \vec{k} \cdot \vec n}\right)}_{=:f_{\mathbf{0}}(\vec{k})}\,\prod_{\xi = {x,y,z}} e^{-i V_o^{1/3}k_\xi/2} \cdot \frac{\sin\left(V_o^{1/3}k_\xi/2\right)}{\left(V_o^{1/3}k_\xi/2\right)}
	\;,\label{alphamomprofile}
\end{align}

\noindent
where the function $f_{\mathbf{0}}(\vec{k})$ is $2\pi/V_o^{1/3}$-periodic in each $\vec{k}$-direction and is thus bounded, say $|f_{\mathbf{0}}(\vec{k})|^2\leq\sum_{\vec n\in\mathbb Z^3}|\overline{\alpha}_{\mathbf{0}}^{(\vec n)}|^2=:C<\infty$. As also expected from the results of Sec.~\ref{sec:truncation}, we see from Eq.~\eqref{alphamomprofile} that all $\vec{k} \in 2\pi\cdot \mathbb{Z}^3/V_o^{1/3}$ modes except $\vec k=\mathbf{0}$ do not contribute to the field momentum profile and are thus removed once spatial homogeneity has been implemented. In particular, the above momentum profile gives again a $|\tilde{\alpha}(\vec k)|^2\sim\sin^2\left(V_o^{1/3}k_\xi/2\right)/\left(V_o^{1/3}k_\xi/2\right)^2$ behaviour similar to that reported in Fig.~\ref{fig:sinx} for the scalar field case. The dominant contribution comes from the $\vec k=\mathbf{0}$ mode with $|\tilde{\alpha}(\mathbf{0})|^2=C/(2\pi)^3$, but there are also contributions from other momenta which are suppressed by order of $\mathcal O\Bigl(\frac{1}{k_\xi V_o^{1/3}}\Bigr)$. The Hamiltonian \eqref{homogeneoustruncatedhamconstraint} is thus the result of a twofold procedure involving both a homogeneous approximation and a mode truncation. Specifically, on the one hand, we ignore everything that happens outside the region $V_o$, i.e. the modes with wavelength greater than $V_o^{1/3}$ (hence $|\vec k|\leq 2\pi/V_o^{1/3}$) which remain inhomogeneous after averaging over the region $V_o$. On the other hand, within the region $V_o$, dynamical fields are approximated by their homogeneous $\vec k=\mathbf{0}$ modes and all the remaining $\vec k\neq\mathbf{0}$ modes are set to zero.~A reasonable objection to the above argument would be that the length scale entering the above discussion of wavevectors and wavelengths was the edge length $L=V_o^{1/3}$ of the cubic fiducial cell in each spatial direction measured w.r.t.~the auxiliary metric $\fidmetric{a}{b}$ rather than the physical spatial metric $q_{ab}$ and, as such, it has no physical meaning. This can be overcome by noticing that the modes wavelength, along which $\alpha(x)$ is $L$-periodic, can be translated into a physical length scale in each spatial direction, say e.g. $x$, given by
\be\label{eq:physwavelength}
\lambda_{phys}^{(x)}(y,z)=\int_{x}^{x+L}\dd x\sqrt{g_{xx}(\vec x)}=\int_{x}^{x+L}\dd x\,|a(\vec x)|=\int_{x}^{x+L}\dd x\,|\alpha(\vec x)|^{\frac{1}{3}}\;,
\ee

\noindent
where in the last equality we used the definition of $\alpha$ given in Eq.~\eqref{diagisotropyvariables}. In the purely homogeneous case $\vec k=\mathbf0$, Eq. \eqref{eq:physwavelength} reduces to (cfr.~Eq.~\eqref{homogandsmearedalpha})
\be\label{eq:lambdaphyshomog}
\lambda_{k_\xi=0}=\beta^{1/3}L|\overline{\alpha}_{\mathbf 0}|^{1/3}=(\text{vol}(V_o))^{1/3}=(\text{vol}_{\fidmetric{}{}}(V_o))^{1/3}\overline{\alpha}^{1/3}=L\alpha^{1/3}\;,
\ee

\noindent
in each spatial direction $\xi=x,y,z$, which is nothing but the edge length of the fiducial cell measured w.r.t. the physical spatial metric $q_{ab}$. The latter provides us with a physical length scale over which spatial homogeneity is imposed and consequently enters also the errors in the mode truncation leading to the homogeneous theory. In particular, the time dependence of $\alpha$ on the r.h.s.~of~\eqref{eq:physwavelength} suggests that the mini-superspace description would be reliable in large fiducial cell regimes -- as it is the case for late-time cosmology -- where significant inhomogeneities occur on scales larger than $V_o$. However, inhomogeneous contributions might become non-negligible for small $V_o$ regimes as e.g. in early-time cosmology.\footnote{As already stressed in the scalar field case of Sec. \ref{sec:truncation}, the above argument about which modes we are truncating and how strong such a truncation/approximation is should be considered as being heuristic at this stage. In order to make a statement about their role for the physics of the smeared field theory quantities  $\alpha(V)$, $\phi(V)$ and their smeared conjugate momenta one would need to study the full field theory dynamics as it was done in Sec. \ref{sec:truncation} for the simple scalar field example.~The complicated non-linear structure of the full gravitational Hamiltonian constraint \eqref{ADMconstraints} or even \eqref{isotropicvectorconstraint} would however prevent us from a straightforward extension to the case of gravity.} We will come back on this point later in Sec.~\ref{outlook:beyondhom} where the first inhomogeneous modes are explicitly included in the above analysis. This conclusion is also in line with previous investigations on the validity of the homogeneous mini-superspace approximation for cosmology based on effective (quantum) field theory arguments (see e.g. \cite{BojowaldTheBKLscenario,BojowaldEffectiveFieldTheory,BojowaldCanonicalderivationof,BojowaldEffectiveEquationsof,BojowaldMinisuperspacemodelsas} and references therein).

As a last comment before proceeding to the quantisation of the homogeneous truncated theory described by the Hamiltonian \eqref{homogeneoustruncatedhamconstraint}, let us notice that boundary terms originating from derivatives of $\chi_V(x)$ have also been neglected in the Hamiltonian \eqref{homogeneoustruncatedhamconstraint}. Such boundary terms are expected not only to encode the already mentioned cross-boundary interactions between modes localised inside and outside the homogeneity region, but we also expect them to play a role about diffeomorphism symmetry in the cosmology setting. Indeed, being the inward normal derivative of $\chi_V(x)$ related to a surface Dirac delta, the boundary terms of the constraints would yield the generating charges of surface symmetries.\footnote{As discussed in Sec. \ref{outlook:beyondhom}, the vector constraint becomes also non-trivial when inhomogeneous modes are included.} Neglecting boundary terms would then be again plausible for large volumes where the averaged bulk physics is expected to not be significantly affected by boundary effects. More subtle would be instead the situation in the small volume regime and one needs to be careful as the truncated homogeneous theory might thus neglect both physically relevant boundary terms and modes. A precise understanding of such lines of thoughts would be interesting and we will reserve a detailed study to future work.

\section{Quantum Cosmology}\label{Sec:QuantumCosmology}

We shall now proceed to study the quantisation of the spatially homogeneous, truncated cosmological model discussed in the previous section or, more precisely, of the full class of canonically inequivalent homogeneous classical theories identified by the different values of $V_o$. Specifically, we will consider a polymer quantisation of the symmetry-reduced model as usually done in LQC \cite{BojowaldLoopQuantumCosmology,AshtekarLoopquantumcosmology:astatusreport,BodendorferAnElementaryIntroduction}, the aim being to study the role played by the fiducial cell $V_o$ at the quantum level.
From now on, we will denote the homogeneity region and its coordinate volume both by $V_o$, where the coordinate volume is simply $V_o =d\,L^3$.
Special focus will lie on the change of the choice of $V_o$ and its implications for quantum dynamics, expectation values, and uncertainty relations. 

\subsection{Quantisation of the Homogeneous Theory}\label{Sec:Quantumrepresentation}

As discussed in the previous section, after imposing homogeneity constraints, the homogeneous local fields restricted to the region $V_o$ are equivalent to their $V_o$-averaged counterparts (cfr. Eqs. \eqref{homogandsmearedalpha}, \eqref{homogandsmearedPalpha}, and similarly for the scalar field sector).~The spatially homogeneous fields $(\alpha=\alpha(x)\bigl|_{x \in V_o}, P_\alpha=P_\alpha(x)\bigl|_{x \in V_o})$ in the gravitational sector and $(\phi=\phi(x)\bigl|_{x \in V_o}, P_\phi=P_\phi(x)\bigl|_{x \in V_o})$ in the matter sector are related to each other via the Hamiltonian constraint \eqref{homogeneoustruncatedhamconstraint}, which encodes the remaining time-reparametrisation gauge freedom. Moreover, the canonical pairs for the gravitational and matter sectors satisfy the Dirac-bracket algebra (cfr. Eq. \eqref{gravityDBxspace})
\begin{equation}\label{eq:GravDB}
	\Poisson{\alpha}{P_\alpha}_D = \frac{1}{V_o} \qquad , \qquad \Poisson{\phi}{P_\phi}_D = \frac{1}{V_o} \;,
\end{equation}

\noindent
i.e., as it was also the case for the scalar field example discussed in Sec. \ref{sec:constraintimplementation} (cfr. Eq. \eqref{eq:Dbracketphixpiy}), after implementing the second class homogeneity constraints strongly by means of the Dirac bracket, we end up with a one-parameter family of classically symmetry-reduced homogeneous theories whose canonical structures are labeled by $V_o$.

In order to quantise the reparametrisation-invariant cosmological model under consideration, we deparametrise the system w.r.t. the clock scalar field $\phi$ and solve the Hamiltonian constraint \eqref{homogeneoustruncatedhamconstraint} for $P_\phi$, thus leading to the true Hamiltonian generating evolution in $\phi$-time
\begin{equation}\label{phiclockHtrue}
	P_\phi(V_o) = \sqrt{\frac{3 \kappa}{2}} V_o \alpha P_\alpha =: H_{true}\;,
\end{equation}

\noindent
where we have only chosen the positive sign of the square root.
This corresponds to the positive frequency modes only.
Usually both kind of modes should be taken into account, however, since our aim here is not to study the complete cosmological model, but rather to focus on the relevance of the fiducial cell, such a restriction would be sufficient and the considerations in the following can be straightforwardly extended to the case in which both positive and negative frequency modes are considered.
The task is now to promote the Dirac brackets \eqref{eq:GravDB} to canonical commutation relations respectively for a weakly continuous representation of the matter sector operators and a weakly discontinuous polymer representation for the gravitational sector. Specifically, the matter sector Hilbert space $\mathscr H_M$ is realised as $\mathscr{H}_M\cong L^2(\mathbb{R},\dd\phi)$, with $\dd\phi$ denoting the Lebesque measure w.r.t to the clock variable $\phi$. On such a Hilbert space, the canonical homogeneous field operators $\hat\phi$, $\hat{P}_\phi$ act as
\be\label{Eq:matterelementartoperators}
\hat{\phi}\,\Psi(\phi) = \frac{\phi}{V_o^{\xi}}\,\Psi(\phi) \qquad , \qquad \hat{P}_\phi \Psi(\phi) = -\frac{i}{V_o^{\zeta}} \partdif{}{\phi} \Psi(\phi) \qquad(\text{$\zeta + \xi = 1$, but else arbitrary})
\ee

\noindent
and satisfy the canonical commutation relations ($\hbar=1$)
\be\label{QC:CCRmatter}
	\left[\hat{\phi}, \hat{P}_\phi\right] = \frac{i}{V_o} \;,
\ee

\noindent
as expected from the classical canonical brackets \eqref{eq:GravDB} via Dirac correspondence principle.\footnote{The arbitrary powers $\xi$ and $\zeta$ ($\xi+\zeta=1$) in Eq.~\eqref{Eq:matterelementartoperators} parametrise the freedom in incorporating the $V_o$-factors in the representation of the
quantum operators to ensure the canonical brackets \eqref{eq:GravDB} to be correctly represented as commutation relations on $\mathscr H_M$. Note that in the analogous relation \eqref{eq:canonicaloprepr} in Sec.~\ref{sec:phiquantisation}, we have chosen $\zeta = 1$ and $\xi = 0$. There, however, one could have been more general as we want to be here, while leaving the final result unaltered.}
The corresponding Schr\"odinger equation for the evolution in $\phi$-clock then reads
\begin{equation}\label{phiclockSE}
	-iV_o^{1-\zeta} \partdif{}{\phi} \Psi(\phi) = \hat{H}_{true} \Psi(\phi) \;,
\end{equation}

\noindent
with the operator $\hat{H}_{true}$ to be defined below.
In order to represent the gravitational sector on a polymer Hilbert space, we impose the Weyl form of the canonical commutation relations \cite{AshtekarMathematicalStructureOf,BojowaldLoopQuantumCosmology,AshtekarLoopquantumcosmology:astatusreport,AshtekarQuantumNatureOf,AshtekarQuantumGravityShadow}
\begin{equation}\label{eq:WCCR}
	\widehat{e^{-i\mu \alpha}} \widehat{e^{-i\nu P_\alpha}} = \widehat{e^{-i\nu P_\alpha}} \widehat{e^{-i\mu \alpha}} e^{-\frac{i \mu \nu}{V_o}} \;,
\end{equation}

\noindent
where we have the units $[\alpha] = [\mu] = 1$ and $[P_\alpha] = [\nu^{-1}] = \text{\textit{length}}^{-3}$.
Choosing the $\alpha$-polarisation, we can then define the exponentiated operators to act as
\be\label{gravelementaryoperators1}
\widehat{e^{-i\mu \alpha}} \Psi(\alpha) = e^{-i \frac{\eta^{\gamma} \mu \alpha}{V_o^\gamma}} \Psi(\alpha)\qquad,\qquad\widehat{e^{-i\nu P_\alpha}} \Psi(\alpha) = \Psi\left(\alpha - \frac{\nu}{\eta^{\gamma} V_o^\delta}\right)\;,
\ee

\noindent
where $\eta = \kappa^{3/2}$ and allows for the freedom to shift the $V_o$ dependence into the representation of $\alpha$ or $P_\alpha$.
Again, we have $\gamma + \delta = 1$, in which case these operators  satisfy the commutation relations \eqref{eq:WCCR}.
These operators are then usually represented on the Hilbert space $\mathscr{H}_{poly}\cong L^2\left(\mathbb{R}_{\text{Bohr}}, \dd\mu_\text{Bohr}\right)$ of square-integrable functions on the Bohr compactification of the real line $\mathbb{R}_{\text{Bohr}}$ w.r.t. its translation-invariant normalised Haar measure \cite{SubinDifferentialAndPseudodifferential} (see also Appendix 28 in \cite{ThiemannBook}), via a weakly continuous and a weakly discontinuous representation in $\alpha$ and in $P_\alpha$, respectively \cite{AshtekarMathematicalStructureOf,BojowaldLoopQuantumCosmology,AshtekarLoopquantumcosmology:astatusreport,AshtekarQuantumNatureOf,AshtekarQuantumGravityShadow}.~Therefore, the operator $\hat{\alpha}$ can be defined via differentiation, that is via the $\mu\to0$ limit of the incremental ratio, while a bare operator $\hat{P}_\alpha$ is not well-defined on $\mathscr{H}_{poly}$ and only the corresponding exponentiated Weyl operator does exist as well-defined operator on $\mathscr{H}_{poly}$.
As it is well known from LQC literature \cite{AshtekarLoopquantumcosmology:astatusreport,BodendorferAnElementaryIntroduction}, the Hamiltonian operator so defined preserves lattices of integer steps, namely 
\begin{equation}\label{eq:nuinteger}
	\nu \in \frac{\lambda}{\eta^\gamma V_o^\delta}\,\mathbb Z\;,
\end{equation}

\noindent
where $\lambda$ is the polymerisation scale and has the units of volume.
Moreover, by choosing the offset for $\alpha$ to be zero, we end up with the kinematical Hilbert space for the gravitational sector given by
\be\label{eq:HilbertGrav}
	\mathscr{H}_{poly} = \overline{\text{span}_{\alpha \in \frac{\lambda}{\eta^\gamma V_o^\delta} \mathbb Z} \left\{\Psi(\alpha) \in \mathbb{C}\right\}}\;,
\ee

\noindent
where the closure is taken w.r.t. the scalar product
\be\label{eq:HilbertGravkinscalarprod}
	\braket{\Psi_1}{\Psi_2} = \sum_{\alpha \in \frac{\lambda}{\eta^\gamma V_o^\delta} \mathbb Z} \Psi_1^*(\alpha) \Psi_2(\alpha)\;.
\ee

\noindent
Note that this is simply the discrete Fourier decomposition of $2\pi V_o^\delta/\eta^\gamma \lambda$-periodic functions in $P_\alpha$.
The elementary operators \eqref{gravelementaryoperators1} thus act on such a Hilbert space as
\begin{equation}\label{eq:polymerrepr}
	\hat{\alpha}\,\Psi(\alpha) = \frac{\eta^\gamma}{V_o^\gamma} \alpha \Psi(\alpha) \qquad , \qquad \widehat{e^{-i\lambda P_\alpha}} \Psi(\alpha) = \Psi\left(\alpha - \frac{\lambda}{\eta^{\gamma} V_o^\delta}\right)\;,
\end{equation}
and satisfy the polymerised commutation relations
\be\label{LQC:polyCCRgrav}
\left[\widehat{e^{-i\lambda P_\alpha}},\hat\alpha\right]=i\reallywidehat{\{e^{-i\lambda P_\alpha},\alpha\}}\;.
\ee

Finally, we can define an ordering and regularisation for the quantum operator corresponding to the Hamiltonian \eqref{phiclockHtrue}.
Using the so-called MMO-ordering \cite{Martin-BenitoFurtherImprovementsIn} and simplifying the expressions for the restriction \eqref{eq:nuinteger}, we get
\begin{equation}
	\hat{H}_{true} = \sqrt{\frac{3 \kappa}{2}} V_o \sqrt{|\hat{\alpha}|} \frac{\reallywidehat{\sin\left(\lambda P_\alpha\right)}}{\lambda} \sqrt{|\hat{\alpha}|} \;,
\end{equation}

\noindent
which annihilates the $\alpha = 0$ state, and preserves positive and negative volume branches as can be readily seen from its explicit action given by\footnote{At the classical level, the sign of volume depends on whether the orientation of the physical triad matches that of the fiducial triad.~In the quantum theory, as can be seen from Eq.~\eqref{eq:Htureaction}, the Hamiltonian operator acts the same way on both positive and negative volume states, resulting in two copies of the same system.~The transformation $\Psi(\alpha)\mapsto\Psi(-\alpha)$ amounts then to a parity-like transformation which does not change the physics of the system under consideration.~This can be implemented by decomposing the state space into the corresponding irreducible representations, that is into either symmetric or anti-symmetric wave functions, and discuss each of them separately.~As there is no qualitative difference in the physics of these sectors, we restrict ourselves to the symmetric representation and work with states satisfying $\Psi(\alpha)=\Psi(-\alpha)$ as typically done in LQC (see e.g.~\cite{AshtekarLoopquantumcosmology:astatusreport} and references therein).}
\begin{equation}\label{eq:Htureaction}
	\hat{H}_{true} \Psi(\alpha) = -\frac{i}{2}\sqrt{\frac{3 \kappa}{2}} \cdot \frac{\eta^\gamma V_o^\delta}{\lambda} \left(\sqrt{|\alpha| \left|\alpha + \frac{\lambda}{\eta^\gamma V_o^\delta}\right|} \Psi\left(\alpha + \frac{\lambda}{\eta^\gamma V_o^\delta}\right) - \sqrt{|\alpha| \left|\alpha - \frac{\lambda}{\eta^\gamma V_o^\delta}\right|} \Psi\left(\alpha - \frac{\lambda}{\eta^\gamma V_o^\delta}\right)\right).
\end{equation}

\noindent
Note again that, due to the presence of $V_o$ in the Dirac bracket \eqref{eq:GravDB}, different values of $V_o$ correspond to different quantum representations. In other words, at the quantum level, the scaling property of the canonical Dirac brackets is shifted into
the quantisation map for the representations of the operators associated to phase space quantities.
Consequently, classical phase space quantities as e.g. $P_\phi$, which do not change if the value of $V_o$ is changed from the value $V_o^{(1)}$ to $V_o^{(2)}$, do transform as operators in the quantum theory according to the following relations\footnote{Note that Eq.~\eqref{eq:Palphaopscaling} is correct as it is written, but contradicts the assumption $\nu \in \lambda \mathbb{Z}$ if $\lambda$ is assumed to be fixed. This is related to the shortcut applied here and skipping the discussion of the kinematical Hilbert space to be the almost-periodic functions $L^2\left(\mathbb{R}_{\text{Bohr}}, \dd \mu_{\text{Bohr}}\right)$ rather than periodic functions. The reason is that, for different $V_o$, different sublattices are preserved within $L^2\left(\mathbb{R}_{\text{Bohr}}, \dd \mu_{\text{Bohr}}\right)$ (cfr. Eq. \eqref{eq:nuinteger}).}
\begin{subequations}
	\begin{align}
		\left.\hat{P}_{\phi}\right|_{V_o^{(1)}} &= \left(\frac{V_o^{(2)}}{V_o^{(1)}}\right)^{\zeta}\left.\hat{P}_{\phi}\right|_{V_o^{(2)}}\qquad,\qquad
		\left.\hat{\phi}\right|_{V_o^{(1)}} = \left(\frac{V_o^{(2)}}{V_o^{(1)}}\right)^{\xi}\left.\hat{\phi}\right|_{V_o^{(2)}}\;,
		\\
		\left.\hat{\alpha}\right|_{V_o^{(1)}} &= \left(\frac{V_o^{(2)}}{V_o^{(1)}}\right)^{\gamma}\left.\hat{\alpha}\right|_{V_o^{(2)}}\qquad,\qquad
		\left.\widehat{e^{-i\nu^{(1)} P_\alpha}}\right|_{V_o^{(1)}} = \left.\widehat{e^{-i \nu^{(2)} P_\alpha}}\right|_{V_o^{(2)}} \quad,\quad \nu^{(1)} = \left(\frac{V_o^{(2)}}{V_o^{(1)}}\right)^\delta \nu^{(2)} \;.\label{eq:Palphaopscaling}
	\end{align}
\end{subequations}

\noindent
As we have seen in the scalar field case (cfr.~Sec.~\ref{sec:phiquantisation}), the Hamiltonian operator has no clear scaling behaviour.
Changing $V_o$ essentially affects the amount of shift into the difference equation \eqref{eq:Htureaction}.
However, for any choice of $V_o$, this is simply a shift to one more higher or smaller value in the $\alpha$-lattice corresponding to the given $V_o$-representation.
This will be important below.
Furthermore, as the Hamiltonian constraint forces $\hat{P}_\phi = \hat{H}_{true}$, it is clear that the choice of $\zeta$ in Eqs. \eqref{Eq:matterelementartoperators}, \eqref{phiclockSE} for the matter sector is not independent of the choice of $\gamma$ and $\delta$ for the gravitational sector as both sides have to behave consistently.
As we use a deparametrised setting, this plays no deeper role here.
In fact, the matter part of the Hilbert space can simply be neglected and the operator $\hat{P}_\phi$ is defined via the constraint on the gravitational Hilbert space.
Similarly, the operator for $\phi$ plays simply the role of a relational clock and is defined via this relation.
With this being said, let us discuss the spectrum of $\hat{H}_{true}$ and how eigenstates transform when the value of $V_o$ is changed. To this aim, let us define the following quantities
\begin{equation}\label{eq:defalphatheta}
	\alpha = \frac{\lambda}{\eta^\gamma V_o^\delta} n \qquad(n \in \mathbb{Z}) \qquad , \qquad \theta = \frac{\lambda}{\eta^\gamma V_o^\delta} \;,
\end{equation}

\noindent
so that the action of the Hamiltonian~\eqref{eq:Htureaction} reads
\begin{equation}\label{Eq:Htruediffeq}
	\hat{H}_{true} \Psi(\alpha) = -\frac{i}{2} \sqrt{\frac{3 \kappa}{2}} \cdot \left(\sqrt{|n||n+1|} \Psi\left(\theta\cdot(n+1)\right) - \sqrt{|n||n-1|} \Psi\left(\theta\cdot(n-1)\right)\right)\;.
\end{equation} 

\noindent
The eigenstates of $\hat{H}_{true}$ are well-studied and can be written in analytic form \cite{Martin-BenitoFurtherImprovementsIn,AshtekarCastinglqcinthe}. However, their explicit expression is not needed to analyse their behaviour when changing the value of $V_o$. In fact, an eigenstate $\Psi_E: \mathbb{Z} \rightarrow \mathbb{C}$ of $\hat H_{true}$ satisfies the equation \cite{Martin-BenitoFurtherImprovementsIn,AshtekarCastinglqcinthe}
\begin{equation}
	-\frac{i}{2} \sqrt{\frac{3 \kappa}{2}} \cdot \left(\sqrt{|n||n+1|} \Psi_E\left(n+1\right) - \sqrt{|n||n-1|} \Psi_E\left(n-1\right)\right) = E \Psi_E(n) \;.
\end{equation}

\noindent
It is then obvious that, according to \eqref{Eq:Htruediffeq}, the functions $\Psi_E(\alpha) = \Psi_E(\theta\cdot n) = \Psi_E(n)$ are eigenstates of the Hamiltonian with eigenvalue $E$.~Suppose now to consider two quantum representations resulting from quantising the classical symmetry-reduced theories associated with two distinct values of $V_o$, say $V_o^{(1)}$ and $V_o^{(2)}$, respectively. In the two quantum representations, the corresponding eigenvalue equations for the Hamiltonian with the two different $V_o$ (hence $\theta$) can be solved and the spectrum is exactly the same providing that their eigenstates are related by
\begin{equation}\label{12eigenstatesidentification}
	\Psi_E^{(1)}\left(\alpha\right) = \Psi_E^{(2)}\left(\left(\frac{V_o^{(1)}}{V_o^{(2)}}\right)^{\delta}\alpha\right) \;,
\end{equation}

\noindent
where the super-scripts ${(1)}$ and $(2)$ denote the eigenstates in the corresponding quantum representations.
Note that it is a priori not necessary to identify the states in the different quantum theories according to the above relation\footnote{At least when almost-periodic functions are used as Hilbert space. The particular subset of $\alpha$-grid and the subsequent only consistent identification of states is however a dynamical property, as the Hamiltonian dictates which grid is preserved.} and one could in principle find other arguments to identify states in the different Hilbert spaces depending on the physical situation one is aiming to describe\footnote{We will further comment on this point in the outlook and concluding discussion of section \ref{sec:furtherfuturedirections}.}.
In the present setting, however, the  requirement \eqref{12eigenstatesidentification} guarantees that the dynamics is equivalent in both representations and is thus independent of $V_o$, as it was the case also in the classical theory.~Consequently, it is possible to define the following isomorphism between the two Hilbert spaces, which preserves $\phi$-dynamics 
\be\label{eq:isomorphygrav}
	\mathscr{I}:\; \mathscr{H}_{poly}^{(1)} \rightarrow \mathscr{H}_{poly}^{(2)} \qquad\text{by}\qquad\Psi^{(1)}(\alpha) \mapsto \Psi^{(2)}(\alpha) = \mathscr{I}\left(\Psi^{(1)}(\alpha)\right):= \Psi^{(1)}\left(\left(\frac{V_o^{(2)}}{V_o^{(1)}}\right)^\delta \alpha\right).
\ee

\noindent
Their expansion coefficients in terms of $\Psi_E^{(1)}$ and $\Psi_E^{(2)}$ will be the same and thus their evolution equivalent.~Eq.~\eqref{eq:isomorphygrav} is the analogue of the mapping \eqref{VoHilbertspaceisomorphism} we found in the study of the quantisation of the homogeneous scalar field theory in Sec.~\ref{sec:phiquantisation}.~The explicit details of the mapping between states in the Hilbert spaces corresponding to the different $V_o$ values depend on the Hamiltonian operator.~Note that, similarly to the scalar field case of Sec.~\ref{sec:phiquantisation} where the $V_o$-factor in the quantum representation of the canonical operators was entirely incorporated in the momentum operator (cfr.~Eq.~\eqref{eq:canonicaloprepr}), now in the gravitational case only the exponent $\delta$ of the $V_o$ factor in the quantum representation \eqref{gravelementaryoperators1} of the momentum operator enters the mapping \eqref{eq:isomorphygrav}.

\subsection{Implications for Uncertainty Relations and Quantum Fluctuations}\label{Sec:cosmHURandfluc}

The isomorphism \eqref{eq:isomorphygrav} between the Hilbert spaces associated to the quantum representations labeled by different values of $V_o$ allows us to make the dynamics of the whole family of $V_o$-labeled quantum theories equivalent. This can be thought of as the quantum analogue of the $V_o$-independence of the classical theory. A quantum theory is however richer than just dynamics as it also involves quantum fluctuations and uncertainty relations. It is then important to study whether and how such quantities depend on $V_o$ and their relation between the different $V_o$-valued quantum representations. To this aim, let us study the scaling behaviour of the expectation values and higher statistical moments of the relevant quantum operators. We start with the elementary operator $\hat\alpha$ for which, recalling the action \eqref{eq:polymerrepr} and the definitions \eqref{eq:defalphatheta}, we find
\begin{align}\label{1to2alphaexpvalue}
	\left<\left.\hat{\alpha}\right|_{V_o^{(1)}}\right>_{\Psi^{(1)}}&:=\braopket{\Psi^{(1)}}{\left.\hat{\alpha}\right|_{V_o^{(1)}}}{\Psi^{(1)}} = \sum_{\alpha \in \frac{\lambda}{\eta^\gamma \left(V_o^{(1)}\right)^\delta} \mathbb Z} \Psi^{(1)*}(\alpha) \frac{\eta^\gamma}{\left(V_o^{(1)}\right)^\gamma} \alpha \Psi^{(1)}(\alpha)
	\notag
	\\
	&=\sum_{\alpha \in \frac{\lambda}{\eta^\gamma \left(V_o^{(1)}\right)^\delta} \mathbb Z} \Psi^{(2)*}\left(\left(\frac{V_o^{(1)}}{V_o^{(2)}}\right)^\delta\alpha\right) \frac{\eta^\gamma}{\left(V_o^{(1)}\right)^\gamma} \alpha \Psi^{(2)}\left(\left(\frac{V_o^{(1)}}{V_o^{(2)}}\right)^\delta\alpha\right)
	\notag
	\\
	&=\sum_{\alpha' \in \frac{\lambda}{\eta^\gamma \left(V_o^{(2)}\right)^\delta} \mathbb Z} \Psi^{(2)*}\left(\alpha'\right) \frac{\eta^\gamma}{\left(V_o^{(1)}\right)^\gamma} \left(\frac{V_o^{(2)}}{V_o^{(1)}}\right)^\delta \alpha' \Psi^{(2)}\left(\alpha'\right)
	\notag
	\\
	&= \frac{V_o^{(2)}}{V_o^{(1)}}\sum_{\alpha' \in \frac{\lambda}{\eta^\gamma \left(V_o^{(2)}\right)^\delta} \mathbb Z} \Psi^{(2)*}\left(\alpha'\right) \frac{\eta^\gamma}{\left(V_o^{(2)}\right)^\gamma} \alpha' \Psi^{(2)}\left(\alpha'\right)
	\notag
	\\
	&= \frac{V_o^{(2)}}{V_o^{(1)}}\braopket{\Psi^{(2)}}{\left.\hat{\alpha}\right|_{V_o^{(2)}}}{\Psi^{(2)}}
	\notag
	\\
	&= \frac{V_o^{(2)}}{V_o^{(1)}} \left<\left.\hat{\alpha}\right|_{V_o^{(2)}}\right>_{\Psi^{(2)}} \;,
\end{align}
where we used the mapping \eqref{eq:isomorphygrav} in the second line, relabeled $\alpha' = \left(V_o^{(1)}/V_o^{(2)}\right)^\delta\alpha$ in the third line, and used the relation $\delta+\gamma=1$ to isolate the factor $V_o^{(2)}/V_o^{(1)}$ in the fourth line. Similarly, for the exponentiated momentum operator whose action is given in the second equation of \eqref{eq:polymerrepr}, we have ($m \in \mathbb{Z}$)

\begin{align}\label{eq:expPalphascaling}
	\left<\left.\widehat{e^{-i\lambda m P_\alpha}}\right|_{V_o^{(1)}}\right>_{\Psi^{(1)}} &= \sum_{\alpha \in \frac{\lambda}{\eta^\gamma \left(V_o^{(1)}\right)^\delta} \mathbb Z} \Psi^{(1)*}(\alpha) \Psi^{(1)}\left(\alpha-\frac{\lambda m}{\eta^{\gamma} \left(V_o^{(1)}\right)^\delta}\right)
	\notag
	\\
	&= \sum_{\alpha' \in \frac{\lambda}{\eta^\gamma \left(V_o^{(1)}\right)^\delta} \mathbb Z} \Psi^{(2)*}(\alpha') \Psi^{(2)}\left(\alpha'-\frac{\lambda m}{\eta^{\gamma} \left(V_o^{(2)}\right)^\delta}\right)
	\notag
	\\
	&=\left<\left.\widehat{e^{-i\lambda m P_\alpha}}\right|_{V_o^{(2)}}\right>_{\Psi^{(2)}}\;.
\end{align}

\noindent
Using then the action \eqref{eq:Htureaction} of $\hat H_{true}$ and relabeling again $\alpha' = \left(V_o^{(1)}/V_o^{(2)}\right)^\delta\alpha$ after invoking the isomorphism \eqref{eq:isomorphygrav}, it is easy to show that the expectation value of the Hamiltonian remains unchanged
\begin{equation}\label{eq:expvalueHtrue12}
	\left<\left.\hat{H}_{true}\right|_{V_o^{(1)}}\right>_{\Psi^{(1)}} = 
	\left<\left.\hat{H}_{true}\right|_{V_o^{(2)}}\right>_{\Psi^{(2)}} \;, 
\end{equation}

\noindent
as it is expected by the mapping \eqref{eq:isomorphygrav} being dynamics-preserving, and is thus independent of the choice of $V_o$. Finally, by means of similar steps, it is straight forward to convince ourselves of the following relations for the higher moments and variances of the operators in Eq. \eqref{eq:polymerrepr}
\begin{subequations}
	\begin{align}
		\left<\left.\hat{\alpha}\right|_{V_o^{(1)}}^n\right>_{\Psi^{(1)}} =\; \left(\frac{V_o^{(2)}}{V_o^{(1)}}\right)^n \left<\left.\hat{\alpha}\right|_{V_o^{(2)}}^n\right>_{\Psi^{(2)}}\qquad&,\qquad
		\left<\left.\widehat{e^{-i\lambda m P_\alpha}}\right|_{V_o^{(1)}}^n\right>_{\Psi^{(1)}} =\;  \left<\left.\widehat{e^{-i\lambda m P_\alpha}}\right|_{V_o^{(2)}}^n\right>_{\Psi^{(2)}}\;,\label{eq:scalingmomentsGR}
		\\
		\Delta_{\Psi^{(1)}} \left.\hat{\alpha}\right|_{V_o^{(1)}} =\; \frac{V_o^{(2)}}{V_o^{(1)}} \Delta_{\Psi^{(2)}} \left.\hat{\alpha}\right|_{V_o^{(2)}} \qquad&,\qquad
		\Delta_{\Psi^{(1)}} \left.\widehat{e^{-i\lambda n P_\alpha}}\right|_{V_o^{(1)}} =\; \Delta_{\Psi^{(2)}} \left.\widehat{e^{-i\lambda n P_\alpha}}\right|_{V_o^{(2)}} \;.\label{eq:scalingvariancesGR}
	\end{align}
\end{subequations}

\noindent
Note that in the above discussion the $\phi$-clock time dependence of the states was left implicit, namely $\ket{\Psi}=\ket{\Psi;\phi}$. Therefore, as the isomorphism \eqref{eq:isomorphygrav} preserves time evolution, the above results for expectation values and variances between different $V_o$-labeled quantum theories hold in a fully dynamical sense.

We are now in the position to investigate whether changing the value of $V_o$ has physical consequences for the smeared observables. To this aim, let us first look at the volume of the fiducial cell itself, i.e. 
\begin{equation}\label{fidvolumeoperator}
	\widehat{\text{vol}(V_o)} := \widehat{\alpha(V_o)} = V_o\,\hat{\alpha}\;,
\end{equation}

\noindent
where in the last equality we used the relation \eqref{homogfullsmearedalpha} between the smeared quantity and the homogeneous local field variable, now promoted to quantum operators.~According to the scaling relations \eqref{eq:scalingmomentsGR}, we find that the expectation value of the operator \eqref{fidvolumeoperator} is actually independent of the value of $V_o$, i.e.
\be\label{eq:volexpvalue}
\left<\reallywidehat{\text{vol}\left(V_o^{(1)}\right)}\right>_{\Psi^{(1)}}=\left<\widehat{\alpha(V_o)}\Bigl|_{V_o=V_o^{(1)}}\right>_{\Psi^{(1)}}=\left<\widehat{\alpha(V_o)}\Bigl|_{V_o=V_o^{(2)}}\right>_{\Psi^{(2)}}= \left<\reallywidehat{\text{vol}\left(V_o^{(2)}\right)}\right>_{\Psi^{(2)}} \;.
\ee

\noindent
Combining this with the canonical commutation relations \eqref{LQC:polyCCRgrav}, the classical canonical brackets \eqref{eq:GravDB}, and the above results \eqref{eq:scalingvariancesGR} for the variances of the elementary operators yields the following uncertainty relations 
\begin{align}\label{eq:volumeUR}
		&\frac{1}{2\,V_o^{(1)}} \biggl|\left<\widehat{\cos\left(\lambda P_\alpha\right)}\bigl|_{V_o^{(1)}}\right>_{\Psi^{(1)}}\biggr| \le \Delta_{\Psi^{(1)}} \hat{\alpha}\bigl|_{V_o^{(1)}} \Delta_{\Psi^{(1)}} \frac{\reallywidehat{\sin\left(\lambda P_\alpha\right)}}{\lambda}\biggl|_{V_o^{(1)}} = \frac{V_o^{(2)}}{V_o^{(1)}} \Delta_{\Psi^{(2)}} \hat{\alpha}\bigl|_{V_o^{(2)}} \Delta_{\Psi^{(2)}} \frac{\reallywidehat{\sin\left(\lambda P_\alpha\right)}}{\lambda}\biggl|_{V_o^{(2)}}\notag\\
		&\Rightarrow\qquad\Delta_{\Psi^{(2)}} \hat{\alpha}\biggl|_{V_o^{(2)}} \Delta_{\Psi^{(2)}} \frac{\reallywidehat{\sin\left(\lambda P_\alpha\right)}}{\lambda}\biggl|_{V_o^{(2)}} \ge \frac{1}{2\,V_o^{(2)}} \biggl|\left<\widehat{\cos\left(\lambda P_\alpha\right)}\bigl|_{V_o^{(2)}}\right>_{\Psi^{(2)}}\biggr| \;,
\end{align}

\noindent
where, to ease a later comparison with existing LQC literature as e.g. \cite{TaverasCorrectionstothe, RovelliWhyAreThe}, we focused on the operator corresponding to the simplest regularisation for the conjugate momentum $P_\alpha$ by combination of its exponentiated version (point holonomies) yielding the sin function polymerisation.\footnote{Motivated by physical inputs or full theory based arguments like general covariance and anomaly-free realisations of the constraint algebra, alternative proposals of polymerisation have been considered in the literature as e.g. \cite{DaporCosmologicaleffectiveHamiltonian,AssanioussiEmergentdeSitter,AssanioussiEmergentdeSitterdetails, AssanioussiPerspectivesonthe,BojowaldCriticalEvaluationof} and references therein. Recalling that $\Delta f(e^{-i\lambda P_\alpha})\approx |f'(\langle e^{-i\lambda P_\alpha}\rangle)|\Delta e^{-i\lambda P_\alpha}$, provided that $f$ is at least twice differentiable and finiteness of the mean and variance, and noticing that the ratio $V_o^{(2)}/V_o^{(1)}$ in \eqref{eq:volumeUR} originates only from the scaling behaviour of $\Delta\hat\alpha$ as the moments and variance of the exponentiated momentum operator do not scale (cfr. Eqs. \eqref{eq:scalingmomentsGR}, \eqref{eq:scalingvariancesGR}), the above discussion of the uncertainty relations should in principle encompass also other polymerisation choices based on sufficiently differentiable functions of point holonomies.} In particular, similarly to the analysis of the scalar field in Sec.~\ref{sec:phiquantisation} (cfr.~Table~\ref{table:comparisonexpvalues}), states saturating the uncertainty relations \eqref{eq:volumeUR} in the quantum theory corresponding to the cell $V_o^{(1)}$ are mapped through \eqref{eq:isomorphygrav} into states saturating them in the quantum theory corresponding to the cell $V_o^{(2)}$.~Moreover, according to the scaling behaviours \eqref{eq:scalingmomentsGR} and \eqref{eq:scalingvariancesGR}, the point around which the states are peaked and their widths will transform under a change of the fiducial cell.~We refer the reader to Appendix~\ref{app:CS} for the explicit discussion of the above scaling properties for coherent states.

In contrast to the classical theory where the homogeneous smeared volume \eqref{homogfullsmearedalpha} of the region $V_o\subset\Sigma$ scales extensively under active physical rescaling of $V_o$ (cfr. discussion below Eqs.~\eqref{homogfullsmearedalpha}-\eqref{homogfullsmearedphi}), at the quantum level the expectation value \eqref{eq:volexpvalue} of the operator for the integrated volume of the cell, and in turn the quantum fluctuations \eqref{eq:volumeUR}, are independent of the choice of $V_o$. More specifically, using eigenstates of $\hat{\alpha}$, i.e. $\Psi = \delta_{\alpha,\alpha_o}$, we find (cfr. Eqs.~\eqref{eq:polymerrepr})
\be\label{eigenstatsmearedalphaexpvalue}
\left<\widehat{\alpha(V_o)} \right>_{\delta_{\alpha,\alpha_o}} = V_o \frac{\eta^\gamma}{V_o^\gamma} \alpha_o \overset{\gamma+\delta=1}{=} V_o^\delta\eta^\gamma\alpha_o\overset{\eqref{eq:defalphatheta}}{=} \lambda n_o \;,
\ee

\noindent
which tells us that, after
imposing the Hamiltonian constraint (and restricting to
positive volumes), we can assign to any fiducial box only volumes in (positive) integer steps of $\lambda$, independently of the choice of $V_o$.
This is physically plausible as $V_o$ is only a coordinate volume and there is no reference to any other physical object.
Only when a state $\Psi$ is chosen, we assign a ``size'' to $V_o$ as we can then compute the expectation value of the smeared operator over the fiducial cell.
The physics is not changed if the coordinate value $V_o$ is changed as only the state determines the volume.\footnote{This is as in the classical theory, where only $\alpha(V_o) = V_o\alpha$ has physical relevance as both $V_o$ and $\alpha$ are coordinate dependent. The value of $\alpha$ is fixed by the initial conditions, which corresponds in the quantum theory to the choice of state. In this sense, the statement of the geometry of the cell to be specified by the choice of state $\Psi $ comes to be the quantum analogue of specifying initial conditions for the classical coordinate-independent quantity $\text{vol}(V_o)$.} Similarly, combining Eqs. \eqref{1to2alphaexpvalue}, \eqref{fidvolumeoperator}, and \eqref{eigenstatsmearedalphaexpvalue}, the equality \eqref{eq:volexpvalue} of the expectation values of the smeared operators comes to be the quantum analogue of the further invariance under active diffeomorphisms rescaling the size of the fiducial cell without deforming it. Indeed, in the classical theory, even though an active rescaling $L_o\mapsto\beta L_o$ of the homogeneity region $V_o$ as a subset of $\Sigma_t$ affects its physical size $L_{phys}=L_oa(t)=L_o\alpha(t)^{1/3}$ as $L_{phys}\mapsto\beta L_{phys}$, the spatially homogeneous solution of the Einstein's equation with scale factor $a/\beta$ has the same initial conditions for $L_{phys}=(\beta L_o)a/\beta$ and leads to equivalent physics. Similarly, at the quantum level, a physical change of the value of $V_o$ from $V_o^{(1)}$ to $V_o^{(2)}$ leads to a corresponding rescaling of the expectation value of the non-smeared quantum operator $\hat\alpha$ w.r.t. the two (eigen)states $\Psi^{(1)}$, $\Psi^{(2)}$ (cfr. Eq. \eqref{eigenstatsmearedalphaexpvalue} and \eqref{1to2alphaexpvalue}) but the corresponding smeared operators \eqref{fidvolumeoperator} have the same expectation value over such states.

Now, unlike it was the case for the classical theory, where the $V_o$-regulator was removed by sending it to infinity after Poisson brackets have been evaluated, which in turn amounts to $\text{vol}(V_o)=\alpha(V_o)\to\infty$ for
any finite initial conditions on $\alpha$, such a limit has no effect on the quantum theory
and does not enlarge the (expectation value of the) total volume of the region $V_o$ in a given
state (cfr.~Eq.~\eqref{eq:volexpvalue}). The quantum counterpart of a hypothetical $V_o\to\infty$ limit would then amount to pick a state $\Psi$ such that $\left<\widehat{\alpha(V_o)}\right>_{\Psi} \rightarrow \infty$, thus geometrically enlarging the homogeneity region.
However, if on the one hand it would in principle be possible for physically relevant observables to have reasonable dynamics, on the other hand all quantum fluctuations would be artificially suppressed in such a limiting case with the theory becoming effectively classical at all scales. To see this, let us consider the operator $\widehat{\text{vol}(V)}$ assigning a volume to an arbitrary subregion $V \subseteq V_o\subset\Sigma$. According to Eq. \eqref{fidvolumeoperator}, this is given by 
\begin{equation}\label{alphaValphaVo}
	\widehat{\text{vol}(V)} = \widehat{\alpha(V)} = V \hat{\alpha} = \frac{V}{V_o} \widehat{\alpha(V_o)}\;.
\end{equation}

\noindent
Note that, due to homogeneity, the ratio $V/V_o$ is purely topological and only counts how often the set $V$ fits into $V_o$ independently of any coordinates or metric.
We can then try to get rid off fiducial structures and send the physical volume of the fiducial cell to infinity.
For regularity, let us only make it sufficiently large which, as discussed above, amounts to choose a state $\Psi$ for which the corresponding expectation value of $\widehat{\alpha(V_o)}$ is large enough.
The expectation value
\begin{equation}\label{expvaluealphaValphaVo}
	\left<\widehat{\alpha(V)}\right>_\Psi = \frac{V}{V_o} \left<\widehat{\alpha(V_o)}\right>_\Psi \;,
\end{equation}

\noindent
of $\widehat{\alpha(V)}$ evaluated in a state where $\left<\widehat{\alpha(V_o)}\right>_\Psi$ is arbitrarily large can be however kept fixed and finite by making the ratio $V/V_o$ comparably small so that $\left<\widehat{\alpha(V)}\right>_\Psi$ could have a reasonable size, as e.g. the size of the universe today. The observable $\widehat{\alpha(V)}$ and the state $\Psi$ are therefore in principle well suited to describe the physics of our universe, at least on large scales. Homogeneity would however be then imposed on a volume much larger than today's universe.
Moreover, since $\widehat{\alpha(V)}$ and $\widehat{\alpha(V_o)}$ are only related by a constant factor (cfr. Eqs. \eqref{alphaValphaVo} and \eqref{expvaluealphaValphaVo}), their evolution is equivalent and, as expected from LQC (see e.g. \cite{BodendorferAnElementaryIntroduction, AshtekarLoopquantumcosmology:astatusreport} and references therein), there exists a minimal finite non-zero value corresponding to a bounce resolving the Big Bang singularity where curvature and matter energy density reach an upper bound. For any $V\subseteq V_o$, the energy density
\be
\rho_\Psi(\phi)=\frac{\left<\widehat{P_\phi(V_o)}\right>_\Psi^2}{\left<\widehat{\text{vol}(V_o)}\right>_\Psi^2}=\frac{\left<\widehat{P_\phi(V)}\right>_\Psi^2}{\left<\widehat{\text{vol}(V)}\right>_\Psi^2}=\frac{\left<\hat{H}_{true}\right>_\Psi^2}{\left<\widehat{\alpha(V)}\right>_\Psi^2}\;,
\ee

\noindent
is in fact independent of the $V_o$-representation as both quantities at the numerator and denominator are extensive (cfr. Eqs. \eqref{homogfullsmearedalpha},\eqref{homogfullsmearedPphi},\eqref{phiclockHtrue}, and \eqref{alphaValphaVo}) thus yielding an intensive ratio. In particular, this holds true for $V=V_o$ due to the relations \eqref{eq:expvalueHtrue12} and \eqref{eq:volexpvalue}. Moreover, as the isomorphism \eqref{eq:isomorphygrav} preserves dynamics, the energy density upper bound $\frac{1}{2\lambda^2}$ at the bounce \cite{BodendorferAnElementaryIntroduction, AshtekarLoopquantumcosmology:astatusreport} is also independent of $V_o$. However, looking now at the uncertainty relations for the volume operator smeared over the subregion $V\subseteq V_o$\footnote{We notice that Eq. \eqref{eq:subcellUR} extends previous results in the literature to the case in which one considers a sub-region $V$ of the fiducial cell $V_o$, see Eqs.~(20), (34) in \cite{RovelliWhyAreThe}, which are then consistently recovered for $V =V_o$.}
\begin{equation}\label{eq:subcellUR}
	\Delta_\Psi \widehat{\alpha(V)} \Delta_\Psi \frac{\widehat{\sin\left(\lambda P_\alpha\right)}}{\lambda} \ge \frac{V}{2\,V_o} \left|\widehat{\cos\left(\lambda P_\alpha\right)}\right| \;,
\end{equation}

\noindent
we see that, starting with a state $\Psi$ for which $\left<\widehat{\alpha(V_o)}\right>_\Psi$
is arbitrarily large and looking then at a small subcell $V$ of $V_o$, the quantum fluctuations of the latter are suppressed by the ratio $V/V_o$. The lower bound for quantum fluctuations would in fact become arbitrary small by simply choosing a state $\Psi$, which is peaked on large fiducial volumes, i.e. a physical situation where the region on which homogeneity is imposed is much larger than the region $V$. This
is physically unreasonable especially if the subregion $V$ we are looking at is Planck sized ($V\subset V_o$ contains only a Planck volume $\left<\widehat{\alpha(V)}\right>_\Psi \sim \ell_p^3$.). In other words, looking at a single finite cell first and then patching multiple identical cells together homogeneously, fluctuations of the single finite cell are present and, as showed by the above argument based on the application of the mapping \eqref{eq:isomorphygrav} between states associated to different $V_o$-values, they are suppressed only at large scales when patching the cells together to form a larger cell as indicated by the ratio $V/V_o$ in \eqref{eq:subcellUR} interpreted as the inverse number of subcells $V$ homogeneously patched together into $V_o$. This is also compatible with the result of \cite{BodendorferCoarseGrainingAs, BodendorferRenormalisationwithsu11} according to which the fluctuations in a large
cell obtained by patching together $N$ subcells grow as $N$, so that the relative fluctuations are vanishing in the infinite cell limit.~Moreover, thinking of $V$ as the volume of Planckian subcells contained in the region of volume $V_o$, the fact that quantum fluctuations are suppressed on large scales where the ratio $V_o/V$ becomes large is compatible with recent investigations of the symmetries in gravitational minisuperspace models with two configuration degrees of freedom  \cite{BenAchourSchroedingerSymmetry}.~There, the central charge of the Schr\"odinger symmetry algebra of such systems turns out to be related precisely to such a ratio between the IR and the UV cut-offs which, in analogy with the hydrodynamic description of quantum many-body systems sharing the same symmetry, was interpreted as the average number of microscopic constituents thus setting the scale for how classical or quantum is the system.

The importance of uncertainty relations for finite small cells has been also emphasised e.g.~in \cite{BojowaldCriticalEvaluationof} (and references therein) where it was argued that, following the collapse of an initially large-scale homogeneous universe, structure forms within the co-moving volume of constant coordinate size $V_o$ so that, when inhomogeneities become appreciable, a smaller region should be selected for the collapse process to be tracked using a homogeneous model, thus progressively reducing the scale of homogeneity.

As a final remark, let us notice that Eq.~\eqref{eq:subcellUR} and the above surrounding discussion is also consistent with the full theory classical commutation relations \eqref{smearedfieldsPB1} which, as discussed in Sec. \ref{sec:setup}, become zero in the $\text{vol}(V)\subset\text{vol}(V') \rightarrow \infty$ limit.
Recall that $P_\alpha(V) = P_\alpha(V_o)$ after imposing homogeneity constraints, which is approximated by the $\sin$-operator on the polymer Hilbert space.
Similarly, we find that the volume spectrum becomes denser as the ratio $V/V_o$ gets smaller, namely (restricting again to
positive volumes)
\begin{equation}
	\left<\widehat{\alpha(V)} \right>_{\delta_{\alpha,\alpha_o}} = \frac{V}{V_o} \left<\widehat{\alpha(V_o)} \right>_{\delta_{\alpha,\alpha_o}} = V \frac{\eta^\gamma}{V_o^\gamma} \alpha_o = \frac{V}{V_o} \lambda n_o\;.
\end{equation}

\noindent
To sum up, the above analysis suggests that the fiducial cell is not only an auxiliary construction intoduced merely to regularise the otherwise divergent spatially non-compact integrals of the classical homogeneous theory, but rather has physical meaning as it provides us with the scale on which homogeneity is imposed, which is a physical requirement.
This region cannot be made arbitrarily large, but rather has to be adapted to the physical circumstances.
 Already at the classical level, even though the dynamics of classical observables is not affected in the $V_o\to\infty$ limit, sending the volume of the homogeneity region to infinity would spoil the canonical structure of the classical theory (cfr. Eqs. \eqref{gravityDBxspace}, \eqref{homogsmearedDB1}, and \eqref{eq:GravDB}). At the quantum level where, as discussed in Sec.~\ref{Sec:Quantumrepresentation}, different quantum representations and Hilbert spaces are identified by the different values of $V_o$, a na\"ive limit which ignores such a $V_o$-dependence of the quantum representation would then spoil the (off-shell) starting point for the canonical commutation relations. This in turn reflects into the fact that, choosing a state $\Psi$ for which the expectation value of the volume assigned to the fiducial cell $V_o$ becomes arbitrarily large, all the proper quantum features of the system as e.g. quantum fluctuations of observables associated to any finite small region $V\subset V_o$ would be suppressed, thus artificially making the quantum description of $V$ effectively classical to an arbitrary precision. In such a limit, a theory of quantum cosmology beyond the effective approximation would not be needed as an effective theory would be sufficient even in the deep quantum gravity regime. As mentioned in the scalar field case (cfr. Sec.~\ref{sec:phiquantisation}), this is due to the fact that a quantum theory is non-local and knows about the correlations in the full region $V_o$.
Having a $V$ which is much smaller than $V_o$, all correlations are negligible w.r.t. those of $V_o$. This however does not mean that these do not play any significant role into the description of the system at all scales, but only that the homogeneous theory provides us with a good approximation on large scales where several smaller cells $V$ are patched together into the macroscopic region $V_o$ over which homogeneity is imposed and the theory becomes effectively classical as homogeneity is imposed over larger scales and the region $V_o$ grows (hence the ratio $V/V_o$ decreases). 
It is thus important in the context of quantum cosmology to carefully evaluate how the state has to be chosen and for which volume homogeneity is imposed.
As long as no experiments for quantum cosmology are available, this will always be a choice allowing more or less quantum correlations. A physically reasonable choice seems to be to use a state $\Psi$ such that $\left<\widehat{\alpha(V_o)}\right>_\Psi \sim$ \textit{volume of the universe today} as the physical state describing our universe. This is then a scale large enough that the homogeneity assumption is valid, but not larger than the scale on which the universe can be observed.

Furthermore, it is important to keep in mind that, as we have discussed in Sec.~\ref{Sec:gravityDBtruncatedtheory}, the construction of the spatially homogeneous and isotropic mini-superspace model relies also on a field modes truncation, the contributions coming from the $\vec k\neq\mathbf 0$ modes being suppressed with $V_o$ (cfr.~Eq.~\eqref{alphamomprofile} and the surrounding discussion at the end of Sec.~\ref{Sec:gravityDBtruncatedtheory}).~It is then important to also keep track of the small momenta, which we neglected here too for the time being to put the emphasis on the role of the fiducial cell regulator already for the description based on the dominant homogeneous $\vec k=\mathbf0$ modes, and estimate whether their contributions have physical significance especially at the small scales close to the bounce.~A preliminary discussion of how one could go beyond homogeneity within the setting developed in this paper is presented in the next section.~A systematic study of the back-reaction of inhomogeneities as~e.g.~in~\cite{Brahma:2021mng} would be however crucial to quantitatively address such questions.~A detailed analysis in the present setting is left for future work.

\section{Going Beyond Homogeneity in Cosmology}\label{outlook:beyondhom}

To move beyond the homogeneous approximation, we need to relax the symmetry assumptions underlying homogeneous theories.~This section will outline how the classical reduction procedure developed so far enables the systematic inclusion of inhomogeneous modes non-perturbatively, at least on larger scales.~Our approach thus complements cosmological perturbation theory, which assumes small corrections to homogeneity while allowing for arbitrarily small inhomogeneities.

As it was the case in Sec.~\ref{sec:setup}, our starting point for the mode expansion is to consider the ADM form of the metric \eqref{eq:ADMmetric} with spatial metric $q_{ab} =a^2(t,x) \delta_{a b}$, which is locally isotropic in the tangent space at each point, but admits neither rotational nor translational globally defined Killing vector fields and is thus neither homogeneous nor isotropic globally.
The expansion in modes allows now to systematically go order by order from the above generic metric ansatz to a homogeneous and isotropic metric as used in cosmology by simply cutting off the modes. Specifically, as shown in Sec.~\ref{Sec:Comsology}, keeping only the zero mode leads to a piecewise FLRW metric. Including then the higher momentum modes in the expansion amounts to take inhomogeneities into account. In particular, the inclusion of the first $\vec k\neq\mathbf 0$ modes would lead to a non-homogeneous metric which is only ``slowly varying in space'' as now small momenta, i.e. large wavelengths, are included on top of the homogeneous $\vec k=\mathbf 0$ mode.~Translated into equations this means that the mode expansion of our dynamical fields, say $\alpha(x)$ for concreteness, now reads as (cfr.~Eqs.~\eqref{fieldpartitioning}, \eqref{alphamodedecomp})
\begin{equation}
	\alpha(x) = \sum_n\chi_{V_n}(x) \alpha^{n}(x) \qquad , \qquad \alpha^{n}(x) = \sqrt{\fidmetric{}{}}\sum_{\vec{k}\in\left\{-1,0,1\right\}^3} \overline{\alpha}^{n}_{\vec{k}}\,\frac{e^{-i\frac{2\pi}{L}\vec k \cdot \vec x}}{\beta L^3}\label{firstinomalphaexpansion}
\end{equation}
from which we see that $\alpha(x)$ keeps its $x$-dependence, but the latter originates only from large wavelengths as only the first $\vec k\neq\mathbf 0$ modes are now included together with the $\vec k=\mathbf 0$ homogeneous mode.\footnote{As already noticed in Sec.~\ref{Sec:gravityDBtruncatedtheory}, the value of $\vec{k} \in \left\{-1,0,1\right\}^3$ is measured w.r.t. the fiducial metric $\fidmetric{a}{b}$ which is an auxiliary construction.~The wavelength in each spatial direction, along which $\alpha(x)$ is $L$-periodic, should be thus translated into a physical length scale measured w.r.t.~$q_{a b}$ as in~Eq.~\eqref{eq:physwavelength}.~However, beyond the fully spatially homogeneous setting where such a physical length scale significantly simplifies (cfr.~Eq.~\eqref{eq:lambdaphyshomog}), this is in general a complicated and non-local function of the metric which can only be computed after the equations of motion for $\alpha(x)$ are solved.~Moreover, the result of \eqref{eq:physwavelength} is in general inhomogeneous,~i.e.~the corresponding wavelength in each direction, say e.g.~$\lambda_{\vec{k}=(1,0,0)}(y,z)$ in $x$-direction, depends on the remaining two spatial directions.~Note also that the presence of the third-root in the expression \eqref{eq:physwavelength} prevents us from simply decomposing the integral therein in a sum over Fourier modes.~The explicit computation leading to a closed-form expression of the physical scale of the allowed inhomogeneities is therefore hard to be carried out in full generality.} The corresponding wave-vectors have three different norms, i.e.
\begin{equation}\label{eq:normwavevectors}
	\|(1,0,0)\| = 1,\dots \quad , \quad \|(1,1,0)\| = \sqrt{2} ,\dots \quad , \quad \|(1,1,1)\| = \sqrt{3},\dots \;,
\end{equation}
where the dots refer to permutations of the components and the possible $\pm$ sign of each of the latter. Thus, the inhomogeneous modes with smallest wave vector (largest wavelength) are $(\pm 1,0,0)$, $(0,\pm 1,0)$ and $(0,0,\pm 1)$.
We could then argue to include only these modes into our discussion. In what follow, nevertheless, we keep the diagonal modes as well.~We can ask then how many degrees of freedom per box $V_n$ are allowed once the above inhomogeneous modes are included.
The set $\left\{-1,0,1\right\}^3$ has $3^3 = 27$ elements, which correspond to $27$ complex coefficients $\overline{\alpha}_{\vec{k}}^{n} \in \mathbb{C}$ per box.
As the volume form $\alpha(x)$ has to be real, the expansion coefficients have to satisfy the reality conditions $(\overline{\alpha}_{\vec{k}}^{n})^{*}=\overline{\alpha}_{-\vec{k}}^{n}$, from which it follows that the zero mode $\overline{\alpha}_{\mathbf0}^{n} \in \mathbb{R}$ is automatically real and the remaining $26$ complex coefficients are related by complex conjugation and momentum inversion.
This halves the number of degrees of freedom for the $\vec k\neq\mathbf0$ coefficients, resulting into $13$ remaining complex ones.
In total, there are therefore $1+(2\cdot13) = 27$ real degrees of freedom per box $V_n$.
This matches with the expansion in $\cos$ and $\sin$ with real coefficients with the Fourier series containing only a sum over the momenta $\vec{k} \in \left\{-1,0,1\right\}^3 / \mathbb{Z}_2$\,.~Finally, note that all $\alpha^{n}(x)$ are $L$-periodic by construction.
As argued in footnote~\ref{ftnte:aeequality}, in the limit $\vec{k} \rightarrow \infty$, we can generate discontinuous jumps to satisfy periodic boundary conditions.~As here the summation runs only over finitely many momenta (in fact only the first smallest non-zero modes), this is always continuous and really $L$-periodic.
Therefore, it is not possible to match the values across the boxes to make the field $\alpha(x)$ continuous.
This would only work if all modes are included and the infinite sum is performed.
Demanding continuity across the boxes would then simply lead to exact copies of one box in all other boxes similarly to the set \textsf{2)} of constraints~\eqref{homogconstraintsmodesalpha}-\eqref{homogconstraintsmodesPphi} in Sec.~\ref{Sec:gravityDBtruncatedtheory}.
As discussed there, this can only be applied in a finite number of boxes at least for the zero modes.
Therefore, even if we arrive at a non-trivial Dirac bracket for all higher modes after implementing these constraints, there would not be continuity as the zero modes are still discontinuous.

With the above premises in mind, we can then introduce the new set of second-class reduction constraints
\begin{align}\label{eq:inhomconstr}
		\Psi_{\vec k}^{\alpha, n}:=\overline{\alpha}^{n}_{\vec k}\approx0\qquad&,\qquad\Psi_{\vec k}^{P_\alpha,n}:=P_{\alpha,\vec k}^{n}\approx0\nonumber
		\\
		\Psi_{\vec k}^{\phi,n}:=\phi_{\vec k}^{n}\approx0\qquad&,\qquad\Psi_{\vec k}^{P_\phi,n}:=\overline{P}_{\phi,\vec k}^{n}\approx0\qquad\qquad\vec k\notin\left\{-1,0,1\right\}^3
	\end{align}
where we applied the same expansion discussed for $\alpha(x)$ also to all other dynamical fields (cfr.~Sec.~\ref{Sec:Comsology}).
According to the above discussion, here we shall not consider the second type of constraints \textsf{2)} in Eqs.~\eqref{homogconstraintsmodesalpha}-\eqref{homogconstraintsmodesPphi} as it does not add anything new to the considerations that will follow.
The construction of the Dirac bracket works exactly as in the homogeneous case discussed in Sec.~\ref{Sec:gravityDBtruncatedtheory}, thus yielding (cfr.~Eqs.~\eqref{PBconstraints1)}-\eqref{modesDB1)})
\begin{equation}
	\Poisson{\overline{\alpha}^{n}_{\vec{k}}}{P_{\alpha, \vec{p}}^{m}}_D = \begin{cases}
		\frac{\delta_{nm} \delta_{\vec{k},\vec{p}}}{\beta L^3} &,\quad \vec k,\,\vec{p} \in \left\{-1,0,1\right\}^3
			\\
			0 &,\quad \text{else}
	\end{cases}
\end{equation}
and similarly for the matter sector, which allows to impose the constraints~\eqref{eq:inhomconstr} strongly in the theory.

\begin{figure}[t!]
	\centering
	\includegraphics[height=5.75cm]{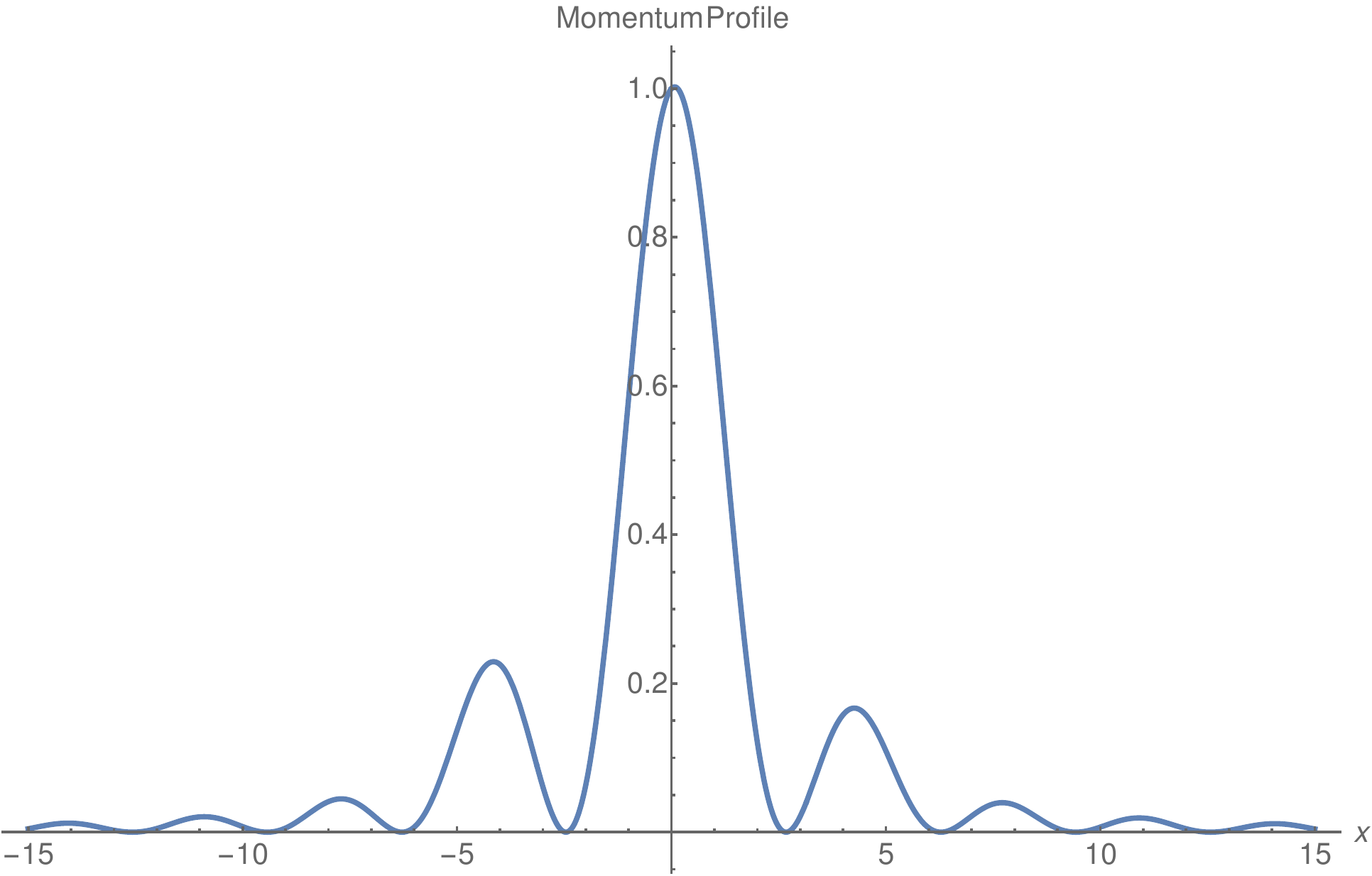}
	\caption{Plot of $\left(\sin(x)/x+0.2\cdot \sin(x)/(x-\pi)+0.3\cdot \sin(x)/(x+\pi)\right)^2$, which is related to $\left|\tilde{\alpha}(k)\right|^2/\left|\tilde{\alpha}(0)\right|^2$ in Eq.~\eqref{eq:momprofileinhom} for $x = L k_\xi/2$. As discussed in the main text, on top of the fully homogeneous zero-mode contribution, the modes $k_\xi\in \left(2\pi/L\right) \mathbb{Z}$ are exactly removed up to $k_\xi=2\pi p_\xi/L$ with $\vec p\in\{-1,0,1\}^3\setminus\{\mathbf0\}$, larger momenta are largely suppressed, but small momenta are still present.
	The functions $f_{\vec{p}}(\vec{k})$ are here approximated as constants.}
	\label{fig:sinx2}
\end{figure}

As in the homogeneous case, we can now ask about the error made in the mode truncation \eqref{firstinomalphaexpansion} by looking at the momentum profile of a continuous Fourier transform.
A straight forward computation yields
\begin{align}
	\tilde{\alpha}(k) :&= \frac{1}{\sqrt{2\pi}^3} \int_{-\infty}^\infty \dd^3 x \, \alpha(x) e^{-i \vec k \cdot \vec x}\\
	&= \frac{1}{\sqrt{2\pi}^3} \int_{-\infty}^\infty \dd^3 x \, \sqrt{\fidmetric{}{}} e^{-i \vec k \cdot \vec x}\left(\sum_{\vec{n} \in \mathbb{Z}^3} \sum_{\vec p \in \left\{-1,0,1\right\}^3} \chi_{V_{\vec{n}}}(x)\,\overline{\alpha}_{\vec{p}}^{\vec{n}}\, \frac{e^{-i\frac{2 \pi}{L} \vec x \cdot \vec p}}{\beta L^3}\right)
	\notag
	\\
	&= \frac{1}{(\sqrt{2\pi}\,L)^3} \sum_{\vec{n} \in \mathbb{Z}^3}\sum_{\vec p \in \left\{-1,0,1\right\}^3} \overline{\alpha}_{\vec{p}}^{\vec{n}} \int_{V_{\vec{n}}} \dd^3 x \, e^{-i \left(\vec k +\frac{2 \pi}{L}\vec p\right) \cdot \vec{x}}
	\notag
	\\
	&= \frac{1}{(\sqrt{2\pi}\,L)^3} \sum_{\vec{n} \in \mathbb{Z}^3}\sum_{\vec p \in \left\{-1,0,1\right\}^3} \overline{\alpha}_{\vec{p}}^{\vec{n}} \prod_{\xi = {x,y,z}}\int_{L n_\xi}^{L(n_\xi+1)} \dd \xi \, e^{-i \left(k_\xi +\frac{2 \pi}{L}p_\xi\right) \cdot \xi}
	\notag
	\\
	&= \frac{1}{\sqrt{2\pi}^3} \sum_{\vec p \in \left\{-1,0,1\right\}^3} \sum_{\vec{n} \in \mathbb{Z}^3} \overline{\alpha}_{\vec{p}}^{\vec{n}} e^{-iL \vec{k} \cdot \vec n}\prod_{\xi = {x,y,z}} e^{-i \frac{Lk_\xi}{2}} \cdot \frac{\sin\left(\frac{Lk_\xi}{2}\right)}{\left(\frac{Lk_\xi}{2}+\pi p_\xi\right)}
	\notag
	\\
	&= \frac{1}{\sqrt{2\pi}^3}\left(f_{\mathbf{0}}(\vec{k})\prod_{\xi = {x,y,z}} e^{-i \frac{Lk_\xi}{2}} \cdot \frac{\sin\left(\frac{Lk_\xi}{2}\right)}{\left(\frac{Lk_\xi}{2}\right)} + \sum_{\vec p \in \left\{-1,0,1\right\}^3\setminus \left\{\mathbf{0}\right\}} f_{\vec{p}}(\vec{k})\prod_{\xi = {x,y,z}} e^{-i \frac{Lk_\xi}{2}} \cdot \frac{\sin\left(\frac{Lk_\xi}{2}\right)}{\left(\frac{Lk_\xi}{2}+\pi p_\xi\right)}\right),\label{eq:momprofileinhom}
\end{align}
where in the last line $f_{\vec{p}}(\vec k):=\sum_{\vec{n} \in \mathbb{Z}^3} \overline{\alpha}_{\vec{p}}^{\vec{n}}\,e^{-iL \vec{k} \cdot \vec n}$ (cfr.~Eq.~\eqref{alphamomprofile}, Sec.~\ref{Sec:gravityDBtruncatedtheory}).~As it was the case also in Sec.~\ref{sec:truncation} for the scalar field and in Sec.~\ref{Sec:gravityDBtruncatedtheory} for homogeneous cosmology, the functions $f_{\vec{p}}(\vec{k})$ are $2\pi/L$-periodic in each $\vec{k}$-direction and can therefore be assumed to be bounded. The first term of the sum is of course the same contribution as in the full homogeneous case (cfr.~Eq.~\eqref{alphamomprofile}), while all the momenta $\vec{k} \in 2\pi\mathbb{Z}^3/L$, except for $\vec k =\mathbf0$, yield vanishing contributions to the momentum profile.
However, as already discussed in the homogeneous theory (cfr.~discussion below Eqs.~\eqref{scalrfieldmomprofile} and~\eqref{alphamomprofile}), there are other non-zero momenta contributing, with the large ones being suppressed by order of $\mathcal O\left(1/k_\xi L\right)$.~The same holds also for the sum in the second term of \eqref{eq:momprofileinhom}, where again integer multiples of $2\pi/L$ do not contribute, except for the momentum $\vec{k} =-2\pi \vec{p}/L$ as expected.~Moreover, the momenta $L \vec{k}/2\pi = \vec{p} \in \left\{-1,0,1\right\}^3$ are dominant in the second term of \eqref{eq:momprofileinhom}, while large momenta are suppressed.~An example of the above momentum profile is reported in Fig.~\ref{fig:sinx2}.~As it was the case also in the homogeneous theory, now with the inclusion of the first small momentum/large wavelength inhomogeneous modes, which lead to the addition of the second term in the momentum profile \eqref{eq:momprofileinhom}, there are surviving non-trivial contributions coming from wavelengths larger than the box size, now given by $L k_\xi/2 \pm \pi p_\xi \in \left[-1,1\right]$.~As already anticipated in previous sections, the relevance of these contributions for the dynamics of the truncated theory would require further insights from the full theory.~In particular, we notice that the explicit momentum profile strongly depends on the choice of functions $f_{\vec{p}}(\vec k)$, which in principle would be determined by the full theory dynamics.

Finally, let us compute the truncated Hamiltonian and vector constraints.
As we have seen previously for the real scalar field in Sec.~\ref{sec:truncation} and for cosmology in Sec.~\ref{Sec:gravityDBtruncatedtheory}, there will be boundary contributions of the form $\partial_a \chi_{V_n}(x)$, which encode the interactions between neighbouring cells.
As before, we will neglect all of them, which truncates the theory and ignores the above discussed modes, especially those with wavelength larger than a single box. The truncated Hamiltonian and vector constraints will thus only contain the modes $\vec k \in 2\pi/L \cdot \left\{-1,0,1\right\}^3$ in addition to the homogeneous $\vec k=\mathbf0$ mode.
Due to the fact that certain $x$-dependencies are now allowed, lapse and shift exhibit a similar $x$-dependence with same periodicity of the modes, i.e.
\begin{equation}
	N(x) = \sum_{n} \chi_{V_n}(x) \sum_{\vec{k} \in \left\{-1,0,1\right\}^3} N_{\vec{k}}^{n} e^{i \frac{2\pi}{L} \vec{k} \cdot \vec{x}} \quad , \quad N^a(x) = \sum_{n} \chi_{V_n}(x) \sum_{\vec{k} \in \left\{-1,0,1\right\}^3} N_{\vec{k}}^{n\,a} e^{i \frac{2\pi}{L} \vec{k} \cdot \vec{x}} \;.
\end{equation}
Inserting these expansions into the gravitational constraints Eqs.~\eqref{isotropicvectorconstraint}-\eqref{isotropichamconstraint} yields
	\begin{align}
		\mathcal{C}_a\left[N^a\right]\bigl|_{V_n} =&\, \beta L^3 \frac{2\pi i}{L}\sum_{\vec{k},\vec{p} \in \{-1,0,1\}^3} N^{n\,a}_{-\vec{k}-\vec{p}} \left(-2\overline{\alpha}_{\vec{k}}^{n} P_{\alpha,\vec{p}}^{(n)}\left(k_a+p_a\right) + p_a\phi_{\vec{p}}^{n} \overline{P}_{\phi,\vec{k}}^{n}\right) \;,\label{eq:vecconstinhom}
		\\
		\mathcal{H}\left[N\right]\bigl|_{V_n} =&\, \sum_{\vec{k},\vec{p},\vec{q} \in \{-1,0,1\}^3} \biggl(\frac{1}{2} N^{n}_{\vec{q}} \overline{P}_{\phi,\vec{k}}^{n} \overline{P}_{\phi,\vec{p}}^{n}\, U^{n}_\alpha\left(\vec{q}+\vec{k}+\vec{p};-1\right)
		\notag
		\\
		&\qquad\qquad\qquad\quad -\frac{4\pi^2}{L^2} N_{\vec{q}}^{n} \phi^{n}_{\vec{k}} \phi^{n}_{\vec{p}} \delta^{a b} k_a p_b U^{n}_\alpha\left(\vec{q}+\vec{k}+\vec{p};\frac{1}{3}\right)
		\notag
		\\
		&\qquad\qquad\qquad\quad+\frac{4\pi^2}{L^2} N^{n}_{\vec{q}} \overline{\alpha}^{n}_{\vec{k}} \overline{\alpha}^{n}_{\vec{p}}\sum_{a=1}^3 \left(k_a p_a -\frac{2}{3} p_a^2\right)U^{n}_{\alpha}\left(\vec{k}+\vec{p}+\vec{q};-\frac{5}{3}\right)
		\notag
		\\
		&\qquad\qquad\qquad\quad-\frac{3}{4} \beta L^3 \overline{\alpha}^{n}_{\vec{k}} P^{n}_{\alpha, \vec{p}} P^{n}_{\alpha, \vec{q}} N^{n}_{-\left(\vec{k}+\vec{p}+\vec{q}\right)}\biggr)\;,\label{eq:Hamconstinhom}
	\end{align}
where again the subscript $V_n$ on the l.h.s. denotes that we restrict to a single box and neglect all boundary contributions coming from derivatives of $\chi_{V_n}(x)$.
Let us also stress that we take care of modes $k \in \left\{-1,0,1\right\}^3$ only and therefore $N^{n}_{\vec{k}} = 0$ if $\vec{k} \notin \left\{-1,0,1\right\}^3$.
In the above expressions, the function $U^{n}_\alpha$ is defined as
\begin{equation}
	U^{n}_\alpha(\vec{k}; \gamma) = \beta \int_{V_n} \dd^3x\, \overline{\alpha}^{n}(x)^\gamma e^{i\frac{2\pi}{L} \vec{k} \cdot \vec{x}} \;,\label{eq:Ualphadef}
\end{equation}
and has no simple closed form explicit expression in terms of the modes unless $\gamma \in \mathbb{N}$.~This reflects the non-linear structure of the Hamiltonian under consideration, and in fact gravity in general, and leads to complicated (volume dependent) interaction terms.~All the terms containing these non-linear expressions become very simple or even vanish in the case where full homogeneity is imposed.~Note that for instance the last term in Eq. \eqref{eq:Hamconstinhom} is a result of $U_\alpha^{n}\left(\vec{k}+\vec{p}+\vec{q}+\vec{l};0\right) = \beta L^3 \delta_{\vec{k}+\vec{p}+\vec{q}+\vec{l},0}$, which is simple to compute.~In order to solve the resulting equations of motion and the corresponding constraints, it will be thus important to understand better these functions $U^{n}_\alpha$.~The best scenario would allow to analytically solve the integral, otherwise one would have to evaluate their relevance and their physical meaning numerically or by means of further simplifying arguments.~A simple observation in this respect is that the contributions $U^{n}_\alpha(\vec k; \gamma < 0)$ grow when $\alpha$ becomes small, i.e.~they are dominant in the high curvature/small volume regimes at early times, but negligible in the large volume regime at late times in the universe evolution.\footnote{Note that it is necessary to assume $\alpha(x) > 0$ and therefore negative $\gamma$ should lead to always finite results.}
Moreover, from the definition \eqref{eq:Ualphadef} it is clear that $U^{n}_\alpha(\vec k; \gamma)$ is essentially the discrete Fourier coefficient of $\alpha^\gamma$, i.e.
\be
U^{n}_\alpha(\vec{k}; \gamma) = \beta L^3 \mathcal{F}_{V_n}\left[\alpha^\gamma\right](\vec k) \;,
\ee
where $\mathcal{F}_{V_n}$ denotes the \textit{discrete} Fourier transform over the volume $V_n$.
As we only want to consider small Fourier modes, it would be consistent to cut off all contributions of $U^{n}_\alpha(\vec k; \gamma)$ with $\vec{k} \notin \left\{-1,0,1\right\}^3$.
Further, for either $\gamma = 0$ or $\alpha$ spatially constant, we have 
\begin{equation}
	U^{n}_\alpha(\vec{k}; \gamma)= \beta L^3 \alpha^\gamma \delta_{\vec{k},0} \;,
\end{equation}
which could be interpreted as a vertex that preserves the momentum.~Considering e.g.~the first term in Eq.~\eqref{eq:Hamconstinhom}, this could be interpreted as an interaction between $N^{n}_{\vec{q}}$ $\overline{P}^{n}_{\phi,\vec{k}}$ and $\overline{P}^{n}_{\phi,\vec{p}}$ with momentum conservation coming from $U^{n}_{\alpha = const.} \propto \delta_{\vec{q} + \vec{k} + \vec{p},0}$.~For non-constant $\alpha$, we might therefore interpret the functions $U^{n}_{\alpha}$ as dynamical vertices, which can absorb or release momentum into the system.

As a last point concerning the analysis of the functions $U^{n}_{\alpha}$, we can further restrict the momenta to only the modes $(\pm1,0,0)$, $(0,\pm1,0)$ and $(0,0,\pm1)$, which have the smallest wave-vector (cfr.~Eq.~\eqref{eq:normwavevectors}).~In this case, Eq. \eqref{firstinomalphaexpansion} reduces to
\be
	\alpha^n(x) = \alpha_\mathbf0 + A_{(1,0,0)} \cos\left(\frac{2\pi}{L} x + \Phi_{(1,0,0)}\right) + A_{(0,1,0)} \cos\left(\frac{2\pi}{L} x + \Phi_{(0,1,0)}\right) + A_{(0,0,1)} \cos\left(\frac{2\pi}{L} x + \Phi_{(0,0,1)}\right),
\ee
with $A_{(1,0,0)}=(\alpha^{n}_{(1,0,0)}-\alpha^{n}_{(-1,0,0)})/2=\text{Re}(\alpha^{n}_{(1,0,0)})$ due to the reality conditions $(\overline{\alpha}_{\vec{k}}^{n})^{*}=\overline{\alpha}_{-\vec{k}}^{n}$, and similarly for $A_{(0,1,0)}, A_{(0,0,1)}$, so that all spatial dependencies are linearly decoupled.~Demanding $\alpha(x)\geq 0$ then simply amounts to
\begin{equation}
	\alpha_\mathbf0 \geq A_{(1,0,0)} + A_{(0,1,0)} + A_{(0,0,1)} \;,
\end{equation}
which tells us that, at the Big Bang where $\alpha_{\mathbf0} \rightarrow 0$, also the inhomogeneity coefficients have to vanish in order to satisfy the above condition.~Using this truncation of the momenta, the expression \eqref{eq:Ualphadef} for $U^{n}_\alpha $ can be further re-written as 
\begin{align}
	U^{n}_\alpha(\vec{k}; \gamma) =&\, \beta \int_{x_0}^{x_0+L} \dd x\, e^{i\frac{2\pi}{L} k_x x} \int_{y_0}^{y_0+L} \dd y\, e^{i\frac{2\pi}{L} k_y y} \int_{z_0}^{z_0+L} \dd z\, e^{i\frac{2\pi}{L} k_z z}\Bigl(\overline{\alpha}_\mathbf 0+ \overline{\alpha}_{(1,0,0)} e^{i\frac{2\pi}{L} x}
	\notag
	\\
	&\quad \quad + \overline{\alpha}_{(-1,0,0)} e^{-i\frac{2\pi}{L} x}+ \overline{\alpha}_{(0,1,0)} e^{i\frac{2\pi}{L} y} + \overline{\alpha}_{(0,-1,0)} e^{-i\frac{2\pi}{L} y}+ \overline{\alpha}_{(0,0,1)} e^{i\frac{2\pi}{L} z} + \overline{\alpha}_{(0,0,-1)} e^{-i\frac{2\pi}{L} z}\Bigr)^\gamma
	\notag
	\\
	=& \frac{\beta L^3}{\left(2\pi i\right)^3 k_x\cdot k_y \cdot k_y} \int_{\partial B_1(0)} \dd s_x \int_{\partial B_1(0)} \dd s_y \int_{\partial B_1(0)} \dd s_z\Bigl(\overline{\alpha}_\mathbf 0+ \overline{\alpha}_{(1,0,0)} s_x^{\frac{1}{k_x}}
	\notag
	\\
	&\quad \quad + \overline{\alpha}_{(-1,0,0)} s_x^{-\frac{1}{k_x}}+ \overline{\alpha}_{(0,1,0)} s_y^{\frac{1}{k_y}} + \overline{\alpha}_{(0,-1,0)} s_y^{\frac{1}{k_y}}\overline{\alpha}_{(0,0,1)} s_z^{\frac{1}{k_z}}  + \overline{\alpha}_{(0,0,-1)} s_z^{-\frac{1}{k_z}}\Bigr)^\gamma \;,
\end{align}
thus involving a complex contour integral over a complex unit circle $\partial B_1(0)$ in each direction.~The latter integrals could be solved by using the methods of complex analysis.~Note that the rewriting in the second equality of the above equation is only possible for $k_x,k_y,k_z \neq 0$.~Further simplifications are then possible when $U^{n}_\alpha$ is restricted to the individual modes $(\pm1,0,0)$, $(0,\pm1,0)$ and $(0,0,\pm1)$.~In this case, indeed, only one of the complex contour integrals would remain, while the others are still real, as e.g.~for the $\vec k=(0,0,1)$ case reported below
\begin{align}
	U^{n}_\alpha(\left(0,0,1); \gamma\right) =&\, \beta \int_{x_0}^{x_0+L} \dd x \int_{y_0}^{y_0+L} \dd y \int_{\partial B_1(0)} \frac{L \dd s_z}{2\pi i}\left(\overline{\alpha}_0+ \overline{\alpha}_{(1,0,0)} e^{i\frac{2\pi}{L} x} \right.\notag\\
	&\quad\left.+ \overline{\alpha}_{(-1,0,0)} e^{-i\frac{2\pi}{L} x}+ \overline{\alpha}_{(0,1,0)} e^{i\frac{2\pi}{L} y}+\overline{\alpha}_{(0,-1,0)} e^{-i\frac{2\pi}{L} y}+ \overline{\alpha}_{(0,0,1)} s_z + \overline{\alpha}_{(0,0,-1)} \frac{1}{s_z}\right)^\gamma.
\end{align}
To sum up, as we have seen in the previous sections, it is possible to solve the gravitational system in the homogeneous setting, where a large amount of symmetries is available.~Using then the mode expansion of the dynamical fields and the implementation of symmetry-reduction via second-class constraints on the mode coefficients, we can make contact with the full theory and loosen these symmetry assumptions by allowing large scale inhomogeneities, i.e.~those which change only ``slowly'' in space.~This corresponds to truncate the discrete Fourier series of the fields at the lowest non zero momenta, but such a truncation could in principle be done also to higher modes.~This would in principle provide us with a complementary approach to perturbation theory, where all modes are involved, but their amplitudes always remain small.~In the present setting, instead, only modes with small momenta are included, but their amplitudes can be non-perturbatively large.~However, this leads to an expression for the Hamiltonian, which is rather complicated as it involves the functions $U_\alpha^{n}\vec k; \gamma)$ defined in \eqref{eq:Ualphadef}, which are in general hard to evaluate thus requiring us to find suitable approximations or truncations in which these functions can actually be explicitly evaluated.~The latter would depend on the physical situation one is aiming to study.~As an example here we considered the situation in which only the six smallest non-zero modes $(\pm1,0,0)$, $(0,\pm1,0)$ and $(0,0,\pm1)$ were included on top of the homogeneous $\vec k=\mathbf 0$ mode.~A further detailed analysis of the procedure presented here would then be needed to better understand eventually the advantages compared to other methods.

One straight forward application is to compute the time evolution of the homogeneous mode of $\alpha$ with the total Hamiltonian $H[N,N^a] = \mathcal{H}\left[N\right] + \mathcal C_a \left[N^a\right]$ given in Eqs.~\eqref{eq:vecconstinhom} and \eqref{eq:Hamconstinhom}.~This gives (all $n$ superscripts are dropped to ease the notation)
\begin{equation}
	\dot{\overline{\alpha}}_\mathbf0 = - \frac{3}{2} N_\mathbf0 \overline{\alpha}_\mathbf0 P_{\alpha,\mathbf0} - \frac{3}{2} \sum_{\vec{k},\vec{p} \in \left\{-1,0,1\right\}^3\setminus\left\{\mathbf0\right\}} \overline{\alpha}_{\vec k} N_{-\vec k - \vec p} P_{\alpha, \vec p} - \frac{4 \pi i}{L} \sum_{\vec{k} \in \left\{-1,0,1\right\}^3 \setminus \left\{\mathbf0\right\}} N^a_{-\vec k} k_a \overline{\alpha}_{\vec k} \;.\label{alpha0EOMinhomog}
\end{equation}
The first term is the homogeneous contribution, which allowed us to interpret $P_{\alpha, 0}$ as related to the Hubble rate (cfr.~Eq.~\eqref{alphaEoM} and its surrounding discussion).~However, as we can see from the r.h.s.~of Eq.~\eqref{alpha0EOMinhomog}, there are higher momentum contributions also to the dynamics of the homogeneous mode of $\alpha$, which we recall is defined as the total volume of the box $V_n$ and is the only configuration degree of freedom for the spatially homogeneous gravitational sector.~Thus, the higher mode terms, which are usually neglected in the symmetry-reduced truncated theory of fully homogeneous cosmology, do contribute to the dynamics of the volume of universe.\footnote{Note that $\text{vol}(V_n) = \beta L^3 \overline{\alpha}^{n}_\mathbf0$ and all higher momentum terms drop out due to the integration.} In order to really judge how relevant these additional contributions are and how they quantitatively affect the validity of the homogeneous description over different scales, it would be necessary to compute all equations of motion and possibly even solve them.~Such a detailed analysis is postponed to future work. 

\section{Summary and Discussion}\label{sec:conclusion}

Motivated by the debate on the interpretation of the fiducial cell introduced to regularise the otherwise divergent integrals in spatially non-compact homogeneous systems, in this paper we presented a systematic procedure for the homogeneous reduction of a classical field theory within the canonical Hamiltonian framework.~We applied it to both a massive scalar field theory and general relativity minimally coupled to a massless real scalar field clock.~This allowed us to explicitly analyse the role the fiducial cell at the various stages of the homogeneous reduction as well as the approximations and truncations involved therein.

Our starting point was to implement spatial homogeneity via the imposition of suitable second-class constraints for the relevant smeared full theory observables defined by averaging the canonical fields over a given spatial region.~The symmetry reduced theory was then obtained by constructing the associated Dirac bracket and solving the constraints strongly in the sense of Dirac's theory of constraints.~A decomposition into continuous Fourier modes of the fields however results into a trivial Dirac bracket.~We thus partitioned the spatial slice into the disjoint union of equal boxes and performed a decomposition into \textit{discrete} Fourier modes in each box individually.~This allowed us to translate the homogeneity constraints into two sets of mutually commuting second-class constraints.~The first kind of constraints forces all modes with non-zero wave-number to vanish, while the second kind sets the zero-modes to be equal across the different boxes.~Importantly, the associated non-trivial Dirac bracket can only be constructed on a finite number $d$ of boxes and becomes singular for $d\to\infty$.~Spatial homogeneity can therefore only be imposed on the finite region resulting from patching together finitely many boxes.~This is precisely the \textit{fiducial cell} $V_o$ whose sizes acquires an explicit physical meaning as the scale over which homogeneity is imposed.~The homogeneous symmetry-reduced theory is then nothing but a theory of such a finite region.

The above procedure further allowed us to highlight what kind of information is neglected in the homogeneous reduction.~The imposition of the homogeneity constraints in fact amounts to neglect all the inhomogeneous $\vec k\neq\mathbf0$ modes with \emph{both} wavelength larger (or equal) and smaller than the fiducial cell size.~Moreover, the fact that the physics encoded into the spatially homogeneous $\vec k=\mathbf 0$ mode within each box is replicated into all the other boxes forming the fiducial cell amounts to neglect boundary terms in the homogeneoeus truncated Hamiltonian.~These boundary terms encode the interactions between neighbouring cells and throwing them away, as typically done in homogeneous minisuerspace models, implicitly assumes that suitable boundary conditions have been imposed.~The description in terms of the homogeneous modes only is then an approximation and the error made depends on the fiducual cell size.~In fact, on top of the dominant homogeneous mode contribution, the momentum profile distribution features contributions from the non-zero modes.~These are suppressed by order of $\mathcal O(1/k_{\xi}L)$ with $L=V_o^{1/3}$.\footnote{In the gravitational setting, as discussed in Sec.~\ref{Sec:gravityDBtruncatedtheory}, this can be translated into a physical length scale at least for the homogeneous case.} The truncation is then not too aggressive for modes with wavelength smaller than the fiducial cell size ($k_{\xi}L\gg1$) but might become rather drastic for sufficiently large wavelengths compared to the cell size ($k_{\xi}L\ll1$).~To judge how large the resulting error is from a dynamical standpoint requires us to solve the dynamics of the non-symmetry-reduced observables.~This can be done explicitly for the case of a free massive scalar field leading to the result that the small momenta/large wavelength modes can be safely neglected when either the mass of the scalar field or the cell size is large, that is $mL \gg 1$.~The same condition served to identify a sector of the full quantum scalar field theory
which is equivalent to the homogeneous theory resulting from classically symmetry reducing first and then quantising.~In particular, the full theory and homogeneous dynamics coincide up to the flux through the boundary which is negligible for sufficiently large $V_o$ as the physics localised in a neighborhood of the boundary is negligible against the bulk physics.~In the gravity case, the non-linear coupled structure of the equations of motions prevents us from a direct comparison of the full theory and symmetry-reduced dynamics.~Nevertheless, our framework naturally lends itself to go beyond the homogeneous setting and the inclusion of some of the inhomogeneous modes indicates that these do contribute to the dynamics.

All in all, the off-shell structures of the classically reduced spatially homogeneous theory such as the truncated Hamiltonian and the Dirac bracket depend on $V_o$ and become singular when the latter is made infinitely large.~As a result, an active rescaling of $V_o$ is not a canonical transformation and one should rather think of the reduced theory as an entire family of canonically inequivalent theories, each corresponding to a different scale over which spatial homogeneity has been imposed.~The scaling behaviour of the Hamiltonian and of the Dirac bracket compensate each another so that the dynamics of classical observables is independent of the fiducial cell.~Thus, after evaluating the Dirac brackets, it is possible to send the fiducial cell size to infinity and still result in sensible physics, at least on large scales where the homogeneous description can be safely trusted.~This is consistent with the on-shell point of view at the full theory level where one could insert a fully homogeneous ansatz directly into the equations of motion.~The physical on-shell results of a \emph{local} classical theory do not depend on the region over which the dynamical fields are approximated by their spatially constant homogeneous modes.

The $V_o$-dependence of the classical symmetry-reduced off-shell structures has important consequences for the quantum theory.~In fact, upon quantization, each of the canonically inequivalent classical theories regularised on a different $V_o$ results into a different quantum representation in terms of operators on a Hilbert space and canonical commutation relations labelled by $V_o$.~As a direct consequence, quantum operators scale differently than their classical counterparts under fiducial cell rescaling.~The $V_o$-rescaling symmetry of classical dynamics may thus be broken at the quantum level.~To investigate this point we constructed an explicit mapping between states in the Hilbert spaces associated to different values of $V_o$ with the same dynamics.~This provided us with an isomorphism between the different $V_o$-labelled homogeneous quantum theories and implements the quantum analogue of an active $V_o$-rescaling.~Dynamically equivalent quantum states have non-trivial transformation behaviour under change of $V_o$ and so do non-local quantum features such as the statistical moments and quantum fluctuations of the relevant quantum operators.~In particular, a state saturating the uncertainty relations in one $V_o$ theory does so also in the other theories as long as the classical value on which it is
peaked and its width scale non-trivially with $V_o$.

An important difference concerns whether the observable under cosideration is smeared over the whole $V_o$ or a subregion $V\subset V_o$:~in the former case, expectation values and quantum fluctuations do not depend on $V_o$;~while, in the latter case, they scale as the (inverse) number $V/V_o$ of regions $V$ homogeneously patched together in $V_o$.~This is of particular relevance for quantum cosmology where the operator of interest is the spatial volume $\hat{\alpha}(V)$ of the region under consideration.~The physical size of the homogeneity region, as given by $\langle\hat{\alpha}(V_o)\rangle_{\Psi}$, is thus determined quantum dynamically by the operator $\hat{\alpha}(V_o)$ and by the quantum state $\Psi$ with no dependence on the coordinate volume as it should be the case in the absence of a fixed background geometry.~The quantum counterpart of a putative $V_o\to\infty$ limit amounts then to chose a state $\Psi$ for which the region of homogeneity is geometrically enlarged, i.e. $\langle\hat{\alpha}(V_o)\rangle_{\Psi}\to\infty$.~Quantum fluctuations over a region $V\subset V_o$ -- whose physical size as given by the (expectation value of) $\hat{\alpha}(V)$ can be operationally thought of as the scale over which one would like the homogeneous description to be probed -- are then suppressed in this limit.~This suppression of quantum fluctuations -- and correspondingly the dynamics of $\hat{\alpha}(V)$ becoming effectively classical -- is meaningful as long as the large scale/volume regimes is concerned where a large number of cells is patched together and a fully homogeneous semi-classical effective description can be safely trusted.~On the contrary, as the quantum fluctuations of the probe region are measured w.r.t. the fluctuations needed to make the fiducial cell homogeneous on a certain scale, they can not be ignored for small volumes where a description in terms of the homogeneous degrees of freedom would require $V_o$ to be small enough so that more and more inhomogeneous
modes with wavelength larger than the fiducial cell can be
neglected within that region.~The states yielding small physical volumes are thus very quantum and the quantum description of $V$ at small scales cannot be made effectively classical to an arbitrary precision.

As a last remark, we would like to comment on the specific relation between the different $V_o$-labelled quantum theories on which our analysis is based.~To make the situation at the quantum level as close as possible to that of the classically symmetry reduced theory, whose on-shell predictions are independent of $V_o$, our criterion for relating the different quantum theories was to identify quantum states with the same dynamics.~As emphasised in the main body of the paper, there is a priori no unique criterion.~Therefore, one could in principle seek for a different mapping for which the resulting scaling behaviours are such that quantum fluctuations remain small when shrinking $V_o$.~However, if such a mapping exists, the dynamics of the states will then be modified under a change of the region $V_o$.~In either case, it is clear that the size of the fiducial cell has non trivial effects at the quantum level and is not just a regulator that can be removed at the end of the day.

\subsection{Future Directions}\label{sec:furtherfuturedirections}

Starting from the analysis presented in this paper and the strategy to impose symmetry-reduction constraints in the canonical theory, various directions for future investigations can be opened.~Let us close this work by summarising some of them.

\textsf{\textbf{Quantum symmetry reduction:}} In the present paper we focused on the implementation of symmetry-reduction constraints at the classical level, and on the quantisation of the classically symmetry-reduced theories so obtained.~A first natural interesting direction would consist then in understanding how the symmetry reduction can be instead performed at the quantum level.~Such a question for the case of a (free) scalar field theory was discussed in Sec.~\ref{sec:symredscalarQFT} where the main steps were also sketched.~Even though we pursued a more intuitive than systematic approach, this was enough to reproduce the scaling behaviour of the classically symmetry-reduced and then quantised theory.~This allowed us also to discuss the limitations of the homogeneous truncation and spell out the condition for its validity ($Lm\gg 1$).~A systematic analysis would however still be required.~Moreover, our main interest relies on applying and eventually extend it to the gravitational case with the general aim of trying to better understand the relation at the quantum level between full LQG and its symmetry reduced sectors describing quantum mini- and midi-superspace models such as cosmology or black holes.~This also connects to already existing literature as e.g.~\cite{BodendorferAnEmbeddingOf, BodendorferQuantumReductionTo,Bodendorfer:2014wea,Bodendorfer:2015qie} in the context of cosmology and black holes.~Specifically, the fact that the canonical analysis for cosmology discussed here is based on variables which are part of the full set of non-isotropic variables introduced in \cite{BodendorferAnEmbeddingOf, BodendorferQuantumReductionTo} to construct an explicit embedding of LQC into a full theory context, the former being understood as a one-vertex (=lattice point) truncation of the latter, provides us with a promising starting point to bridge between symmetry-reduced and full theory context also at the the quantum level.~Of particular interest in this direction would be to first consider interacting field theories and eventually study the relation to renormalisation.~In this respect, the dependence of coupling constants on the fiducial cell as found in Sec.~\ref{sec:quantuminteractions} should be further investigated and clarified from a full QFT perspective.~As anticipated in the previous section, intriguing questions in this direction are for instance: Is the fiducial cell related to a running coupling and renormalisation? How important are the small modes in the interacting theory?~We will come back on this point later in this section.

\textsf{\textbf{Beyond homogeneous approximation:}} Another direction, which is of interest already at the classical level, consists of using the formalism of symmetry reduction presented in this paper to go beyond the homogeneous approximation in a controlled way, as it was sketched for the cosmological setting in Sec.~\ref{outlook:beyondhom}.~As discussed there, the inclusion of the next to homogeneous modes allows us to study a possibly non-perturbative description of the system, the latter being only allowed to inhomogeneously change on large scales.~This leads already to a rather complicated structure of the Hamiltonian and a possible analytic treatment of it and its solutions remains to be explored.~This would also serve as a preparation for the study of the quantisation of the system as a non-perturbative toy model beyond homogeneity.~In particular, our main interest would reside in the application of LQG quantisation techniques, and hence to construct holonomies in terms of the Ashtekar connection as well as to study any other polymerisation scheme which might eventually become available in such a setting.~Due to the fact that the model is restricted to a fixed box, it can still formally be described as a particle-mechanical system on top of which only a few slowly varying field theory degrees of freedom are considered.~This might then allow to face the problem of a midi-superspace model in a controlled way.

\textsf{\textbf{Mode expansion for spherically-symmetric or slowly-time dependent settings:}}~Our classical symmetry reduction procedure focused on a specific mode expansion based on a starting metric ansatz and leading to flat FLRW-cosmology.~Another interesting direction would then consist in extending the present framework to other kinds of geometries and sets of symmetries as e.g. spherically symmetric systems by means of a suitable mode expansion.~For example, an expansion in terms of spherical harmonics would be characterised by spherically symmetric zero modes, while the higher modes would allow to go systematically beyond the perfectly spherically-symmetric setting with potential application to black holes, maybe even rotating ones, or non-spatially flat cosmologies.~A throughout study of how different kinds of mode expansions should be adopted to describe certain (classes of) geometries would be then desirable.

Another direction concerns slowly time dependent systems.~In the context of black holes, for instance, one usually approximates the black hole to be static outside of the collapsing shell so that similar techniques can in principle be applied to decompose the non-compact directions into finite boxes and expand the time dependence of the metric in terms of the homogeneous modes.~But what kind of mode expansion would be suited in this setting?~The discrete Fourier modes used in this paper would not as these would force the fields to be $L$-periodic and therefore would also force the process so described to be periodic in the fiducial time interval.~A particularly promising candidate instead is provided by Legendre polynomials, which can be defined on a compact interval and would allow to capture also non periodic physics.~The leading order is homogeneous (thus modelling a static scenario), and the next one is linearly changing.~It would be interesting to check whether this turns out to be sufficient to model Hawking radiation.~Allowing in fact only the next-to-homogeneous modes to contribute to the dynamics of the system would only take care of slow physical processes.~High frequency effects coming from the collapse of matter would thus be ignored.~On the other hand, Hawking radiation is a very slow process, at least in the solar mass black hole stage, and could be captured by the next-to-homogeneous modes while truncating out of the description at the same time more complicated phenomena related to collapse. 

\textsf{\textbf{Small momentum/large wavelength modes and boundary terms:}} As already stressed in the main body of the paper, the homogeneous theory results form a twofold approximation consisting both in setting to zero the inhomogeneous modes within the fiducial cell, and in truncating the remaining inhomogeneous modes with wavelength larger than the size of the fiducial cell.~A detailed investigation of the truncated small momentum modes and their contributions to dynamics at different scales is crucial to study the validity of the homogeneous approximation on a quantitative basis.~To this aim, Einstein's equations should be solved for the non-homogeneous metric \eqref{eq:ADMmetric} and the dynamics of the volume averaged observables compared with the predictions of symmetry reduced homogeneous models.~As was already shown in Sec.~\ref{outlook:beyondhom}, in fact, the higher modes contribute to the homogeneous dynamics.

This also leads to the question of boundary terms, which were systematically neglected in the present analysis, and their possible physical relevance.~One can try to regularise these and include their contribution to the dynamics thus leading to a lattice-like regularisation of the full theory.~The resulting model should be then compared to previous proposals for lattice LQC with interacting homogeneous and isotropic cells as e.g. discussed in a slightly different context in \cite{Wilson-EwingLatticeLoopQuantum}.~More generally, in fact, one might ask how lattice QFT techniques could be imported to the gravitational systems considered here.~In the cosmological case, the study of boundary terms arising for instance at the codimension 2 boundary of the homogeneity region $V_o$ or even at the interface of the different subcells $V_n$ patched together therein would be particularly interesting from the point of view of diffeomorphism symmetry.~In particular, it would be interesting to investigate whether they might be related to a local breaking of diffeomorphism symmetry -- e.g. due to the specific choice of modes or to the fact that boundary terms are usually neglected in homogeneous descriptions -- and how the latter symmetry might be eventually restored once these are properly taken into account.~This might also provide a complementary point of view on the study of dynamical symmetries in gravitational minisuperspace models such as cosmology and black holes recently investigated in several works \cite{BenAchour:2019ufa,BenAchour:2020njq,BenAchour:2020xif,BenAchour:2020ewm,Achour:2021lqq,Achour:2021dtj,Geiller:2020xze,Geiller:2021jmg,Geiller:2022baq,BenAchourSchroedingerSymmetry} and possibly give insights on the geometric origin of such symmetries.~No boundaries in fact have been explicitly included in the analysis so far.~As already mentioned in Sec.~\ref{Sec:cosmHURandfluc}, the fact that the application of our mapping between the quantum theories associated with different values of $V_o$ to the quantum fluctuations of smaller subcells within the region $V_o$ results into the fluctuations being suppressed as the ratio $V_o/V$ increases resonates with the observation in  \cite{BenAchourSchroedingerSymmetry} that the central charge of the Schr\"odinger symmetry algebra of the system, interpreted as the average number of microscopic cells, is given by the very same ratio between the IR and the UV cut-offs which sets the scale for how classical or quantum is the system.~We find these similarities encouraging and pointing towards something important underlying them to be understood, also in relation to the role of such scales from the point of view of coarse graining and renormalisation when changing the observation scale of the system (see also the last paragraph below).

The boundary terms are also strictly related to the fact that we used discontinuous modes.~It might be useful to use different modes, which are not fully homogeneous, but continuous.~What would the resulting physics be and how would the boundary terms change?~This is closely related to the remaining momentum profile.~In the setting discussed in this work, as we have seen in Secs.~\ref{sec:truncation},~\ref{Sec:gravityDBtruncatedtheory},~and~\ref{outlook:beyondhom}, only integer momenta do not contribute and higher momenta are polynomially suppressed.~One could then try to do a more elaborated ansatz better approximating a momentum distribution with only small momenta.~A possibility would be a very sharp Gaussian profile or maybe a combination of $\tanh$ to approximate the characteristic function.~This might also provide an alternative way to go beyond homogeneity.~The physical applicability of certain modes of course needs to be judged on a case by case basis.

\textsf{\textbf{Coarse graining and renormalisation:}} Finally, we would like to point out some similarity between the strategy presented here for imposing homogeneity constraints and coarse graining.~More specifically, this concerns the imposition of the second kind of constraints demanding the homogeneous zero modes of different (sub)cells to be smoothly matched across the cells (cfr.~Sec.~\ref{Sec:gravityDBtruncatedtheory},~Eqs.~\eqref{homogconstraintsmodesalpha}-\eqref{homogconstraintsmodesPphi}).~In fact, when considering multiple cells patched together into a bigger cell, the fully homogeneous analysis discussed in the present paper demands the zero modes in different cells to be equal while setting to zero all the inhomogeneous modes.~This essentially amounts to replicate the physics in one cell, say $V_1$, into all the other cells $V_n$,~$\sqcup_n V_n=V_o$.~As a consequence, in the resulting family of homogeneous quantum theories labelled by the different values of $V_o$, the dynamics is preserved under the mapping \eqref{eq:isomorphygrav} between the different Hilbert spaces and no renormalisation of the Hamiltonian is involved.~The present setting can be therefore extended in the following ways.~First of all, still neglecting interactions between the cells which in our framework should be encoded in the boundary terms at the interfaces of the individual cells, the inclusion of inhomogeneous modes with wavelength larger than a single cell size would require us to impose different gluing conditions for the modes so that we expect the dynamics for states in the quantum theories of a single cell or many cells not to be the same anymore and, consequently, the mapping between different $V_o$-valued Hilbert spaces to be modified.~Second, the interactions between neighbouring cells should be also included.~In this respect, it would be interesting to compare the resulting analysis with previous work on perturbations around homogeneous cosmological spacetimes as initiated in \cite{Wilson-EwingLatticeLoopQuantum}, and systematically study the regime of validity of perturbative treatments of inhomogeneities at the quantum level. 

As a preliminary step in this direction before moving to the gravitational setting, these thoughts could be developed for the case of an interacting scalar field theory.~In fact, as discussed in Sec.~\ref{sec:quantuminteractions}, the resulting $V_o$-dependence of the coupling constant seems to suggest a sort of ``running coupling'' behaviour when interactions are considered.~It would be therefore insightful to establish a relation between the interacting scalar field discussed in this work and the lattice QFT approach.~Of course, in the end we would be interested in quantum gravity applications.~In this regards, it is worth mentioning that questions about coarse graining and renormalisation have gained increasing attention in the LQG and related literature, both from a full theory and a symmetry-reduced perspective.~For the case of cosmology, for instance, there has been recent effort e.g. in \cite{BodendorferCoarseGrainingAs,BodendorferRenormalisationwithsu11,BodendorferStateRefinementsAnd,BodendorferPathIntegralRenormalization,Bodendorfer:2020nku}.~From the full theory side, instead, such directions have been investigated both in the context of Hamiltonian renormalisation \cite{LangHamiltonianRenormalisationI,LangHamiltonianRenormalisationII,BahrPropertiesofthe,LangHamiltonianRenormalisationIII,LangHamiltonianRenormalisationIV,LiegenerHamiltonianRenormalisationV,Thiemann:2020cuq,ThiemannRenormalisationWavelets,ThiemannHamiltonianRenormalisationVI,ThiemannHamiltonianRenormalisationVII} and spin foam path-integral \cite{Dittrich:2014mxa,Steinhaus:2014jvq,Bahr:2016hwc,Dittrich:2016tys,Bahr:2017klw,Steinhaus:2018aav,Bahr:2018gwf,Steinhaus:2020lgb} (see also \cite{CarrozzaFlowinginGFT} and references therein for a review on renormalisation in the group field theory formalism).~As in principle the cells $V_n$ can be arbitrarily small, this can be seen as a lattice regularisation of the field theory.~There might be then also the possibility to go beyond the homogeneous approximation at the quantum level and the gravitational RG-flow might eventually be studied.~In particular, it would be interesting to see whether a single vertex truncation in the sense of a full theory embedding as in \cite{BodendorferAnEmbeddingOf} for the different cells might result into some notion of coarse graining and renormalisation in a full theory context.

The various directions mentioned above are of course not unrelated to each other, and making progress in one of them can give valuable insights also for the others.~The setup developed in this work should then offer interesting starting points for studying these questions in future investigations.

\section*{Acknowledgements}

The work of FMM was supported by funding from Okinawa Institute of Science and Technology Graduate University.~FMM's research at Western University is also supported by Francesca Vidotto's Canada Research Chair in the Foundation of Physics, and NSERC Discovery Grant ``Loop Quantum Gravity: from Computation to Phenomenology''.~Western University is located in the traditional territories of Anishinaabek, Haudenosaunee, L\=unaap\'eewak and Chonnonton Nations.~This project/publication was made possible in part through the support of the ID\# 62312 grant from the John Templeton Foundation, as part of the Second Phase of \href{https://www.templeton.org/grant/the-quantum-information-structure-of-spacetime-qiss-second-phase}{\textit{The Quantum Information Structure of Spacetime}} Project (QISS).~The work of JM was made possible through the support of the ID\# 61466 grant from the John Templeton Foundation, as part of the First Phase of \href{https://www.templeton.org/grant/the-quantum-information-structure-of-spacetime-qiss}{\textit{The Quantum Information Structure of Spacetime}} Project (QISS).~The opinions expressed in this project/publication are those of the author(s) and do not necessarily reflect the views of the John Templeton Foundation.

\appendix

\section{Geometry Independent Mode Decomposition}\label{app:modedecomposition}

In the main text the homogeneity constraints were implemented by decomposing the dynamical fields in terms of modes.
This seems to require a notion of background geometry to define a $L^2$-pairing, a scalar product, and thus a notion of orthogonal modes.
However, in this appendix we would like to show how such a mode decomposition can be formulated without any reference to the geometry of the spacial slice and how such a formulation is related to the expansions presented in the main text.
To fix a generic notation to be applied to both the scalar field and the cosmology case considered in this work, let us introduce the local fields $Q(x)$ and $P(x)$, where $Q$ is a scalar and $P$ a scalar density.
In the case of a scalar field theory considered in Sec.~\ref{sec:scalarfieldsetup}, $Q$ plays the role of the scalar field itself and $P$ of the corresponding conjugate momentum.
In the cosmological setting of Sec. \ref{sec:setup}, $P$ corresponds to the field $\alpha$ as both are density-1 objects, while $Q$ can be thought of as the momentum $P_\alpha$.

\noindent
The strategy can then be summarised in the following main steps:
\begin{enumerate}
	\item We want to decompose the field $Q(x)$ -- a scalar density of weight 0 -- and the field $P(x)$ -- a scalar density of weight 1 -- in terms of modes\footnote{The modes are assumed to be complex valued although in our case we discuss real-valued fields $Q$, $P$ only. However, even if $f_n$ is complex valued, the fields being real-valued can be guaranteed by choosing proper reality conditions for the coefficients.} $f_n:\Sigma_t \rightarrow \mathbb{C}$, where $n$ labels the modes.
	In the following, we assume the range of all $n$ to be countable.
	Further, we assume the modes $f_n$ to be compactly supported and smooth, i.e. $f_n \in C^{\infty}_c\left(\Sigma_t\right)$.
	These form a basis\footnote{To be precise a Schauder basis, for which it is assumed that a notion of topology and convergence can be defined.} of $C^\infty\left(\Sigma_t\right)$.
	
	\item In addition to the countably infinite functions $f_n \in C^\infty_c(\Sigma_t)$, we need their dual forms, which are linear continuous maps $F_n: C^\infty(\Sigma_t) \rightarrow \mathbb{C}$.
	The latter are elements of the topological dual space of $C^\infty(\Sigma_t)$ and are uniquely determined by the conditions
	\be\label{eq:orthogonalitycompleteness}
	F_n[f_m] = \delta_{nm} \qquad,\qquad\sum_{n=1}^{\infty}F_n[g] f_n = g\qquad \forall\; g \in C^\infty(\Sigma_t)\;,
	\ee
	
	\noindent
	to which we refer to as orthonormality and completeness, respectively.
	
	\item The canonical fields can be then expanded without loss of information as 
	\begin{equation}\label{eq:expansniongeneral}
		Q(x) = \sum_{n=1}^\infty F_n[Q] f_n(x) \qquad , \qquad P[f]:= \int_{\Sigma_t} \dd^3x\, P(x) f(x) = \sum_{n=1}^\infty P[f_n;t] F_n[f] \;,
	\end{equation}
	
	\noindent
	with expansion coefficients
	\begin{equation}
		\tilde{Q}_n = F_n\left[Q\right] \quad , \quad \tilde{P}_n = P[f_n] \;.
	\end{equation}

	\noindent
	Note that the integral measure $\dd^3x P(x)$ is well-defined and does not require a reference to a background metric as $P(x)$ is a density of weight $1$, i.e. $P \dd^3x$ is a volume form by its own.
	
	\item Assuming the two fields $Q$ and $P$ to satisfy the standard commutation relations, i.e. 
	$$
	\Poisson{Q(x)}{P(y)} = \delta(x-y) \;,
	$$
	
	\noindent
	leads the smeared fields $F[Q]$ and $P[g]$, $F:\,C^\infty(\Sigma_t)\rightarrow \mathbb{C}$, $g \in C^\infty_c(\Sigma_t)$, to satisfy
	$$
	\Poisson{F[Q]}{P[g]} = F[g] \;.
	$$
	
	\noindent
	Note that we have to smear the field $Q$ by a functional, while $P$ as a density of weight $1$ can naturally act on a function.
	Therefore, the expansion coefficients will satisfy
	\begin{equation}
		\Poisson{\tilde{Q}_n}{\tilde{P}_m} = F_n\left[f_m\right] = \delta_{nm} \;.
	\end{equation}
\end{enumerate}

Let us note that once the functions $f_n$ are chosen, their dual functionals $F_n$ are uniquely fixed due to the conditions~\eqref{eq:orthogonalitycompleteness} with no reference to the geometry of the spatial slice $\Sigma_t$ needed.~However, when a fixed background geometry is available, the functionals $F_n$ can be explicitly written in terms of the functions $f_n$ via $L^2$-pairing defined with respect to the induced spatial metric $q_{ab}$ as\footnote{The space $C^\infty(\Sigma_t,\mathbb{C}\text{\,or\,}\mathbb{R})$ is an infinite-dimensional differential manifold equipped with compact-open topology. Tangent and cotangent spaces can be then defined at each point of that manifold as well as their $L^2$-dual pairing, the latter defined w.r.t. the (fixed) induced metric on the spatial slice.}
\begin{equation}\label{L2pairingdualmodes}
	\braket{f}{g} := \int_{\Sigma_t} \dd^3x \, \sqrt{q}f^*(x) g(x)\;,
\end{equation}

\noindent
where $q := \det(q_{ab})$. In such a case, there is then a natural isomorphism given by the identification
\be\label{eq:L2isomorphism}
	i:\;C^\infty\left(\Sigma_t\right)^\star \longrightarrow C^\infty\left(\Sigma_t\right)\qquad\text{by}\qquad F \longmapsto f \quad , \quad F[g] =: \braket{f}{g} \quad \forall g \in C^\infty\left(\Sigma_t\right) \;.
\ee

\noindent
The problem of finding the set of functions $f_n$ and theirs duals $F_n$ reduces then to only find an orthogonal basis $f_n$, i.e. $\braket{f_n}{f_m} = N_n\cdot \delta_{nm}$, where $1 \neq N_n \in \mathbb{R}$ does not enforce the basis to be normalised.
The functionals $F_n$ can then be induced by means of
\begin{equation}
	F_n[g] := \int_{\Sigma_t} \dd^3x \sqrt{q} \,\frac{f_n^*(x)}{N_n} g(x) \qquad \forall g \in C^\infty(\Sigma_t)\;,
\end{equation}

\noindent
and thus $i\left(F_n\right) = f_n/N_n$.
The system $(f_n,F_m)$ fulfils the above orthogonality and completeness relations \eqref{eq:orthogonalitycompleteness}.
With this identification, we can then reconstruct an expansion of the local field $P(x)$ as\footnote{Here $P[\bullet]$ is the functional constructed out of the local density $P(x)$. We did not use different symbols to avoid overloading the notation.} 
\begin{equation}\label{eq:localdensity}
	\overline{P}(x) =  i\left(P[\bullet]\right)(x) = \frac{P(x)}{\sqrt{q}}\,.
\end{equation}

\noindent
Note that the isomorphism $i$ is complex-linear, i.e. $i\left(\lambda F\right) = \lambda^* i\left(F\right)$.
This leads to
\begin{equation}
	i\left(P\right) = \sum_n P[f_n]^* i\left(F_n\right) = \sum_n P[f_n]^* \frac{f_n}{N_n} \;.
\end{equation}

\noindent
In contrast, it is possible to expand the scalar function $\overline{P}$ in such modes as
\begin{equation}
	\overline{P} = \sum_n F_n\left[\,\overline{P}\,\right] f_n\;,
\end{equation}

\noindent
which in turn leads to identifying the expansion coefficients as
\begin{equation}
	\overline{P}_n := F_n\left[\,\overline{P}\,\right] = \frac{P[f_n]^*}{N_n} = \frac{\tilde{P}_n^*}{N_n} \;,
\end{equation}

\noindent
thus yielding the local expansion
\begin{equation}\label{eq:localexpansion}
	P(x) = \overline{P} \sqrt{q} = \sqrt{q} \sum_n \frac{\tilde{P}_n^*}{N_n} f_n(x)\,.
\end{equation}

Let us now explicitly discuss the above formal steps for the case of the scalar field considered in Sec.~\ref{Sec:Warmupscalarfield} where $Q(x) = \phi(x)$ and $P(x) = \pi(x)$.
As modes we chose box-wise Fourier modes, i.e.
\begin{equation}
	f^n_{\vec k}(x) = \chi_{V_n}(x) e^{+i\frac{2\pi}{L} \vec{k}\cdot \vec{x}} \;,
\end{equation}

\noindent
where $\vec{k}\cdot \vec{x} := k_1 x^1 + k_2 x^2 + k_3 x^3$ and the spatial slice was decomposed into a countable number of boxes $V_n$ with edge-length $L$.
As also emphasised in the main text, this is a purely topological construction and only states that the edges of the box runs from $x = x_n$ to $x = x_n + L$\footnote{For the case of a scalar field on Minkowski background of Sec. \ref{sec:scalarfieldsetup}, the coordinate edge length $L$ is equal to the physical edge length $L = \int_{x_n}^{x_n+L}\dd x \sqrt{q_{xx}}$.}.
The dual modes are implicitly determined by the conditions~\eqref{eq:orthogonalitycompleteness}, which become
\begin{equation}\label{eq:explicitconddualmodes}
	F_{\vec{k}}^n\left[f_{\vec{p}}^m\right] = \delta_{n m} \delta_{\vec k ,\vec p} \qquad , \qquad g = \sum_n \sum_{\vec k \in \mathbb{Z}^3} f^n_{\vec{k}} F_{\vec{k}}^n\left[g\right] \qquad \forall\; g \in C^\infty\left(\Sigma_t\right) \;.
\end{equation}

\noindent
The expansions of the fields $\phi$ and $\pi$ work as in Eq.~\eqref{eq:expansniongeneral} above and thus requires no explicit notion of background metric.~Using however the fact that there is a fixed background metric, Minkowski in the specific case under cosideration now so that the iduced metric on the spacial slice is simply $q_{ab} = \delta_{ab}$, we can explicitly write the dual modes $F^n_{\vec{k}}$ in terms of the modes $f_{\vec{k}}^n$ themselves as
\begin{equation}
	F_{\vec{k}}^n[g] = \frac{1}{L^3}\int_{\Sigma_t} \dd^3 x \sqrt{q} \left(f^n_{\vec{k}}\right)^* g(x) = \frac{1}{L^3} \int_{V_n} \dd^3 x \,e^{-i\frac{2\pi}{L} \vec{k} \cdot \vec{x}} g(x)\;.
\end{equation}

\noindent
This leads to the mode decomposition of the main text (cfr. Eqs.~\eqref{eq:expansion}, \eqref{eq:piexpansion}), i.e.\footnote{The last line uses the expression~\eqref{eq:localexpansion}, where the reality condition $\left(\tilde{\pi}^n_{\vec{k}}\right)^* = \tilde{\pi}^n_{-\vec{k}}$ and a re-labelling $\vec{k} \mapsto -\vec{k}$ were used.} 
\begin{subequations}
	\begin{align}
		\phi(x) &= \sum_n \sum_{\vec{k} \in \mathbb{Z}^3} \tilde{\phi}_{\vec{k}}^n f^n_{\vec{k}} (x) = \sum_n \sum_{\vec{k} \in \mathbb{Z}^3} \tilde{\phi}_{\vec{k}}^n \chi_{V_n}(x) e^{+i\frac{2\pi}{L} \vec{k} \cdot \vec{x}} \qquad,\qquad \tilde{\phi}_{\vec{k}}^n= F_{\vec{k}}^n[\phi] = \frac{1}{L^3} \int_{V_n} \dd^3x\sqrt{q}\, \phi(x) e^{-i\frac{2\pi}{L} \vec{k} \cdot \vec{x}}
		\\
		\pi[g] &= \sum_n \sum_{\vec{k} \in \mathbb{Z}^3} \tilde{\pi}_{\vec{k}}^n F^n_{\vec{k}}[g] = \sum_n \sum_{\vec{k} \in \mathbb{Z}^3} \tilde{\pi}_{\vec{k}}^n \frac{1}{L^3} \int_{V_n} \dd^3x\sqrt{q}\, e^{-i\frac{2\pi}{L} \vec{k} \cdot \vec{x}} g(x)\quad,\quad\tilde{\pi}_{\vec{k}}^n= \pi\left[f^n_{\vec{k}}\right] = \int_{V_n} \dd^3 x\, \pi(x) e^{+i\frac{2\pi}{L} \vec{k} \cdot \vec x} \,
		\\
		&\text{i.e.}\qquad \overline{\pi}(x) := i\left(\pi[\bullet]\right) = \frac{\pi(x)}{\sqrt{q}} = \sum_n \sum_{\vec{k}\in\mathbb{Z}^3} \tilde{\pi}^n_{\vec{k}} \,\chi_{V_n}(x) \frac{e^{-i\frac{2\pi}{L} \vec{k} \cdot \vec{x}}}{L^3} \;.\label{eq:pibarexpansion}
	\end{align}
\end{subequations}

\noindent
Let us emphasise here that the last line Eq.~\eqref{eq:pibarexpansion} is the expansion of the scalar quantity $\overline{\pi}(x) = \pi(x)/\sqrt{q}$ (cfr. Eq.~\eqref{eq:localdensity}).
While the expansion of the smeared density $\pi[f]$ is possible without the knowledge of $\sqrt{q}$, this is not possible for $\overline{\pi}$, the latter can only be constructed when $q$ is known.
However, as $q_{ab} = \delta_{ab}$ for the specific case under consideration, $\sqrt{q} = 1$, and $\pi(x)$ and $\overline{\pi}(x)$ have the same values as evaluated in the same Cartesian chart so that their difference becomes detectable only by studying their transformation behaviour.

The situation is different in the gravity or cosmological case where no background metric is available, the metric itself being part of the canonical variables and as such it is not suited to be used in a mode expansion.
As discussed in Sec. \ref{sec:setup} of the main text, working in the diagonal metric gauge and imposing local isotropy, the description of the system can be simplified by introducing the canonical variables $\alpha(x)$ and $P_\alpha(x)$ defined in Eq. \eqref{diagisotropyvariables} as the only gravitational degrees of freedom.
In the present notation, we can identify $P(x) = \alpha(x)$ and $Q(x) = P_\alpha$ as indicated by their density properties.
Starting then again with the modes 
\begin{equation}
	f^n_{\vec k}(x) = \chi_{V_n}(x) e^{+i\frac{2\pi}{L} \vec{k}\cdot \vec{x}} \;,
\end{equation}

\noindent
the dual modes are uniquely fixed by the conditions~\eqref{eq:explicitconddualmodes}.
Note that the parameter $L$ is purely topological as it only defines what the coordinate range of a certain volume $V_n$ is.
The physical length is now given by $\int_{x_n}^{x_n+L} \dd x \,\sqrt{q_{xx}(x)}$, which is a dynamical quantity. The canonical fields can be then expanded as
\begin{subequations}
	\begin{align}
		\alpha[g] &= \sum_n \sum_{\vec{k}\in\mathbb{Z}^3} \alpha[f_{\vec{k}}^n] F_{\vec{k}}^n[g] \quad \forall g \in C^\infty\left(\Sigma_t\right) \quad,\quad\tilde{\alpha}_{\vec{k}}^n = \alpha[f_{\vec{k}}^n] := \int_{\Sigma_t} \dd^3 x \, \alpha(x) \chi_{V_n}(x) e^{+i\frac{2\pi}{L} \vec{k}\cdot \vec{x}} \;,
		\\
		P_\alpha(x) &= \sum_n \sum_{\vec{k} \in \mathbb{Z}^3} F_{\vec{k}}^n\left[P_\alpha\right] f_{\vec{k}}^n \;,
	\end{align}
\end{subequations}

\noindent
which completes the mode decomposition, at least formally.~Such an abstract prescription is in fact not very practical as no expansion of the local quantities is explicitly available.~As we have no fixed background metric at our disposal, there is no canonical way to deduce the explicit expressions for $F_{\vec{k}}^n$ from the modes $f_{\vec{k}}^n$.
However, as customarly done in the LQC literature (see e.g. \cite{AshtekarLoopquantumcosmology:astatusreport} and references therein for details), a fiducial metric can be introduced on the non-compact spatial slice partitioned into the disjoint union of boxes $V_n$.
This is the point where different options are possible and the decomposition becomes ambiguous.
As in the main text, we can introduce a fiducial metric $\fidmetric{a}{b}$ whose coordinate axes are associated with the local triads along the edges of the cell (cfr. Eq.~\eqref{eq:deffidmetric})
\begin{equation}\label{flatfiducialmetric}
\fidmetric{a}{b} \dd x^a \dd x^b = \beta^\frac{2}{3} \delta_{a b} \dd x^a \dd x^b \;,
\end{equation}

\noindent
which is completely artificial and has no physical meaning.
The parameter $\beta$ labels the freedom in choosing a flat metric.
Nevertheless, it allows us to write the dual modes explicitly as
\begin{equation}
	F_{\vec{k}}^n[g] = \frac{1}{\text{vol}_{\fidmetric{}{}}(V_n)}\int_{\Sigma_t} \dd^3 x \sqrt{\fidmetric{}{}}\, f_{\vec{k}}^n(x)^* g(x) = \frac{1}{\beta L^3} \int_{V_n} \dd^3 x\,\beta e^{-i\frac{2\pi}{L} \vec{k} \cdot \vec{x}} g(x)\;,
\end{equation}

\noindent
where $\text{vol}_{\fidmetric{}{}}(V_n) := \int_{V_n} \dd^3x \, \sqrt{\fidmetric{}{}}$ is the volume of $V_n$ measured w.r.t the fiducial metric.
Note that, due to the $1/\beta$ prefactor, which cancels the $\beta$ coming from the determinant of $\fidmetric{}{}$ and is needed to get orthonormality, the above expression for $F_{\vec{k}}^n$ does not depend on the additional parameter $\beta$.
Therefore, the dual modes $F_{\vec{k}}^n$ are independent of this fiducial metric, while its decomposition in terms of $L^2$-pairing and the isomorphism $i$ given in Eq.~\eqref{eq:L2isomorphism} are not.
This becomes explicit as 
\begin{equation}
	i(F_{\vec{k}}^n) = \frac{f_{\vec{k}}^n}{\text{vol}_{\fidmetric{}{}}(V_n)} = \chi_{V_n}(x) \,\frac{e^{-i\frac{2\pi}{L} \vec{k} \cdot \vec{x}}}{\beta L^3}\;.
\end{equation}

Finally, to get an expansion of the local quantity $\alpha(x)$, we need to construct first the scalar $\overline{\alpha} := i\left(\alpha[\bullet]\right)=\alpha/\sqrt{\fidmetric{}{}}$ according to the identification of Eq.~\eqref{eq:localdensity} (cfr. Eq.~\eqref{eq:pibarexpansion}).
This leads to the decomposition of the main text (cfr. Eqs.~\eqref{alphamodedecomp}, \eqref{eq:Palphamodedecomp}), i.e.\footnote{Again, in comparison with Eq.~\eqref{eq:localexpansion}, we used the re-labelling $\vec{k} \mapsto -\vec{k}$ and the reality condition $(\tilde{\alpha}^n_{-\vec{k}})^*=\tilde{\alpha}^n_{\vec{k}}$.}
\begin{subequations}
	\begin{align}
		\alpha(x) =&\, \sqrt{\fidmetric{}{}} \overline{\alpha}(x) =\sqrt{\fidmetric{}{}}\sum_{n} \sum_{\vec{k} \in \mathbb{Z}^3} \tilde{\alpha}^n_{\vec{k}} \, \chi_{V_n}(x) \frac{e^{-i\frac{2\pi}{L} \vec{k} \cdot \vec{x}}}{\beta L^3} \;,
		\\
		P_\alpha(x) =&\, \sum_n \sum_{\vec{k} \in \mathbb{Z}^3} \tilde{P}^{n}_{\alpha,\vec{k}} \,\chi_{V_n}(x)\,e^{+i\frac{2\pi}{L} \vec{k} \cdot \vec{x}} \qquad,\qquad\tilde{P}^{n}_{\alpha,\vec{k}} :=\, F_{\vec{k}}^n\left[P_\alpha\right] = \frac{1}{\beta L^3} \int_{V_n} \dd^3 x\, \beta e^{-i\frac{2\pi}{L} \vec{k} \cdot \vec{x}} P_\alpha(x) \;.
	\end{align}
\end{subequations}

\noindent
As spacetime is dynamical, these local expressions can only be written down once a (fiducial) metric is specified, as e.g. the $\fidmetric{a}{b}$ given in \eqref{flatfiducialmetric}, the latter being a purely auxiliary construction at this stage.

\section{Coherent States and their Scaling Properties}\label{app:CS}

In this appendix we shall detail the explicit computation of the main properties and the scaling behaviours for coherent states in the $V_o$-labelled polymer quantum theories of spatially homogeneous and isotropic cosmology discussed in Sec.~\ref{Sec:cosmHURandfluc}.~In the volume representation of LQC, these are given by
\be\label{eq:CSlqc}
\ket{\Psi_{(\alpha',P_\alpha',\sigma)}}=\mathcal N \sum_{\alpha\in\mathbb Z}e^{-(\alpha-\alpha')^2\frac{\ell^2}{2\sigma^2}}e^{-i\lambda P_{\alpha}'(\alpha-\alpha')}\ket{\ell\alpha}\;,
\ee

\noindent
where $\mathcal N$ is a normalisation constant, $\ell$ is a chosen lattice spacing, and $\sigma>0$ is the width of the Gaussian\footnote{More precisely, recalling that solutions to the Hamiltonian constraint do not lie in the kinematical Hilbert space, but rather in its algebraic dual, \eqref{eq:CSlqc} is the \textit{shadow state} on the regular lattice $\ell\alpha$ ($\alpha\in\mathbb Z$) of the physical coherent state selected by the triplet $(\alpha',P_\alpha',\sigma)$ \cite{AshtekarMathematicalStructureOf,AshtekarQuantumGravityShadow,WillisPhd}.}.~In the present setting, $\ell=\frac{\lambda}{\eta^\gamma V_o^\delta}$ as determined by the regular lattice preserved by the action of the Hamiltonian operator (cfr.~Eqs.~\eqref{eq:nuinteger},~\eqref{eq:Htureaction} and the surrounding discussion).~Note that, expressing $\lambda$ in terms of $\ell$ and $V_o$ in \eqref{eq:CSlqc}, the above state can be rewritten as
\be\label{CSlqc2}
\ket{\Psi_{(\alpha',P_\alpha',\sigma)}}=\mathcal N \sum_{\alpha\in\ell\mathbb Z}e^{-\frac{(\alpha-\alpha'\ell)^2}{2\sigma^2}}e^{-i P_{\alpha}'(\alpha-\alpha'\ell)\eta^\gamma V_o^\delta}\ket{\alpha}\;,
\ee
which is the discrete polymer analogue of the Gaussian state \eqref{eq:gaussianstate} considered for the scalar field case at the end of Sec.~\ref{sec:phiquantisation}, where we chose $\gamma=0,\delta=1$ in the quantum representations of the elementary operators.

To construct a coherent state, we also have to specify the width $\sigma$ of the Gaussian, that is the tolerance for quantum fluctuations of the volume.~As discussed in \cite{AshtekarQuantumGravityShadow,WillisPhd}, coherent states in the polymer quantum representation are in good agreement with the standard coherent state of Schr\"odinger representation provided that the lattice spacing $\ell$ is much smaller than the scale $\sigma$ defining our tolerance ($\ell\ll\sigma$).~This is usually the case in LQC where the lattice spacing is taken to be of order of the the Planck scale where the effects of the spatial discreteness of the quantum geometry are expected to arise.~Given the above expression $\ell\sim\lambda/V_{o}^\delta$ and recalling that $\lambda\sim\ell_{Pl}$ \cite{AshtekarLoopquantumcosmology:astatusreport,AshtekarMathematicalStructureOf,AshtekarQuantumNatureOf}, this is a good approximation for large $V_o$ values.~Moreover, it remains so under fiducial cell rescaling.~To see this, let us consider states of the form \eqref{CSlqc2} in the quantum theories associated with two values $V_o^{(1)}$ and $V_o^{(2)}$ of $V_o$.~The corresponding regular lattice spacings $\ell^{(k)}$, $k=1,2$ are related by
\be\label{latticescaling}
\ell^{(1)}=\frac{\lambda}{\eta^\gamma V_o^{(1)\,\delta}}=\left(\frac{V_o^{(2)}}{V_o^{(1)}}\right)^\delta\ell^{(2)}\;,
\ee
so that, in agreement with the isomorphism \eqref{eq:isomorphygrav} between the two quantum theories, we have
\begin{align}\label{lqcCSscaling}
\Psi^{(2)}_{(\alpha',P_\alpha',\sigma^{(2)})}(\alpha)&=\mathcal N\,e^{-\frac{\left(\alpha-\alpha'\ell^{(2)}\right)^2}{2\sigma^{(2)\,2}}}e^{-i P_{\alpha}'(\alpha-\alpha'\ell^{(2)})\eta^\gamma V_o^{(2)\delta}}\nonumber\\
&=\mathcal N\,e^{-\left[\alpha \left(\frac{V_o^{(2)}}{V_o^{(1)}}\right)^\delta-\alpha'\ell^{(1)}\right]^2\left(\frac{V_o^{(1)}}{V_o^{(2)}}\right)^{2\delta}\frac{1}{2\sigma^{(2)\,2}}}e^{-i P_{\alpha}'\left[\alpha \left(\frac{V_o^{(2)}}{V_o^{(1)}}\right)^\delta-\alpha'\ell^{(1)}\right]\eta^\gamma V_o^{(1)\delta}}\nonumber\\
&=\Psi^{(1)}_{(\alpha',P_\alpha',\sigma^{(1)})}\left(\left(\frac{V_o^{(2)}}{V_o^{(1)}}\right)^\delta\alpha\right)\;,
\end{align}
where
\be\label{CSscalingsigma}
\sigma^{(1)}=\left(\frac{V_o^{(2)}}{V_o^{(1)}}\right)^\delta\sigma^{(2)}\;.
\ee
Thus, both the lattice spacing and the width of the Gaussian scale with the fiducial cell in such a way that the ratio $\sigma/\ell$ is independent of $V_o$.~This means that, if for instance $V_o^{(1)}$ was large enough for $1\ll\sigma^{(1)}/\ell^{(1)}$, then $1\ll\sigma^{(2)}/\ell^{(2)}$ no matter what $V_o^{(2)}$ is.~As the requirement $\ell\ll\sigma$ will play an important role in the sequel for studying the properties of the states \eqref{eq:CSlqc}, the fact that it is not affected by $V_o$ ensures that the same steps and approximations can be performed in theories corresponding to different $V_o$.

Let us then use the coherent states to compute the expectation values and fluctuations of the relevant operators, and discuss their scaling properties along the way.
\begin{itemize}
    \item Norm of the state:
    \be\label{normCSlqc1}
        \braket{\Psi_{(\alpha',P_\alpha',\sigma)}}{\Psi_{(\alpha',P_\alpha',\sigma)}}=|\mathcal N|^2\sum_{\alpha\in\ell\mathbb Z}e^{-\frac{(\alpha-\alpha'\ell)^2}{\sigma^2}}=|\mathcal N|^2\sum_{n\in\mathbb Z}e^{-\frac{(n-\alpha')^2\ell^2}{\sigma^2}}\;.
    \ee
    
    \noindent
    Following \cite{AshtekarMathematicalStructureOf,WillisPhd}, the series on the r.h.s. of \eqref{normCSlqc1} can be handled using Poisson summation formula
    \be\label{Poissonsum}
    \sum_{n\in\mathbb Z}g(x+n)=\sum_{n\in\mathbb Z}e^{2i\pi nx}\int_{-\infty}^{+\infty}g(y)e^{-2i\pi yn}\dd y\;.
    \ee
    with $x=0$ and $g(y)=e^{-(y-\alpha')^2\ell^2/\sigma^2}$.~Thus,
    \begin{align}\label{normCSlqc2}
        \|\Psi_{(\alpha',P_\alpha',\sigma)}\|^2&=|\mathcal N|^2\sum_{n\in\mathbb Z}\int_{-\infty}^{+\infty}e^{-\frac{(y-\alpha')^2\ell^2}{\sigma^2}}e^{-2i\pi yn}\dd y\nonumber\\
        &=|\mathcal N|^2e^{-\frac{\alpha'^2\ell^2}{\sigma^2}}\sum_{n\in\mathbb Z}e^{\left(\alpha'-i\pi n\frac{\sigma^2}{\ell^2}\right)^2\frac{\ell^2}{\sigma^2}}\int_{-\infty}^{+\infty}e^{-\left[y-\left(\alpha'-i\pi n\frac{\sigma^2}{\ell^2}\right)\right]^2\frac{\ell^2}{\sigma^2}}\dd y\nonumber\\
        &=|\mathcal N|^2\frac{\sqrt{\pi}\sigma}{\ell}\sum_{n\in\mathbb Z}e^{-\pi^2n^2\frac{\sigma^2}{\ell^2}}\;,
        \end{align}
        and, using $\sigma/\ell\gg1$ to truncate the series after the second term, we get
        \be\label{eq:CSnormapprox}
        \|\Psi_{(\alpha',P_\alpha',\sigma)}\|^2\approx|\mathcal N|^2\frac{\sqrt{\pi}\sigma}{\ell}\left(1+2e^{-\pi^2\frac{\sigma^2}{\ell^2}}\right)\;,
        \noindent
        \ee
    \item Expectation value of $\hat\alpha$:
    
    Recalling the action \eqref{eq:polymerrepr} of $\hat\alpha$, we have
    \begin{align}
        \langle\hat\alpha\rangle_{\Psi_{(\alpha',P_\alpha',\sigma)}}&=\frac{|\mathcal N|^2}{\|\Psi_{(\alpha',P_\alpha',\sigma)}\|^2}\sum_{\alpha\in\ell\mathbb Z}\frac{\eta^\gamma}{V_o^\gamma}\alpha\,e^{-\frac{(\alpha-\alpha'\ell)^2}{\sigma^2}}\nonumber\\
        &=\frac{|\mathcal N|^2}{\|\Psi_{(\alpha',P_\alpha',\sigma)}\|^2}\frac{\ell\,\eta^\gamma}{V_o^\gamma}\sum_{n\in\mathbb Z}n\,e^{-\frac{(n-\alpha')^2\ell^2}{\sigma^2}}\;,
\end{align}
which, using Poisson summation formula \eqref{Poissonsum} with $x=0$ and $g(y)=y\,e^{-(y-\alpha')^2\ell^2/\sigma^2}$, yields
\begin{align}\label{CSalphaEV}
        \langle\hat\alpha\rangle_{\Psi_{(\alpha',P_\alpha',\sigma)}}&=\frac{|\mathcal N|^2}{\|\Psi_{(\alpha',P_\alpha',\sigma)}\|^2}\frac{\ell\,\eta^\gamma}{V_o^\gamma}\sum_{n\in\mathbb Z}\int_{-\infty}^{+\infty}y\,e^{-(y-\alpha')^2\frac{\ell^2}{\sigma^2}}e^{-2i\pi yn}\dd y\nonumber\\
        &=\frac{|\mathcal N|^2}{\|\Psi_{(\alpha',P_\alpha',\sigma)}\|^2}\frac{\ell\,\eta^\gamma}{V_o^\gamma}e^{-\frac{\alpha'^2\ell^2}{\sigma^2}}\sum_{n\in\mathbb Z}e^{\left(\alpha'-i\pi n\frac{\sigma^2}{\ell^2}\right)^2\frac{\ell^2}{\sigma^2}}\int_{-\infty}^{+\infty}y\,e^{-\left[y-\left(\alpha'-i\pi n\frac{\sigma^2}{\ell^2}\right)\right]^2\frac{\ell^2}{\sigma^2}}\dd y\nonumber\\
        &=\frac{|\mathcal N|^2}{\|\Psi_{(\alpha',P_\alpha',\sigma)}\|^2}\frac{\sqrt{\pi}\sigma}{\ell}\frac{\ell\,\eta^\gamma}{V_o^\gamma}\sum_{n\in\mathbb Z}e^{-\pi^2n^2\frac{\sigma^2}{\ell^2}}\left(\alpha'-i\pi n\frac{\sigma^2}{\ell^2}\right)\nonumber\\
        &=\frac{\ell\,\eta^\gamma}{V_o^\gamma}\alpha'\nonumber\\
        &=\frac{\lambda}{V_o}\alpha'\;,
    \end{align}
where, in going from the third to the fourth line, we used the fact that the second term in the round bracket yields a vanishing contribution in the sum over ${n\in\mathbb Z}$, and the expression \eqref{normCSlqc2} for $\|\Psi_{(\alpha',P_\alpha',\sigma)}\|^2$.~In the last equality, we used the expression $\ell=\frac{\lambda}{\eta^\gamma V_o^\delta}$ and $\gamma+\delta=1$.~Thus, for a given value of $V_o$, the coherent state \eqref{eq:CSlqc} is peaked around the point $\lambda\alpha'/V_o$.~The $V_o$-dependence in \eqref{CSalphaEV} is compatible with the $V_o$-dependence of the $\alpha$-lattice and, more specifically, with the result \eqref{eigenstatsmearedalphaexpvalue} in the main text according to which the expectation value of the smeared volume operator $\widehat{\alpha(V_o)}=\widehat{\text{vol}(V_o)}=V_o\,\hat\alpha$ over the entire cell takes values in integer steps of $\lambda$, independently of the choice of $V_o$.~Moreover, this is in agreement with the scaling behaviour derived in Sec.~\ref{Sec:cosmHURandfluc} (cfr.~Eq.~\eqref{1to2alphaexpvalue}) by means of the mapping \eqref{eq:isomorphygrav}.~Indeed, from Eq.~\eqref{CSalphaEV} we see that the expectation value of the operator $\hat\alpha$ taken over the states \eqref{eq:CSlqc} in the quantum theories associated to the fiducial cell $V_o^{(1)}$ and $V_o^{(2)}$ are related to each other as
\be\label{CSalphaEVscaling}
\langle\hat\alpha|_{V_o^{(1)}}\rangle_{\Psi^{(1)}_{(\alpha',P_\alpha',\sigma^{(1)})}}=\frac{V_o^{(2)}}{V_o^{(1)}}\langle\hat\alpha|_{V_o^{(2)}}\rangle_{\Psi^{(2)}_{(\alpha',P_\alpha',\sigma^{(2)})}}\;,
\ee
and the expectation values of the corresponding smeared volume operator $\widehat{\alpha(V_o)}=\widehat{\text{vol}(V_o)}$ is independent of $V_o$ (cfr.~\eqref{eq:volexpvalue}).
    \item Expectation value of polymerised momentum:
    
    Recalling the action \eqref{eq:polymerrepr} of $\widehat{e^{\pm i\lambda P_\alpha}}$, we have
    \begin{align}\label{CSexpPalphaEV}
        \langle\widehat{e^{\pm i\lambda P_\alpha}}\rangle_{\Psi_{(\alpha',P_\alpha',\sigma)}}&=\frac{|\mathcal N|^2}{\|\Psi_{(\alpha',P_\alpha',\sigma)}\|^2}\sum_{n\in\mathbb Z}e^{-(n-\alpha')^2\frac{\ell^2}{2\sigma^2}}e^{i\lambda P_\alpha'(n-\alpha')}e^{-(n\mp1-\alpha')^2\frac{\ell^2}{2\sigma^2}}e^{-i\lambda P_\alpha'(n\mp1-\alpha')}\nonumber\\
        &=\frac{|\mathcal N|^2}{\|\Psi_{(\alpha',P_\alpha',\sigma)}\|^2}e^{-\frac{\ell^2}{2\sigma^2}}e^{\pm i\lambda P_{\alpha}'}e^{\frac{\alpha'\ell^2}{\sigma^2}}e^{\frac{-\alpha'^2\ell^2}{\sigma^2}}\sum_{n\in\mathbb Z}e^{-\left[n^2-2n(\alpha'\pm\frac{1}{2})\right]^2\frac{\ell^2}{\sigma^2}}\nonumber\\
        &\approx e^{-\frac{\ell^2}{4\sigma^2}}e^{\pm i\lambda P_{\alpha}'}\;,
        \end{align}
        where we used again Poisson summation formula \eqref{Poissonsum} for the series, evaluated the resulting integral with similar steps as in \eqref{normCSlqc2}, and truncated the resulting expression by using $\sigma/\ell\gg1$.~Therefore,
        \begin{align}\label{CSsinEV}
        \left\langle\reallywidehat{\frac{\sin(\lambda P_\alpha)}{\lambda}}\right\rangle_{\Psi_{(\alpha',P_\alpha',\sigma)}}&=\frac{1}{2i\lambda}\left\langle\widehat{e^{i\lambda P_\alpha}}-\widehat{e^{-i\lambda P_\alpha}}\right\rangle_{\Psi_{(\alpha',P_\alpha',\sigma)}}\nonumber\\
        &\approx \frac{\sin(\lambda P_\alpha')}{\lambda}\,e^{-\frac{\ell^2}{4\sigma^2}}=\frac{\sin(\lambda P_\alpha')}{\lambda}\left(1-\mathcal O\left(\ell^2/\sigma^2\right)\right)\;.
        \end{align}
    Recalling from Eqs.~\eqref{latticescaling} and \eqref{CSscalingsigma} that the ratio $\ell/\sigma$ does not scale with $V_o$, we see that \eqref{CSsinEV} is independent of $V_o$ in agreement with with the scaling behaviour derived in Sec.~\ref{Sec:cosmHURandfluc} (cfr.~Eq.~\eqref{eq:expPalphascaling}) by means of the mapping \eqref{eq:isomorphygrav}.~Moreover, in the low curvature regime $\lambda P_\alpha'\ll1$\footnote{We recall that $P_\alpha$ is proportional to the Hubble rate $\frac{\dot a}{a}$ (cfr.~Eq.~\eqref{alphaEoM} and surrounding discussion) and that the on-shell value of the Ricci scalar is proportional to $(\frac{\dot a}{a})^2$.~Moreover, the polymerisation scale $\lambda$ is related to the Planck scale so that $\lambda P_\alpha'\ll1$ far from the Planck regime.~We also notice that the scaling property \eqref{latticescaling} of $\ell=\frac{\lambda}{\eta^\gamma V_o^\delta}$ ensures the polymerisation scale $\lambda$ controlling the onset of quantum effects to be independent of $V_o$ with quantum effects becoming relevant in the regime $\lambda P_\alpha\sim1$.}, the polymerised momentum operator reduces to the standard momentum and \eqref{CSsinEV} reduces to $P_\alpha'$ at leading order in $\ell\ll\sigma$ as expected.  

    \item Fluctuations and uncertainty relations:
    
    Recalling the action \eqref{eq:polymerrepr} of $\hat\alpha$, using Poisson summation formula \eqref{Poissonsum}, and similar steps as in \eqref{CSalphaEV}, we have
    \begin{align}\label{CSalpha2EV}
        \left\langle\hat\alpha^2\right\rangle_{\Psi_{(\alpha',P_\alpha',\sigma)}}&=\frac{|\mathcal N|^2}{\|\Psi_{(\alpha',P_\alpha',\sigma)}\|^2}\left(\frac{\ell\,\eta^\gamma}{V_o^\gamma}\right)^2\sum_{n\in\mathbb Z}n^2\,e^{-\frac{(n-\alpha')^2\ell^2}{\sigma^2}}\nonumber\\
        &=\frac{|\mathcal N|^2}{\|\Psi_{(\alpha',P_\alpha',\sigma)}\|^2}\left(\frac{\ell\,\eta^\gamma}{V_o^\gamma}\right)^2\sum_{n\in\mathbb Z}\int_{-\infty}^{+\infty}y^2e^{-(y-\alpha')^2\frac{\ell^2}{\sigma^2}}e^{-2\pi iyn}\dd y\nonumber\\
        &=\frac{|\mathcal N|^2}{\|\Psi_{(\alpha',P_\alpha',\sigma)}\|^2}\frac{\lambda^2}{V_o^2}\sum_{n\in\mathbb Z}e^{-\pi^2n^2\frac{\sigma^2}{\ell^2}}\int_{-\infty}^{+\infty}y^2e^{-\left[y-\left(\alpha'-i\pi n\frac{\sigma^2}{\ell^2}\right)\right]^2\frac{\ell^2}{\sigma^2}}\dd y\nonumber\\
        &=\frac{|\mathcal N|^2}{\|\Psi_{(\alpha',P_\alpha',\sigma)}\|^2}\frac{\lambda^2}{V_o^2}\frac{\sqrt{\pi}\,\sigma}{\ell}\left[\left(\alpha'^2+\frac{\sigma^2}{2\ell^2}\right)\sum_{n\in\mathbb Z}e^{-\pi^2n^2\frac{\sigma^2}{\ell^2}}-\frac{\pi^2\sigma^4}{\ell^4}\sum_{n\in\mathbb Z}n^2e^{-\pi^2n^2\frac{\sigma^2}{\ell^2}}\right]\nonumber\\
        &\approx\frac{\lambda^2}{V_o^2}\left(\alpha'^2+\frac{\sigma^2}{2\ell^2}\right)\;,
    \end{align}
    where in the last line we used $\ell\ll\sigma$ to truncate the series at leading order.~From \eqref{latticescaling} and \eqref{CSscalingsigma}, we see that \eqref{CSalpha2EV} scales as $\langle\hat\alpha^2|_{V_o^{(1)}}\rangle_{\Psi^{(1)}_{(\alpha',P_\alpha',\sigma^{(1)})}}=\left(\frac{V_o^{(2)}}{V_o^{(1)}}\right)^2\langle\hat\alpha^2|_{V_o^{(2)}}\rangle_{\Psi^{(2)}_{(\alpha',P_\alpha',\sigma^{(2)})}}$ and, combining it with \eqref{CSalphaEV}, we find
    \be\label{CSalphavar}
    \Delta_{\Psi_{(\alpha',P_\alpha',\sigma)}}\hat\alpha^2=\left\langle\hat\alpha^2\right\rangle_{\Psi_{(\alpha',P_\alpha',\sigma)}}-\left\langle\hat\alpha\right\rangle_{\Psi_{(\alpha',P_\alpha',\sigma)}}^2\approx\frac{\lambda^2}{V_o^2}\frac{\sigma^2}{2\ell^2}\;,
    \ee
    which scales as
    \be\label{CSalphavarscaling}
    \Delta_{\Psi^{(1)}_{(\alpha',P_\alpha',\sigma^{(1)})}}\hat\alpha^2|_{V_o^{(1)}}=\left(\frac{V_o^{(2)}}{V_o^{(1)}}\right)^2\Delta_{\Psi^{(2)}_{(\alpha',P_\alpha',\sigma^{(2)})}}\hat\alpha^2|_{V_o^{(2)}}
    \ee
    in agreement with the scaling behaviours \eqref{eq:scalingmomentsGR} and \eqref{eq:scalingvariancesGR} in the main text.
    
    As for the fluctuations of the momentum operator, we have
    \be\label{CSsin2EV1}
    \left\langle\reallywidehat{\frac{\sin(\lambda P_\alpha)}{\lambda}}^{\,2}\right\rangle_{\Psi_{(\alpha',P_\alpha',\sigma)}}=-\frac{1}{4\lambda^2}\left\langle\widehat{e^{i\lambda P_\alpha}}^2+\widehat{e^{-i\lambda P_\alpha}}^2-2\widehat{e^{+i\lambda P_\alpha}}\widehat{e^{-i\lambda P_\alpha}}\right\rangle_{\Psi_{(\alpha',P_\alpha',\sigma)}}\;.
    \ee
    The third term in \eqref{CSsin2EV1} simply yields $\frac{1}{2\lambda^2}$.~Using the action \eqref{eq:polymerrepr} of the exponentiated momentum operator and repeating analogous steps as in \eqref{CSexpPalphaEV}, the first and second terms yield
    \be
    \left\langle\widehat{e^{\pm i\lambda P_\alpha}}^{2}\right\rangle_{\Psi_{(\alpha',P_\alpha',\sigma)}}\approx e^{-\frac{\ell^2}{\sigma^2}}e^{\pm 2i\lambda P_{\alpha}'}\;.
    \ee
    Therefore,
    \begin{align}\label{CSsin2EV2}
    \left\langle\reallywidehat{\frac{\sin(\lambda P_\alpha)}{\lambda}}^{\,2}\right\rangle_{\Psi_{(\alpha',P_\alpha',\sigma)}}&\approx\frac{1}{2\lambda^2}\left(1-\cos(2\lambda P_\alpha')e^{-\frac{\ell^2}{\sigma^2}}\right)\nonumber\\
    &=\frac{\sin^2(\lambda P_\alpha')}{\lambda^2}+\frac{1}{2\lambda^2}\frac{\ell^2}{\sigma^2}\cos(2\lambda P_\alpha')+\mathcal O\left(\frac{\ell^4}{\sigma^4}\right)\;.
    \end{align}
    From \eqref{CSsinEV} and \eqref{CSsin2EV2} then it follows that
    \be\label{CSsin2var}
        \Delta_{\Psi_{(\alpha',P_\alpha',\sigma)}}\reallywidehat{\frac{\sin(\lambda P_\alpha)}{\lambda}}^{\,2}=\left\langle\reallywidehat{\frac{\sin(\lambda P_\alpha)}{\lambda}}^{\,2}\right\rangle_{\Psi_{(\alpha',P_\alpha',\sigma)}}-\left\langle\reallywidehat{\frac{\sin(\lambda P_\alpha)}{\lambda}}\right\rangle^{2}_{\Psi_{(\alpha',P_\alpha',\sigma)}}\approx\frac{1}{2\lambda^2}\frac{\ell^2}{\sigma^2}\cos^2(\lambda P_\alpha')\;,
    \ee
    where we used $\sigma\gg\ell$ to truncate the series after the second term.~Finally, using \eqref{CSalphavar} and \eqref{CSsin2var}, we have
    \be\label{CSUR}
    \Delta_{\Psi_{(\alpha',P_\alpha',\sigma)}}\hat{\alpha}\,\Delta_{\Psi_{(\alpha',P_\alpha',\sigma)}}\reallywidehat{\frac{\sin(\lambda P_\alpha)}{\lambda}}\approx\frac{1}{2V_o}\left|\cos(\lambda P_\alpha')\right|\;,
    \ee
    which, together with (cfr.~\eqref{CSexpPalphaEV})
    \be
    \left\langle\reallywidehat{\cos(\lambda P_\alpha)}\right\rangle_{\Psi_{(\alpha',P_\alpha',\sigma)}}=\frac{1}{2}\left\langle\widehat{e^{i\lambda P_\alpha}}+\widehat{e^{-i\lambda P_\alpha}}\right\rangle_{\Psi_{(\alpha',P_\alpha',\sigma)}}\approx\cos(\lambda P_\alpha')\;,
    \ee
    tells us that the states \eqref{eq:CSlqc} saturate the uncertainty relations with good approximation as $\sigma\gg\ell$.~Moreover, we see that the $V_o$-scaling behaviour \eqref{eq:volumeUR} of the uncertainty relations is explicitly reproduced for the states \eqref{eq:CSlqc} and they remain saturated in the quantum theories associated to different cells.
\end{itemize}

To sum up, combining the results \eqref{CSalphaEV} and \eqref{CSsinEV}, we see that the state $\Psi_{(\alpha',P_\alpha',\sigma)}$ is peaked around the point $(\frac{\lambda}{V_o}\alpha',P_\alpha')$ as long as $\sigma\gg\ell$ and $\lambda P_\alpha'\ll1$.~The requirement of the lattice spacing being much smaller than the scale of tolerance for quantum fluctuations is not affected by a fiducial cell rescaling (the ratio $\sigma/\ell$ does not scale under a change of fiducial cell $V_o^{(1)}\to V_o^{(2)}$).~This ensures the relative dispersions $\frac{\Delta\hat\alpha}{\langle\hat\alpha\rangle}$ to be small (cfr.~Eqs.~\eqref{CSalphaEV},~\eqref{CSalphavar}) and the uncertainty relations \eqref{CSUR} to be saturated for such states in theories corresponding to different cells.~However, in agreement with the mapping $\mathscr{I}:\; \mathscr{H}_{poly}^{(1)} \to \mathscr{H}_{poly}^{(2)}$ given in the main text (cfr.~Eq.~\eqref{eq:isomorphygrav}), the parameters of the coherent states \eqref{eq:CSlqc} in the quantum theories corresponding to the cells $V_o^{(1)}$ and $V_o^{(2)}$ are related as (see also \cite{CorichiCoherentSemiclassicalStates,Corichi:2011sd})
\be
(\alpha'^{(1)},P_\alpha'^{(1)},\sigma^{(1)})\longmapsto (\alpha'^{(2)},P_\alpha'^{(2)},\sigma^{(2)})=\left(\frac{V_o^{(1)}}{V_o^{(2)}}\alpha'^{(1)},P_\alpha'^{(1)},\frac{V_o^{(1)}}{V_o^{(2)}}\sigma^{(1)}\right)\;.
\ee
The states \eqref{eq:CSlqc} will thus remain sharply peaked around the same point only if $\frac{V_o^{(1)}}{V_o^{(2)}}\sim1$.~It should be noticed though that the expectation value $\langle\widehat{\alpha(V_o)}\rangle_{\Psi}$ of the smeared volume operator $\widehat{\alpha(V_o)}=\widehat{\text{vol}(V_o)}=V_o\,\hat\alpha$, the physical observable corresponding to the volume of the cell, does not depend on $V_o$ (cfr.~Eqs.~\eqref{CSalphaEVscaling} and \eqref{eq:volexpvalue}).


\end{document}